\documentclass[aps,prd,twocolumn,amsmath,showpacs,amssymb,superscriptaddress,nofootinbib,longbibliography,eqsecnum,preprintnumbers]{revtex4-1}
\usepackage{graphicx}
\usepackage{bm}
\usepackage{amssymb,amsmath}
\usepackage{mathrsfs}
\usepackage{braket} 
\usepackage{latexsym}
\usepackage{color}
\usepackage{float}
\usepackage[normalem]{ulem} 
\usepackage{dcolumn}
\usepackage[colorlinks=true,citecolor=blue,urlcolor=blue]{hyperref}
\usepackage[usenames,dvipsnames]{xcolor}

\allowdisplaybreaks

\newcommand{\Caltech}{\affiliation{Theoretical Astrophysics 350-17, California Institute of Technology, Pasadena, CA 91125}}
\newcommand{\CITA}{\affiliation{Canadian Institute for Theoretical Astrophysics, 60 St. George Street, Toronto, ON, M5S 3H8 Canada}}

\newcommand{\ba}{\begin{align}}
\newcommand{\ea}{\end{align}}

\newcommand{\lm}{_{\ell m}}
\newcommand{\bma}{\begin{pmatrix}}
\newcommand{\ema}{\end{pmatrix}}

\newcommand{\Rb}{\tilde{\mathcal R}}
\newcommand{\TBH}{\tilde{\mathcal T}_{\rm BH}}
\newcommand{\RBH}{\tilde{\mathcal R}_{\rm BH}}
\newcommand{\K}{\tilde{\mathcal K}}
\newcommand{\RW}{\tilde{\mathcal R}_{\rm W}}

\newcommand{\psiup}{\tilde \psi_{\rm up}}
\newcommand{\psiin}{\tilde \psi_{\rm in}}

\newcommand{\psiecho}{\tilde \psi_{\rm echo}}
\newcommand{\gref}{\tilde g_{\rm ref}}
\newcommand{\gBH}{\tilde g_{\rm BH}}
\newcommand{\ZinfBH}{Z^{\infty}_{\rm BH}}
\newcommand{\ZhBH}{Z^{\rm H}_{\rm BH}}
\newcommand{\Zref}{Z^{\infty}_{\rm ref}}
\newcommand{\Zecho}{Z_{\rm echo}}

\begin{document}

\title{A recipe for echoes from exotic compact objects}

\author{Zachary Mark} \Caltech
\author{Aaron Zimmerman} \CITA
\author{Song Ming Du} \Caltech
\author{Yanbei Chen} \Caltech

\date{\today}

\begin{abstract}
Gravitational wave astronomy provides an unprecedented opportunity to test the nature of black holes and search for exotic, compact alternatives. 
Recent studies have shown that exotic compact objects (ECOs) can ring down in a manner similar to black holes, but can also produce a sequence of distinct pulses resembling the initial ringdown. 
These ``echoes'' would provide definite evidence for the existence of ECOs. 
In this work we study the generation of these echoes in a generic, parametrized model for the ECO, using Green's functions.
We show how to reprocess radiation in the near-horizon region of a Schwarzschild black hole into the asymptotic radiation from the corresponding source in an ECO spacetime.
Our methods allow us to understand the connection between distinct echoes and ringing at the resonant frequencies of the compact object.
We find that the quasinormal mode ringing in the black hole spacetime plays a central role in determining the shape of the first few echoes. 
We use this observation to develop a simple template for echo waveforms.
This template preforms well over a variety of ECO parameters, and with improvements may prove useful in the analysis of gravitational waves.
\end{abstract}

\preprint{LIGO-P1700145}


\maketitle

\section{Introduction}

The existence of event horizons is one of the most astonishing predictions of General Relativity.  
Horizons generically \cite{Thorne:1972ji} form during the gravitational collapse of classical matter and are expected to be common occurrences in our universe. 
Observations of black holes are undergoing a revolution, with the advent of gravitational wave astronomy \cite{Abbott:2016blz,Abbott:2016nmj,TheLIGOScientific:2016pea,LIGO:GW170104} and the promise of very-long-baseline radio observations of supermassive black holes by the Event Horizon Telescope \cite{Falcke:1999pj,Johnson:2015iwg}.
While black holes are consistent with all electromagnetic and gravitational wave observations to date \cite{TheLIGOScientific:2016src,Yagi:2016jml,Yunes:2016jcc,TheLIGOScientific:2016pea,LIGO:GW170104}, no experiment has been able probe spacetime near the event horizon \cite{Eckart:2017bhq,Abramowicz:2002vt,CardosoReview}.
Moreover, the event horizon is at  the heart of the BH information paradox \cite{Unruh:2017uaw}, and the role of black holes in a quantum theory of gravity is an open question. 

These puzzles have inspired proposals for horizonless alternatives to black holes including gravastars \cite{Mazur:2001fv}, boson stars \cite{Schunck:2003kk}, wormholes \cite{Morris:1988tu}, fuzzballs \cite{Mathur:2005zp} and others \cite{Barcelo:2010vc, Barcelo:2014cla,Holdom:2016nek}. 
Many of these exotic compact objects (ECOs) can be ruled out on theoretical grounds.
ECOs with angular momentum often suffer from a superradiant instability, although this instability can quenched by tuning the compactness and other parameters describing the ECO \cite{Maggio:2017ivp,Hod:2017cga}. Cardoso et al.~\cite{Cardoso:2014sna} have conjectured that any ECO with an unstable photon orbit may suffer from nonlinear instabilities.

While the gravitational wave astronomy has the potential to probe black holes (BHs) like never before \cite{Yagi:2016jml}, distinguishing BHs from highly compact ECOs will be difficult. 
The problem is that astrophysical processes are usually insensitive to the spacetime geometry near the horizon, and highly compact ECOs behave very similarly to BHs \cite{Abramowicz:2002vt}.
Attempts to distinguish merging BHs from merging ECOs using inspiral waveforms are plagued by the strong equivalence principal, which means that  the properties of extended self-gravitating bodies only appear in the  equations of motion at high post-Newtonian order. 
Nonetheless, several promising studies \cite{Cardoso:2017cfl,Maselli:2017cmm} predict tidal distortion and tidal heating effects will allow LISA \cite{AmaroSeoane:2012km} to distinguish merging black holes from highly compact, merging ECOs (see also e.g.~\cite{Pani:2010em,Krishnendu:2017shb} for tests incorporating inspirals). 

Spacetime near the event horizon has an especially interesting effect on the ringdown waveform of the merging objects. 
Standard tests of the nature of the final merged object call for the black hole's resonant frequencies \cite{Nollert, Berti2009}, known as quasinormal mode (QNM) frequencies, to be extracted from the ringdown portion of the waveform and compared to theoretical calculations \cite{Dreyer2004,Berti:2005ys,Pani:2013pma,Berti:2016lat,Yang:2017zxs}. Working in the test particle limit, Cardoso et al.~\cite{Cardoso:2016rao} pointed out that in the case of highly compact wormholes, the ringdown of the final ECO is initially nearly identical to that of a BH  despite the fact that QNM spectrum is radically changed \cite{Ching:1995rt, Nollert:1998ys,Pani:2009ss}. 
A naive application of the QNM based tests would be fooled by a highly compact ECO.   

However, Cardoso et al.~\cite{Cardoso:2016rao} also realized that the later portion of the ringdown of highly compact ECOs contains a train of decaying echo pulses. 
The time delay between the echoes is related to the ECO compactness while the decay and shape of each pulse encodes the reflective properties of the ECO. 

Further work established that this picture was robust across many different ECO models with many different test particle sources, but breaks down for less compact ECOs, which sometimes have ringdowns consistent with the resonant frequencies of the ECO \cite{Cardoso:2016oxy,Price:2017cjr}.
Price and Khanna conjectured that the echoes can be considered as a superposition of the resonant modes of the ECO \cite{Price:2017cjr}. 
Volkel and Kokkaotas \cite{Volkel:2017kfj} then provided a method for inferring the exact details of the ECO model from the ECO modes. 
Namely, they demonstrated that the effective scattering potential experienced by the gravitational waves could be approximately reconstructed with a knowledge of ECO spectrum.

Recently, it has been proposed that LIGO has observed echoes in the binary black hole waveforms \cite{Abedi:2016hgu, Abedi:2017isz}. While there has been much skepticism in the community \cite{Ashton:2016xff}, such tests will only become more definitive as LIGO accumulates binary merger observations. 

Most of the past studies have been in the context of a particular ECO model, using specific orbits for the merging objects.
The goal of this work is explicitly relate waveforms from  black holes to waveforms from  ECOs.
We study evolution of test scalar fields as a proxy for gravitational perturbations, which allows us to replace a generic ECO with simple reflecting boundary conditions in a BH spacetime.
We use this formalism to show that the ECO waveform can be understood either as a superposition of echo pulses or as a superposition of ECO modes and illustrate the types of behavior that can arise. 
We investigate which features of the BH waveforms shape the first few echoes, leading to a simple template for the ECO waveform.

In Sec.~\ref{sec:wavebc} we review the basic equations obeyed by the scalar field. 
We parameterize (completely) the influence of the ECO on scalar waves in the exterior vacuum region by a complex frequency-dependent reflectivity (a slight generalization of the models used in \cite{Maggio:2017ivp,Hod:2017cga,Nakano:2017fvh}). 
In Sec.~\ref{sec:ECOvsBH} we relate the ECO and BH waveform by determining the relationship between the ECO and BH Green's function. 
We find that the ECO waveform can be constructed from the BH waveform and a reprocessed version of the waveform observed on the BH horizon.
In Sec.~\ref{sec:Echoes} we show how the extra piece of the ECO waveform can be expressed as sum of echoes.
In Sec.~\ref{sec:ECOModes} we discuss the relationship between the ECO QNMs and the BH QNMs and study the ECO mode spectrum numerically for two particular ECO models. In Sec.~\ref{sec:SingleMode} and Sec.~\ref{sec:EchoInt}, we show how the difference between the ECO waveform and the BH waveform can be expressed as a superposition of ECO modes. 
In Sec.~\ref{sec:Gen} we determine general properties of the individual echoes and develop a simple template for the ECO waveform.
We also study the energy in the ECO waveform, discovering a simple relationship to the energy in the black hole waveforms reaching infinity and passing through the horizon. 

During the final stages of this work, we learned of the work of Nakano et al.~\cite{Nakano:2017fvh}, who discussed gravitational perturbations in the Kerr spacetime and arrived at a similar expression for ECO waveforms by different means.

\section{Waves near a compact object}

\subsection{Wave Equation and Boundary Conditions}
\label{sec:wavebc}

We focus on static, spherically symmetric exotic compact objects.
In this setting, an ECO consists of an exterior Schwarzschild spacetime patched to a spherically symmetric interior metric at an areal radius $r=r_0$.

We study a massless scalar field $\Phi(x^\mu)$ that obeys the sourced, curved spacetime wave equation,
\begin{align}
\Box \Phi = - \rho \,.
\end{align}
If we define the scalar $\psi (x^\mu) = r \Phi$ and decompose this scalar into frequency and spherical harmonics \cite{Casals:2015oaa},
\begin{align}
\psi (x^\mu) &= \int_{-\infty}^{\infty} \frac{d\omega}{2\pi}\sum_{\ell, m} \tilde \psi_{\ell m}(\omega,r) Y_{\ell m}(\theta, \phi)e^{-i\omega t} \,, \\
\rho (x^\mu) &= \int_{-\infty}^{\infty} \frac{d\omega}{2\pi}\sum_{\ell, m} \tilde \rho_{\ell m}(\omega,r) Y_{\ell m}(\theta, \phi)e^{-i\omega t} \,,  
\end{align}
then the wavefunctions $\tilde \psi_{\ell m}$ obey the following radial equation, 
\begin{align}
\label{eq:RW}
\frac{d^2 \tilde \psi\lm}{dx^2} &+\left(\omega^2-f V \right) \tilde \psi\lm= \tilde S \,, \\
\tilde S(\omega,x) & \equiv - r(x) f \rho\lm(\omega,x) \,.
\end{align}
Here $x$ is the usual tortoise coordinate, defined through
\begin{align}
\frac{dx}{dr} & = \frac{1}{f(r)} \,,
\end{align}
while the metric component $f(r)$ and the potential $V(r)$ depend on the particular spacetime.
In the exterior, Schwarzschild portion of the spacetime,
\begin{align}
f & = 1 - \frac{2M}{r} \,, & V & = \frac{\ell(\ell +1)}{r^2} + \frac{2M}{r^3}\,, 
\end{align}
and we treat $f$ and $V$ as implicit functions of $x$ through $r(x)$, with 
\begin{align}
x = r + 2M\ln\left(\frac{r-2M}{M}\right)\,.
\end{align}
From here we suppress the harmonic indices $(\ell,m)$.

The scalar field $\tilde \psi$ obeys an outgoing wave boundary condition $\tilde \psi \sim e^{i\omega x}$ as $x\to \infty$. 
In addition, it obeys a boundary condition inside the ECO, such as regularity at $r=0$. 
For wormholes, one would instead insist that the waves were outgoing at null infinity on the other side of the throat.

When the ECO is very compact, $r_0/(2M)-1 \ll 1$, and all sources are restricted to reside in the Schwarzschild portion of the spacetime, we may replace the second boundary condition with a reflecting boundary condition at the ECO surface $r_0$. Namely, near the ECO the potential is small, $V\approx 0$, and $\tilde \psi$ is a linear combination of ingoing and outgoing waves $e^{\pm i\omega x}$. Therefore near the ECO surface $x_0 = x(r_0)$, we must have
\begin{align}
\label{eq:refBC}
\tilde \psi \propto e^{-i\omega (x-x_0)}+\Rb(\omega) e^{i\omega (x-x_0)} \,.
\end{align}
for some frequency dependent reflectivity $\Rb(\omega)$.

\begin{figure}[t]
\includegraphics[width =0.98 \columnwidth]{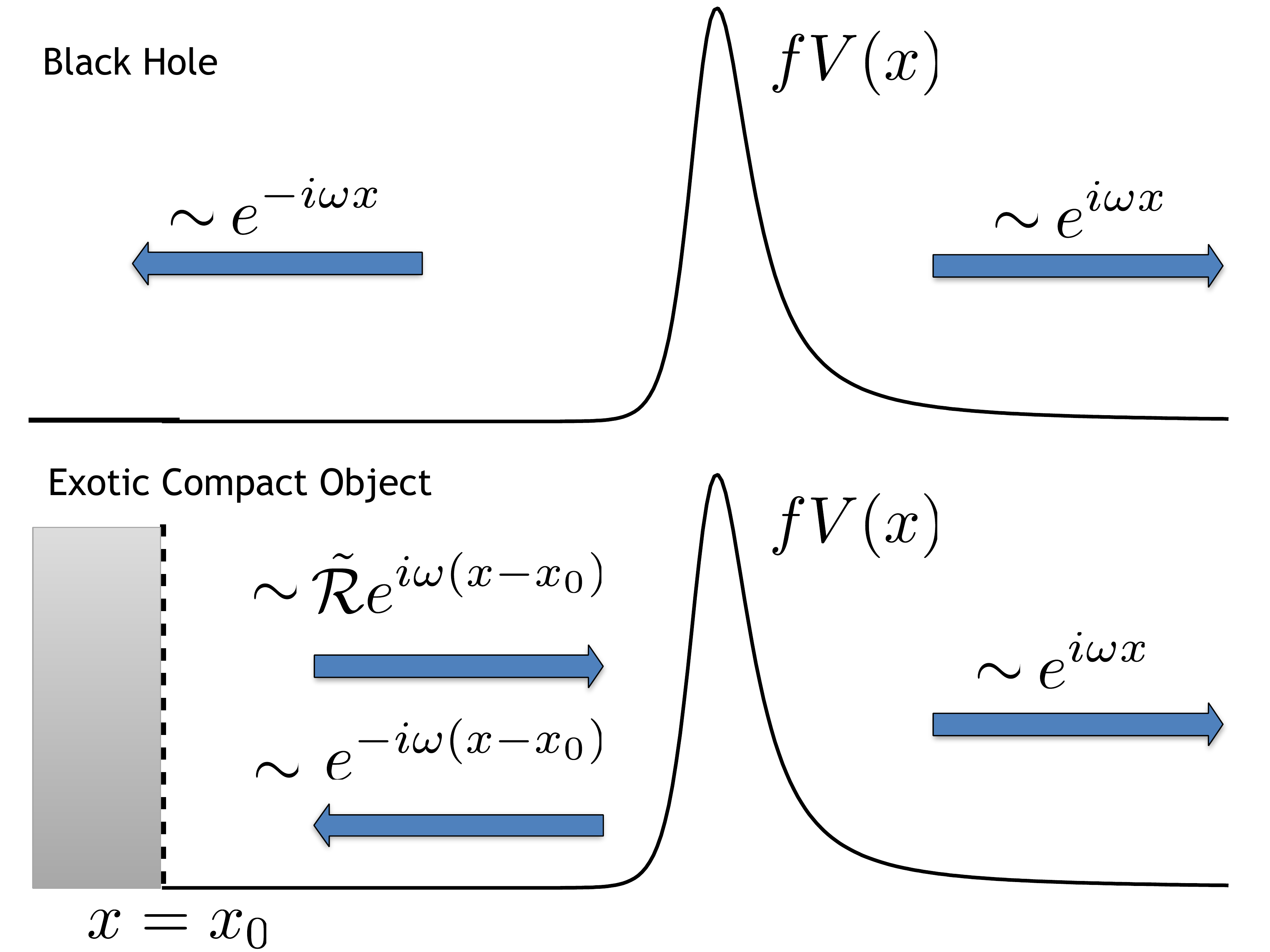}
\caption{Top: The boundary conditions for waves propagating on a black hole spacetime. Bottom: The reflecting boundary conditions for the waves in the exterior of an ECO.
}
\label{fig:BCs}
\end{figure}

With this insight, we can study wave emission and propagation in the ECO spacetime using a Schwarzschild BH equipped with a reflecting boundary, as shown in Fig.~\ref{fig:BCs}.
This perspective is useful since it allows us to reprocess the emission by test particles in a BH spacetime into the corresponding emission in the ECO spacetime, by taking the reflecting boundary into account. From here on we can focus on BH spacetimes, and compare wave propagation with the usual boundary conditions at the horizon to the case of a reflecting boundary. 

\subsection{Generating ECO waveforms from BH waveforms}

\label{sec:ECOvsBH}

We are interested in computing the scalar waves seen by distant observers in a BH spacetime with a reflecting boundary.
For this we wish to construct the scalar radial Green's function $\gref(x,x')$, which obeys the scalar wave equation with a delta function source,
\begin{align}
\label{eq:RWGF}
\frac{d^2 \gref}{dx^2} + \left( \omega^2 - f V \right) \gref = \delta(x-x') \,,
\end{align}
and the reflecting boundary condition \eqref{eq:refBC}.
With the Green's function, we can compute the field produced by sources $\tilde S$ through integration,
\begin{align}
\label{eq:Sourcedpsi}
\tilde \psi(x) & = \int_{-\infty}^{\infty} dx' \, \gref(x,x') \tilde S(x')\,.
\end{align}
We compute $\gref$ for sources outside the reflecting boundary, $x'>x_0$.

To compute $\gref$ we first recall how the scattering of waves works in the usual Schwarzschild spacetime \cite{Frolov:1998wf}.
Consider the two linearly independent, homogeneous solutions $\psiin$,
\begin{align}
\psiin & \sim 
\left \{
\begin{array}{ll}
A_{\rm out}(\omega) e^{i \omega x} + A_{\rm in}(\omega) e^{-i\omega x}\,, & x \to \infty \,, \\
e^{-i\omega x}\,, & x \to - \infty \,, \\
\end{array} \right. \label{eq:psiin}
\end{align}
which is purely outgoing at the horizon, and $\psiup$,
\begin{align}
\psiup & \sim 
\left \{
\begin{array}{ll}
e^{i\omega x} \,, & x \to \infty \,, \\
B_{\rm out}(\omega) e^{i \omega x} + B_{\rm in}(\omega) e^{-i\omega x}\,, & x \to - \infty \,, \\
\end{array} \right. \label{eq:psiup}
\end{align}
which is purely outgoing at infinity.

The effective potential $V$ provides a scattering barrier for waves in the BH spacetime. 
For waves incident from infinity, inspection of $\psiin$ shows that the reflection amplitude is $A_{\rm out}/A_{\rm in}$ and the transmission amplitude is $1/A_{\rm in}$.
For our purpose, it is more convenient to consider the problem of reflection and transmission of waves incident on $V$ from the left. By inspecting $\psiup$ we find that the reflection and transmission amplitudes for waves from the left are 
\begin{align}
\label{eq:RTAmplitudes}
\RBH (\omega) & = \frac{B_{\rm in}}{B_{\rm out}} \,, & \TBH(\omega) & = \frac{1}{B_{\rm out}} \,.
\end{align}
The relationship between these and the usual reflection and transmission amplitudes can be derived by noting that $B_{\rm out} = A_{\rm in}$ and $B_{\rm in} = - A_{\rm out}^*$ \cite{Frolov:1998wf} .

The Green's function for Schwarzschild, $g_{\rm BH}(x,x')$, also obeys Eq.~\eqref{eq:RWGF}, but with an ingoing boundary condition at the horizon and an outgoing boundary condition at infinity.
In terms of the homogeneous solutions, it is
\begin{align}
\gBH =& \frac{\psiin(x_<) \psiup(x_>)}{W_{\rm BH}} \,, \label{eq:gBH}
\end{align}
where we have defined $x_> = \max (x,x')$, $x_< = \min (x,x')$, and the Wronskian $W_{\rm BH} = 2 i \omega B_{\rm out}$ of $\psiin$ and $\psiup$.

Since $\gBH$ and $\gref$ both obey Eq.~\eqref{eq:RWGF}, we can construct $\gref$ by adding a homogenous solution of the scalar equation, times a free function of $x'$, to $\gBH$.
The homogenous solution must have the correct boundary condition as $x \to \infty$, and so we use $\psiup(x)$.
Meanwhile, the free function in $x'$ is fixed by ensuring that $\gref$ obeys the correct reflecting boundary condition,
\begin{align}
\gref(x,x') \propto e^{-i\omega (x-x_0)}+\Rb(\omega) e^{i\omega (x-x_0)} \,.
\end{align}
This gives
\begin{align}
\label{eq:Gref}
\gref(x,x') & = \gBH(x,x') + \K \, \frac{\psiup(x) \psiup(x')}{W_{\rm BH}} \,,\\
\label{eq:EchoTransfer}
\K(\omega) & \equiv \frac{\TBH \Rb e^{-2 i \omega x_0}}{1 - \RBH \Rb e^{-2 i \omega x_0}} \,.
\end{align}
This is our first key result.
It shows that wave propagation in the presence of the reflecting barrier is the same as in a BH spacetime, with an additional component controlled by the transfer function $\K$, which contains all the dependence on the reflectivity $\Rb$. 

With the Green's function in hand, we can compute the waves seen by distant observers.
Again it is useful to first consider a BH spacetime with the usual boundary conditions.
We define the amplitudes of waves seen by distant observers $\ZinfBH$ and of waves at the horizon $\ZhBH$ through
\begin{align}
\tilde \psi_{\rm BH}(x) & \sim 
\left \{
\begin{array}{ll}
\ZinfBH(\omega) e^{i \omega x} \,, & x \to \infty \,, \\
\ZhBH(\omega) e^{-i\omega x}\,, & x \to - \infty \,. \\
\end{array} \right. \label{eq:ZBHdef}
\end{align}
In terms of a given source $\tilde S$ with support outside $x_0$, Eqs.~\eqref{eq:Sourcedpsi}, \eqref{eq:psiup} and \eqref{eq:gBH} imply 
\begin{align}
\ZinfBH & = \int_{-\infty}^\infty dx' \frac{\psiin(x') \tilde S(x')}{W_{\rm BH}} \,, \\
\label{eq:ZH}
\ZhBH & = \int_{-\infty}^\infty dx' \frac{\psiup(x') \tilde S(x')}{W_{\rm BH}} \,.
\end{align}
With our definitions, $\ZinfBH$ is simply related to the waveform measured by asymptotic observers in terms of the retarded time $u = t-x $,
\begin{align}
\psi^{\infty}_{\rm BH} (u) = \int_{-\infty}^{+\infty} \frac{d\omega}{2\pi} \ZinfBH e^{-i\omega u} \,.
\end{align}
Similarly, in terms of the advanced time $v= t+x$, the waveform at the BH horizon is the Fourier conjugate to $\ZhBH$,
\begin{align}
\psi^{\rm H}_{\rm BH}(v) & =  \int_{-\infty}^{+\infty} \frac{d\omega}{2\pi} \ZhBH e^{-i\omega v} \,. \label{eq:ZFT}
\end{align}

Having defined these amplitudes, in the presence of the reflecting boundary we can use $\gref$ from Eq.~\eqref{eq:Gref} in Eq.~\eqref{eq:Sourcedpsi} to compute the asymptotic amplitude associated with scalar waves $\tilde \psi$,
\begin{align}
\tilde \psi & \sim \Zref e^{i \omega x} \,, & x & \to \infty \,. \label{eq:Zrefdef}
\end{align}
We find that
\begin{align}
\label{eq:Zinf}
\Zref & = \ZinfBH + \K \ZhBH \,.
\end{align}
This is our second key result.
It shows that the waveform seen by distant observers can be understood as the sum of the usual emission in a BH spacetime, along with an additional signal $\K \ZhBH$.
This additional emission arises from the reflection of the radiation which would normally enter the horizon, but is reprocessed by the transfer function $\K$.
The power of Eq.~\eqref{eq:Zinf} is that is allows us to compute the total asymptotic waveform in and ECO spacetime from the corresponding waveforms observed near infinity and the horizon in a BH spacetime, given a particular choice of $\Rb$ and $x_0$.

\begin{figure}[t]
\includegraphics[width = 1.0 \columnwidth]{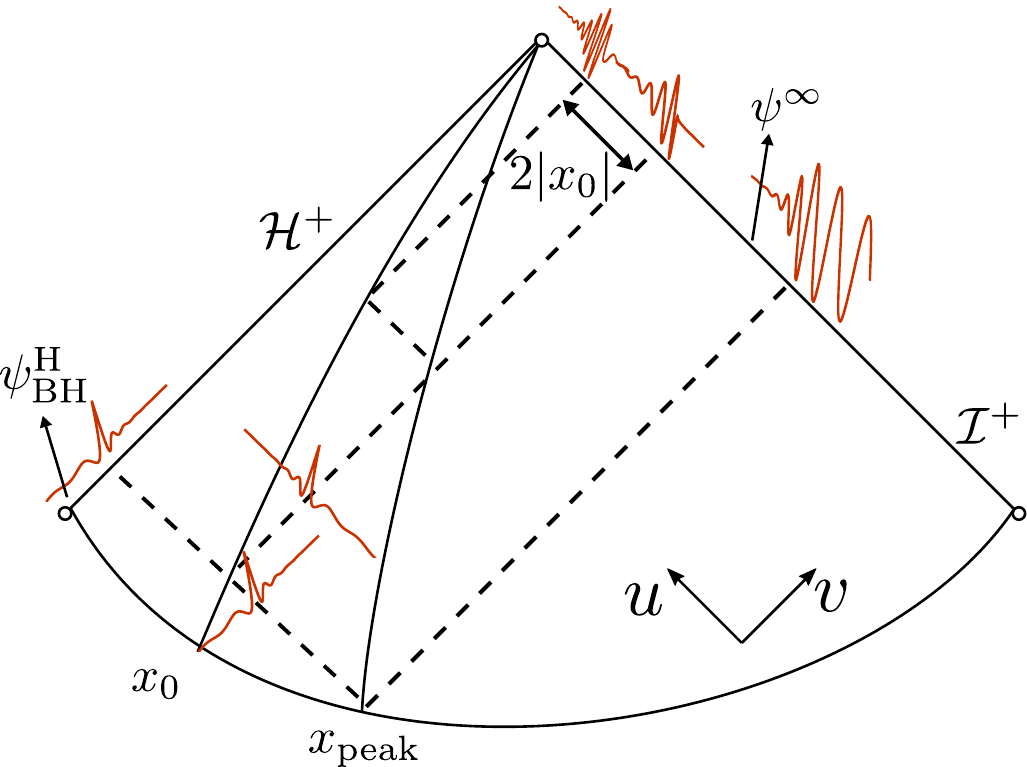}
\caption{A conformal diagram illustrating the production of echoes. The waveform that impinges on the reflecting boundary at $x_0$ is approximately the same as the waveform that reaches the horizon in the BH spacetime, $\psi^{\rm H}_{\rm BH}(v)$. Repeated partial reflections between $x_0$ and the peak of the potential $x_{\rm peak}$ result in an asymptotic waveform $\psi^\infty(u)$ made up of a main burst followed by echoes. Each echo is a reprocessed version of the waveform on the horizon $\psi^{\rm H}_{\rm BH}(v)$.}
\label{fig:CausalDiagram}
\end{figure}

We gain further insight into the nature of the additional emission by expanding $\K$ as a geometric series,
\begin{align}
\K & = \TBH \Rb e^{-2i \omega x_0} \sum_{n=1}^{\infty} (\RBH \Rb)^{(n-1)} e^{-2i(n-1)\omega x_0 } \,. \label{eq:Kgeo}
\end{align}
This shows that the additional signal takes the form of a series of terms, each reprocessing the waves that impinge on the boundary with a different transfer function. 
As we show in Sec.~\ref{sec:Echoes}, in many circumstances each term in this sequence results in a distinct pulse. 
Figure~\ref{fig:CausalDiagram} illustrates the propagation of the echoes on a conformal diagram.
The first term is the result of the primary reflection of $\psi^{\rm H}_{\rm BH}$ off of the boundary at $x_0$, which generates a factor of $\Rb $ along with a phase factor $ 2 i \omega x_0$. 
The phase factor corresponds to a time delay between the first pulse and the main burst due the pulse's extra round trip journey between the boundary at $x_0$ and the peak of the scattering potential $V$ at $x_{\rm peak}\approx 0$. 
When the pulse reaches the potential barrier, it is partially transmitted, contributing the final factor of $\TBH$.

The successive terms are ``echoes'' of this first reflection which bounce an integer number of times between the potential barrier, contributing a factor of $\RBH$, and the reflecting boundary, contributing a factor of $\Rb$, before transmitting through the potential barrier with an additional propagation delay.
Note that while the precise propagation delay of each pulse depends on the phases of $\TBH$, $\RBH$, and generically $\Rb$, the delay between echoes is constant starting with the second echo.
With this picture in mind, we define the difference between the waveform and the corresponding BH waveform to be the echo amplitude
\begin{align}
\Zecho & = \K \ZhBH \,.
\end{align}

Meanwhile, we can also consider the entire transfer function $\K$ given in Eq.~\eqref{eq:EchoTransfer}.
This function possesses its own set of resonances, and there is a complementary perspective where the waves propagating towards the reflecting boundary excite the modes of a resonant cavity between the boundary and potential barrier.
We discuss this perspective in Sec.~\ref{sec:ECOExcitation}.

\section{Examples of Echoes}
\label{sec:Echoes}

In this section we illustrate the reprocessing of the horizon waveform $\psi^{\rm H}_{\rm BH}$ using two simple examples: a spacetime with a frequency independent reflectivity $\Rb$ and a wormhole spacetime.
We show that the additional waves appear as a sequence of echoes when the boundary is far from the peak of the potential barrier, but this behavior is lost for boundaries closer to the peak.

\subsection{Individual echoes}

\begin{figure}[t]
\includegraphics[width =1 \columnwidth]{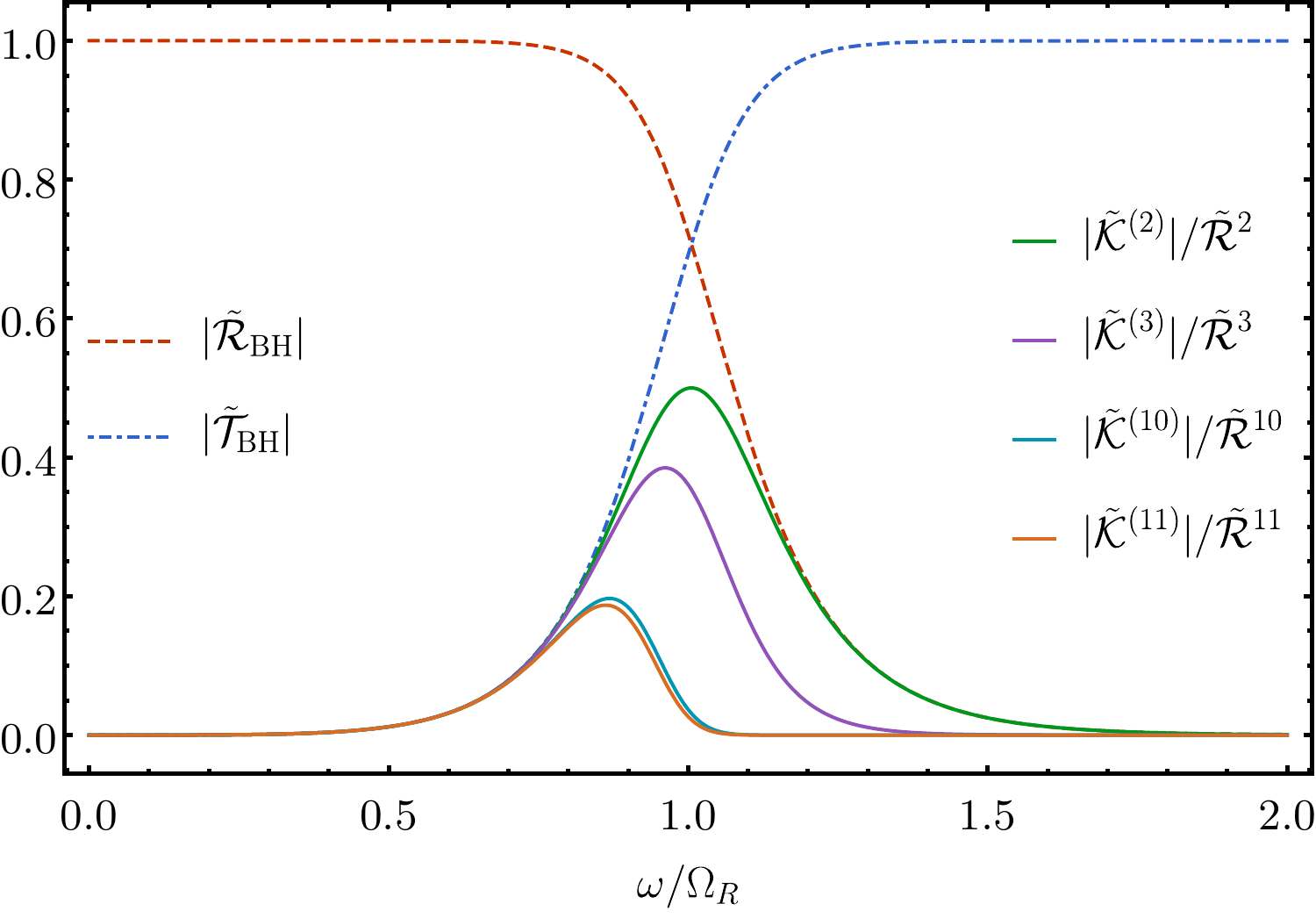}
\caption{
The frequency domain $\ell =2$ black hole reflectivity $|\RBH|$ and transmissivity $|\TBH|$. We also plot the magnitude of the rescaled transfer functions $|\K^{(n)}|/\Rb^n$ for a boundary with constant reflectivity, for $n=2,3,10$ and $11$. 
}
\label{fig:ConstantRtransfer}
\end{figure}

The picture of successive echoes is made even more apparent by working in the time domain.
The waveform seen by distant observers is determined through $\Zref$ by
\begin{align}
\psi^\infty(u) & =  \int_{-\infty}^{\infty} \frac{d\omega}{2\pi} \Zref e^{-i\omega u} 
= \psi^{\infty}_{\rm BH}(u) + \psi_{\rm echo}(u)\,, \\
\psi_{\rm echo}(u) & \equiv \int_{-\infty}^{\infty} \frac{d\omega}{2\pi}   \K \ZhBH e^{-i\omega u}  \,,
\end{align}
where we have denoted the additional waveform due to the reflecting boundary $\psi_{\rm echo}$.
For understanding the echoes, it is useful to further split $\psi_{\rm echo}= \sum_n \psi_{\rm echo}^{(n)}$ into contributions $\psi_{\rm echo}^{(n)}$ from each term in Eq.~\eqref{eq:Kgeo} for $\K$,
\begin{align}
&\psi_{\rm echo}^{(n)}(u)  \equiv \int_{-\infty}^{+\infty} \frac{d\omega}{2\pi}\K^{(n)}\ZhBH e^{-i\omega u}  \,, \\
\label{eq:Kn}
&\K^{(n)}(\omega)\equiv(\TBH \Rb)(\RBH \Rb)^{(n-1)}e^{-2i\omega x_0 n} \,,
\end{align}
which are defined in terms of transfer functions $\K^{(n)}$ for each echo. 

In the time domain, the reflection and transmission amplitudes are given by response functions
\begin{align}
\mathcal R_{\rm BH}(t) = & \int \frac{d\omega}{2\pi} \, \RBH(\omega) e^{-i \omega t} \,,
\end{align}
and similarly for $\mathcal T_{\rm BH}(t)$, $\mathcal R(t)$, and $\mathcal{K}(t)$.

To derive the expression for the echoes, recall that multiplication of two functions $\tilde f(\omega)$ and $\tilde g(\omega)$ in the frequency domain corresponds to convolution $(f*g)$ in the time domain, where
\begin{align}
(f*g)(t) = \int_{-\infty}^{\infty} d\tau f(t - \tau) g(\tau)\,.
\end{align}
With this notation the first echo is 
\begin{align}
\psi_{\rm echo}^{(1)}(u) & = [\mathcal{K}^{(1)} *\psi^{\rm H}_{\rm BH}](u)  
\nonumber \\&
=  [(\mathcal T_{\rm BH} * \mathcal R)* \psi^{\rm H}_{\rm BH}](u + 2 x_0),
\end{align}
where $\mathcal K^{(n)}$ is the Fourier conjugate to $\K^{(n)}$, $\psi^{\rm H}_{\rm BH}$ is the Fourier conjugate to $\ZhBH$, and recall that $x_0$ is negative for boundaries near the horizon.
For the successive echoes,
\begin{align}
\psi_{\rm echo}^{(n)}(u) =&
[\mathcal{K}^{(n)} *\psi^{\rm H}_{\rm BH}](u)
\nonumber \\
=& [(\mathcal T_{\rm BH} * \mathcal R) * (\mathcal R_{\rm BH} * \mathcal R) * \dots 
\notag \\ & 
* (\mathcal R_{\rm BH} * \mathcal R) * \psi^{\rm H}_{\rm BH}](u + 2nx_0) \,. \label{eq:echon}
\end{align}
where there are $n-1$ convolutions of $(\mathcal R_{\rm BH} * \mathcal R)$ with $\psi^{\rm H}_{\rm BH}$.

\begin{figure}[t]
\includegraphics[width =1.0 \columnwidth]{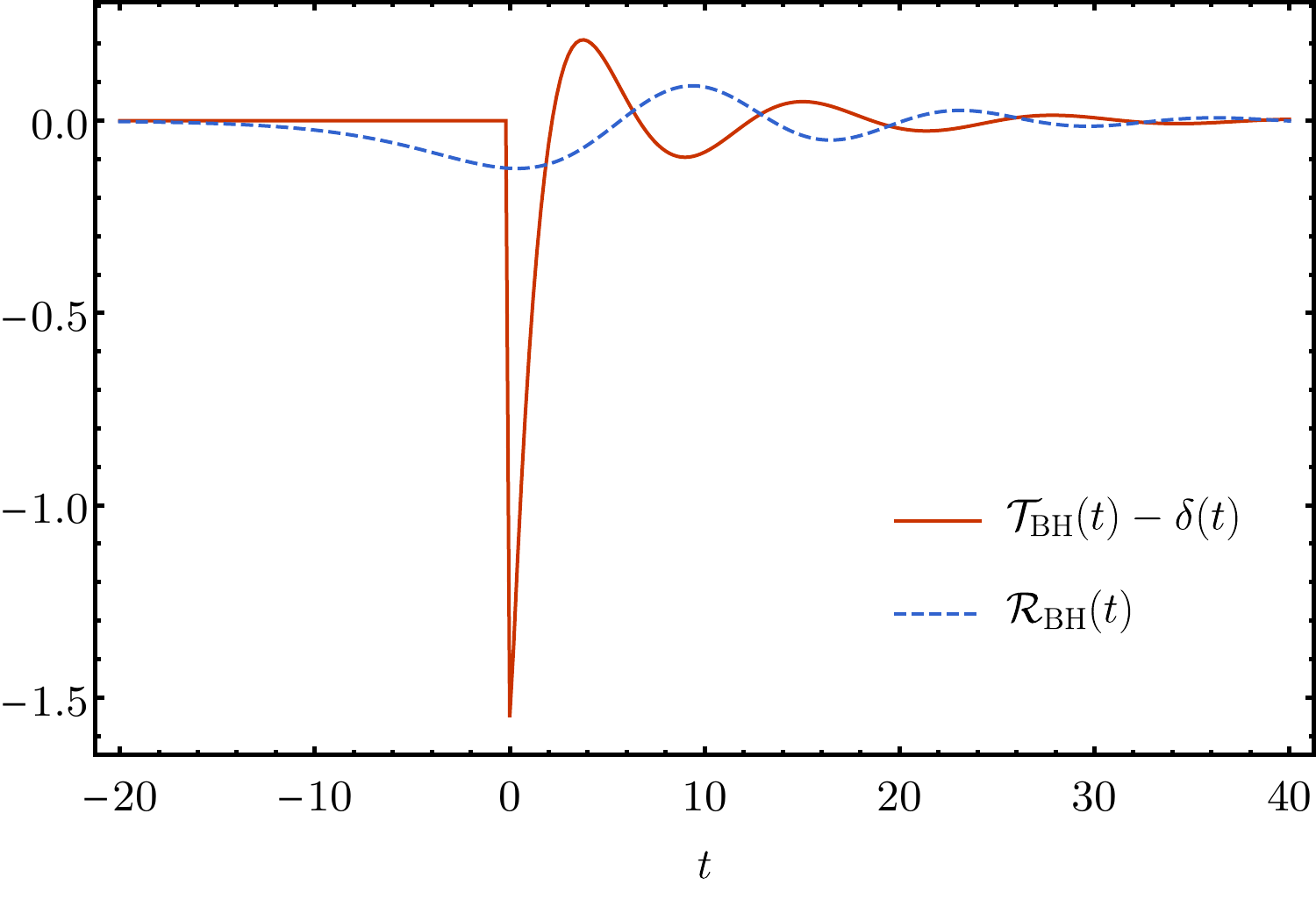}
\caption{ The $\ell =2$ scalar reflectivity and transmissivity of the potential barrier, calculated numerically in the time domain.}
\label{fig:RTtimeDom}
\end{figure}

We calculate the BH response functions $\mathcal R_{\rm BH}$ and $\mathcal T_{\rm BH}$ both in the time and frequency domain using numerical methods described in Appendix \ref{sec:RTcalc}.
The blue and red dashed curves in Fig.~\ref{fig:ConstantRtransfer} show $\RBH$ and $\TBH$ in the frequency domain for the $\ell =2$ scalar wave equation\footnote{
From their definitions,  $\TBH=1/B_{\rm out}$ and $\RBH=B_{\rm in}/B_{\rm out}$ possess resonances (poles) at the complex BH QNM frequencies \cite{Berti:2009kk}; however these resonances do not manifest themselves as clearly separated peaks on the real $\omega$ axis since the width of the QNM resonances is large compared to their spacing.
}. 
As expected \cite{Caponthesis,wheeler1972magic,Frolov:1998wf}, at low frequencies compared to the size of the potential peak $(M\omega)^2 \ll V_p$, waves are completely reflected,
\begin{align}
&|\TBH(\omega) |\to 0,& &|\RBH(\omega) |\to 1,
\end{align}
while at high frequencies $(M\omega)^2 \gg V_p$ waves are completely transmitted
\begin{align}
&\TBH(\omega) \to 1,& &|\RBH(\omega) |\to 0,
\end{align}
The transition between the two regimes occurs at approximately the real part of the $\ell = 2$ fundamental BH QNM frequency  
\begin{align}
M\Omega = M\Omega_R + i M\Omega_I \approx 0.48-0.10i \,,
\end{align}
since $V_p\approx (M\Omega_R)^2$.

Figure \ref{fig:RTtimeDom} shows $\mathcal R_{\rm BH}$ and $\mathcal T_{\rm BH}$ in the time domain. Both response functions ring down at the BH QNM frequency $\Omega$. 
As is explained in the appendix, the high frequency behavior for $\TBH$ implies that in the time domain $\mathcal T_{\rm BH}(t)$ contains a $\delta(t)$ singularity at $t=0$, which is subtracted off in the figure.

Using the echo response functions computed from $\TBH$ and $\RBH$, we now study the echo morphology from a variety of ECOs.
When presenting numerical results, we use units so that the mass of the BH spacetime is unity, $M=1$, and when we discuss a particle with scalar charge $q$ we also set $q = 1$.

\subsection{Frequency Independent Reflectivity}
\label{sec:FIechos}

The simplest type of boundary condition in this model is a frequency independent reflectivity $\Rb$.  In this case, the echoes have a straightforward dependence on the ECO parameters $\mathcal \Rb$ and $x_0$. 
The reflectivity factors out of the response functions $\mathcal K^{(n)}$ and controls the size of each echo, without contributing any phase factors. 
Thus the majority of the time delay between echoes is due to the phase $2\omega x_0$, corresponding to a round trip journey from the potential peak near $x\approx 0$ and the boundary at $x_0$, with only  a small contribution from the BH scattering coefficient $\RBH$. 

The shape of each echo is described by the rescaled response functions
\begin{align}
\label{eq:ShiftedEchoResponse}
e^{2i\omega x_0 n}\K^{(n)}(\omega)/\Rb^n = \TBH (\omega)\RBH(\omega) ^{(n-1)} \,,
\end{align}
which we show in Fig.~\ref{fig:ConstantRtransfer}.
Recall that $|\TBH|$ is approximately zero low frequencies and approximately one at large frequencies, while the opposite is true for $|\RBH|$.
This behavior produces a small window of frequencies where the second echo response function is nonzero. 
The third echo response comes from the multiplying the second echo response function by $\RBH$; this results in a smaller slightly shifted window of frequencies. This pattern repeats with each subsequent response function. However, as the window shifts to the left, $|\RBH|\to 1$ and so the change in absolute value of the transfer functions slows, so that there is very little difference between 10th and 11th echoes.

\begin{figure}[t]
\includegraphics[width=0.98\columnwidth]{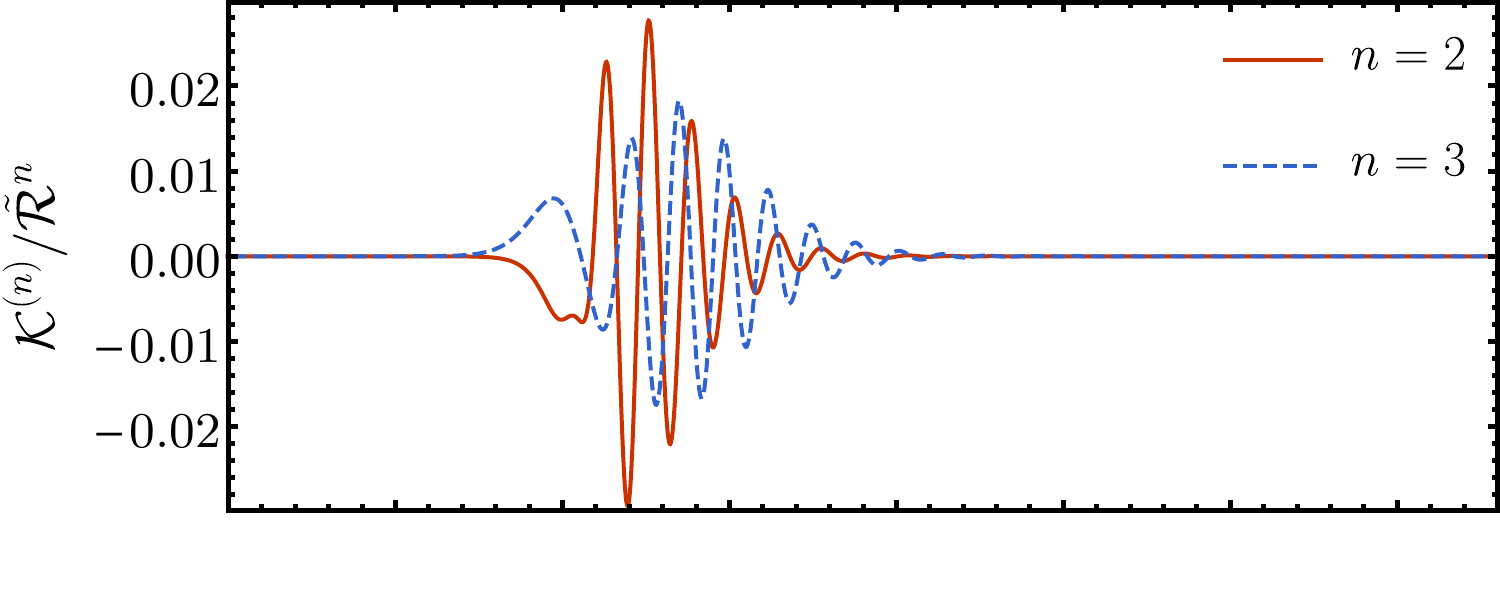}\\
\vspace{-14.5pt}
\includegraphics[width=0.98\columnwidth]{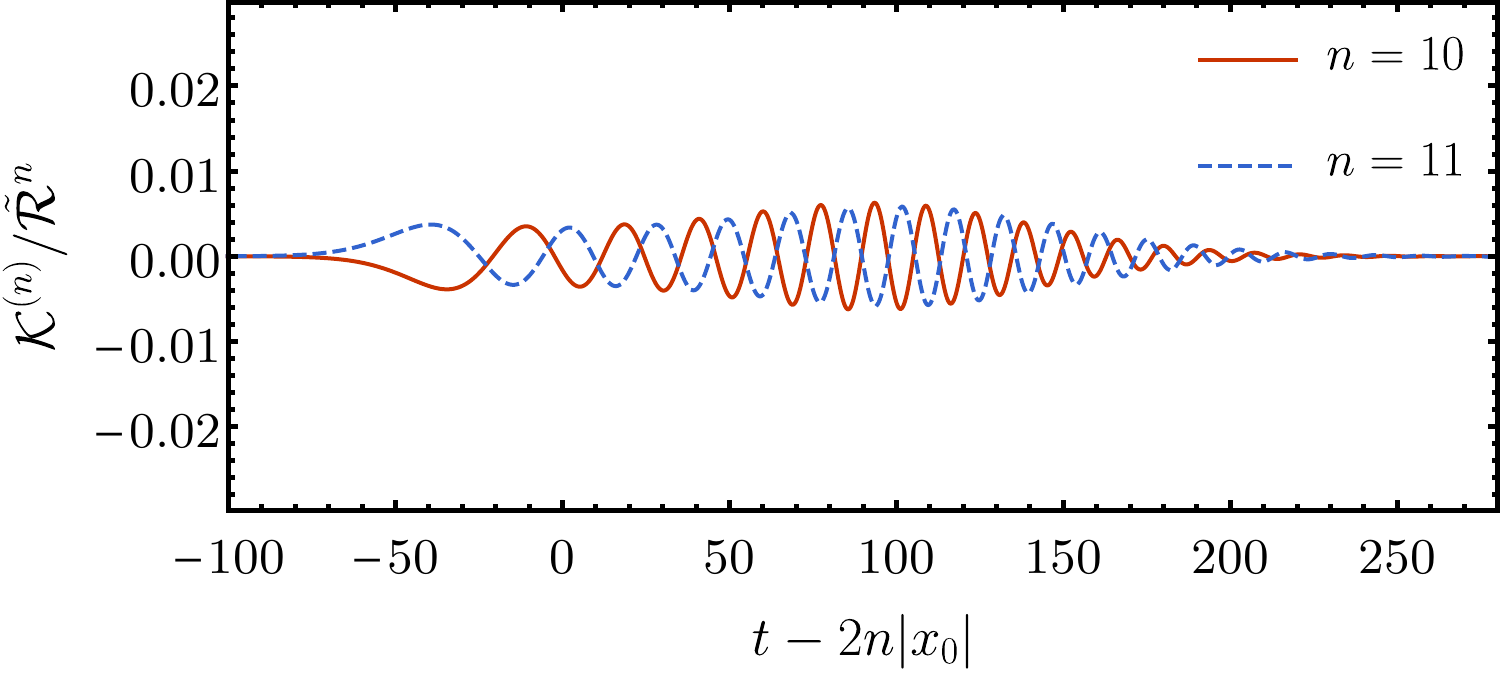}
\caption{
The constant reflectivity $\ell = 2$ echo response functions $\mathcal K^{(n)}$ for $n=2$ and 3 (top) and $n=10$ and 11 (bottom). We divide the response functions by $\Rb^n$ to rescale them and time shift each by $2 n |x_0|$ so they overlap.
}
\label{fig:Echotransfertime}
\end{figure}

In the time domain, the rescaled response functions in Eq.~\eqref{eq:ShiftedEchoResponse} are time shifted to remove the delay between echoes due to the factor of $e^{2i\omega x_0 n}$. Figure \ref{fig:Echotransfertime} shows the rescaled and shifted time domain echo response functions, obtained by numerically performing the convolutions on $\mathcal T_{\rm BH}$ and $\mathcal R_{\rm BH}$.
Each transfer function goes to zero at early times and is a decaying sinusoid at late times. The complex frequency of the sinusoid is nearly the fundamental QNM frequency $\Omega$ for the first few echoes, while for later echoes the decay time gets longer and the oscillation frequency gets slightly smaller. 

Similar trends are seen in the echoes themselves.
The waveforms at both infinity and on the horizon depend on our particular choice of sources and initial data.
As an illustration throughout this paper, we consider the echoes produced by a test particle with unit scalar charge following an orbit that we refer to as the ISCO plunge orbit. 
This orbit is a geodesic that spirals inward from the innermost stable circular orbit (ISCO), with the ISCO energy and angular momentum, and reaches the horizon at an advanced time $v_{\rm H}$.
We select this orbit since it is a reasonable model for the ringdown portion of the scalar waveform for orbits that have been circularized prior to reaching the ISCO radius, by a mechanism such as radiation reaction \cite{Hadar:2009ip}.
We use a numerical Green's function to generate the waveform from this source, which we subsequently window at early times so it smoothly starts from zero. 
Details on the entire procedure are found in Appendix~\ref{sec:Numerics}.

Since our method is to reprocess waveforms from BH spacetimes, our formalism cannot capture the emission in an actual ECO spacetime after the particle passes $x_0$.
Namely, Eq.~\eqref{eq:Gref} for $\gref$ can only be used when the source is outside $x_0$, but we use Eq.~\eqref{eq:Gref} for all source locations.
Using a particular ECO model, this additional radiation could be added directly to our waveforms, with only a small remaining inaccuracy due to the suppressed emission in our waveforms as the particle travels from $x_0$ to the horizon. 

\begin{figure}[t]
\includegraphics[width = 1 \columnwidth]{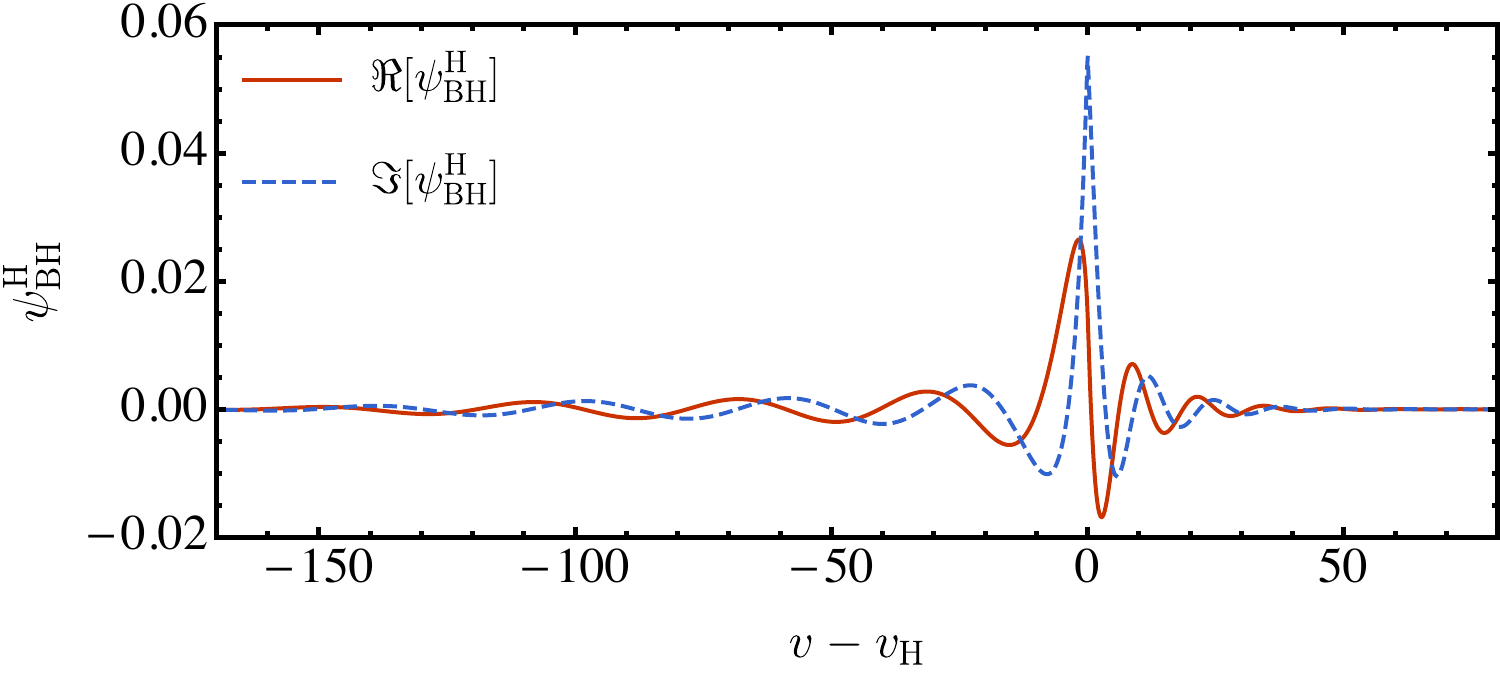}\\
\includegraphics[width = 1 \columnwidth]{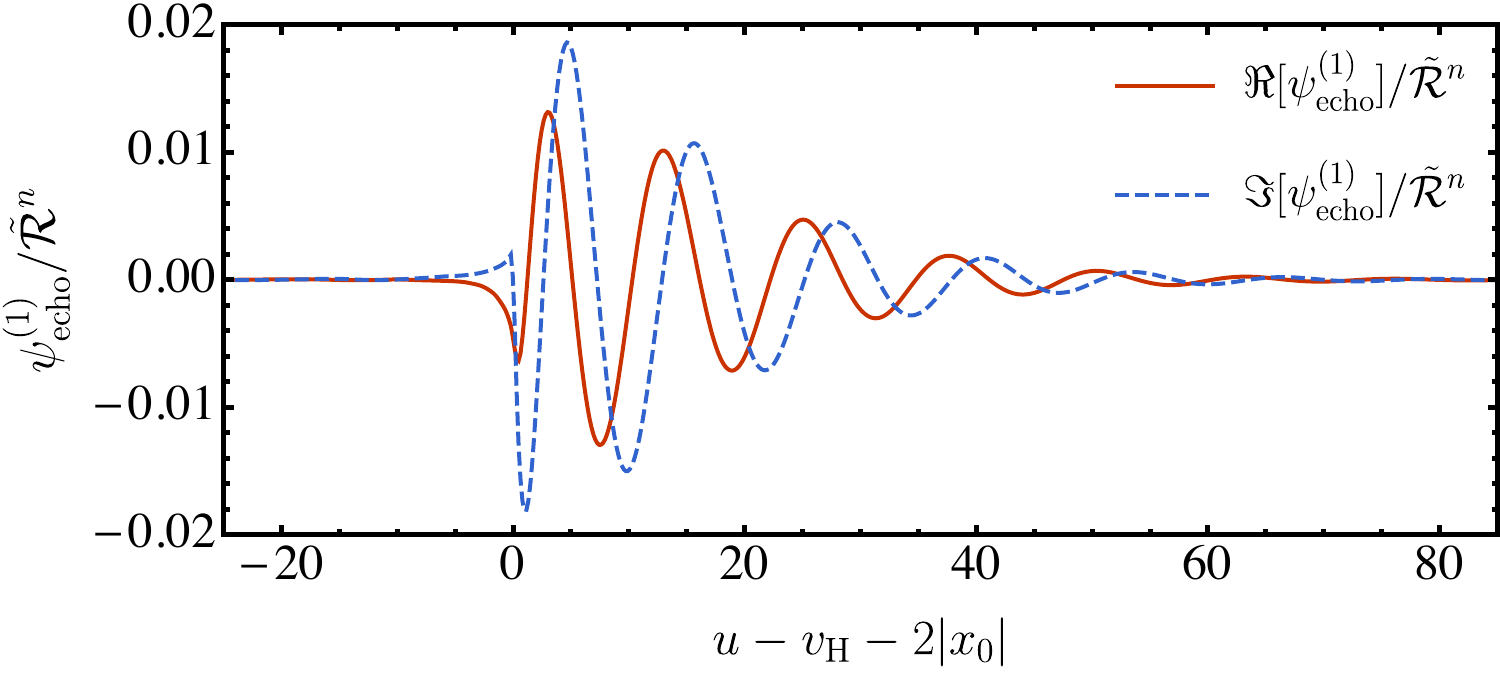}
\caption{
Top: The $(\ell,m)=(2,2)$ waveform on the horizon $\psi_{\rm BH}^{\rm H}$, as produced by a test charge following the ISCO plunge orbit.
Bottom: The corresponding first echo $\psi_{\rm echo}^{(1)}$, rescaled and shifted in time, for a frequency-independent reflectivity.
}
\label{fig:Echotime}
\end{figure}

\begin{figure}[t]
\includegraphics[width = 1 \columnwidth]{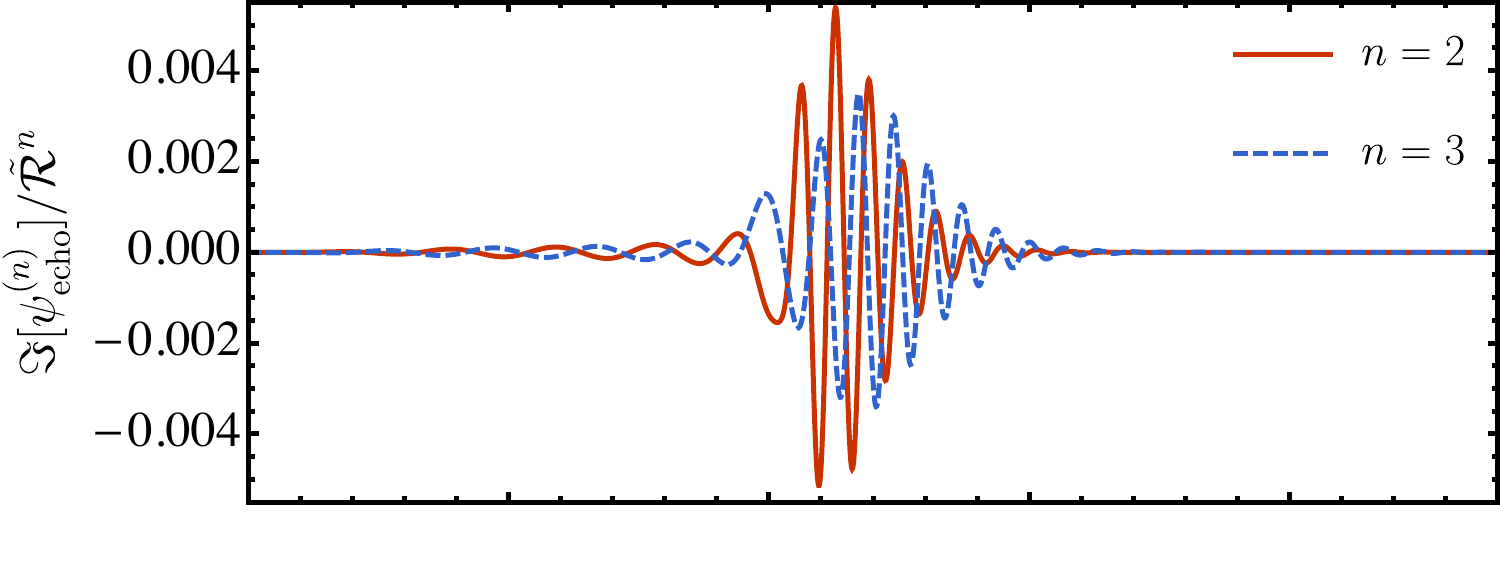}\\
\vspace{-11.5pt}
\includegraphics[width = 1 \columnwidth]{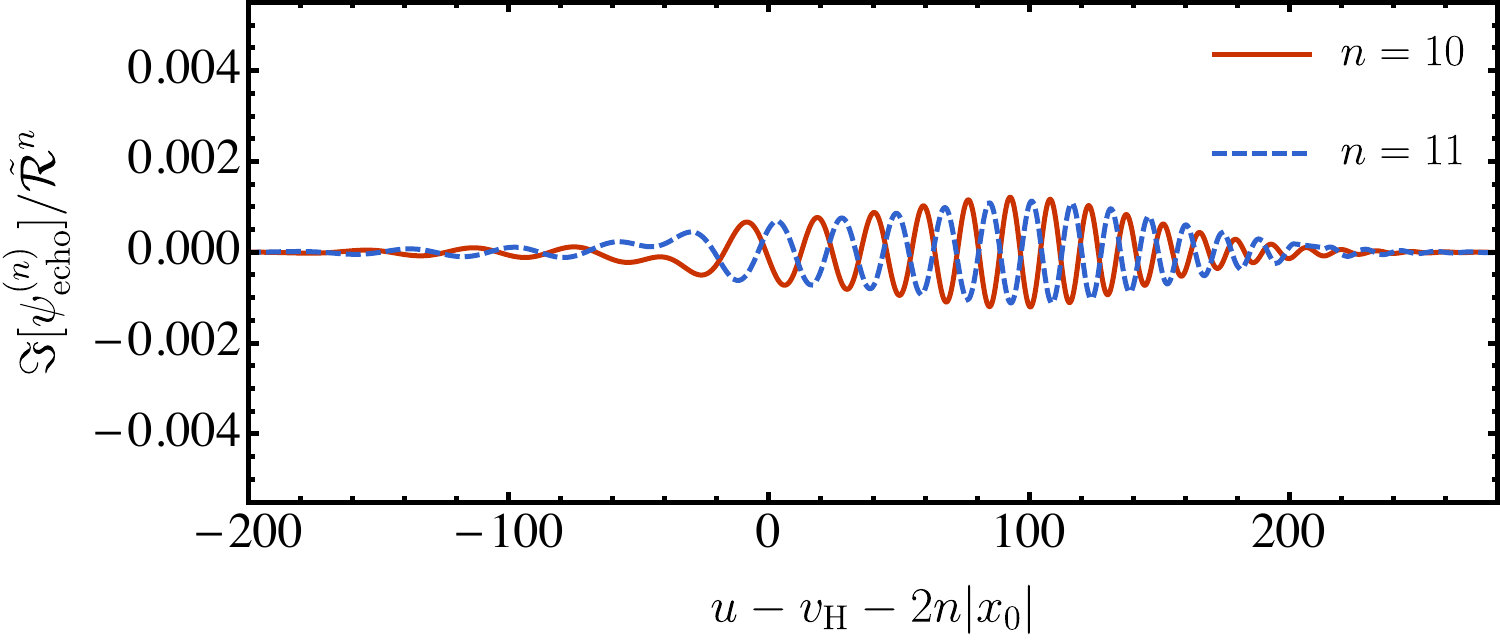}
\caption{
The $(\ell,m)=(2,2)$ echoes for a frequency independent reflectivity $\Rb$. The source is a test charge following the ISCO plunge orbit. We show the imaginary part of each echo, rescaled by $\Rb^n$ and shifted in time to overlap.
Top: The second and third echoes.
Bottom: The tenth and eleventh echoes. At this stage, successive echoes change only slightly in duration and amplitude.
}
\label{fig:Echotime2}
\end{figure}

Figures \ref{fig:Echotime} and \ref{fig:Echotime2} show the $(\ell,m)=(2,2)$ horizon waveform and select echoes in the time domain from the ISCO plunge.
At early times the horizon waveform frequency is $\omega=m\Omega_{\rm ISCO}$, where $\Omega_{\rm ISCO}$ is the ISCO orbital frequency, and at late times there is a ringdown at the fundamental BH QNM frequency. 
The echoes also display a highly suppressed oscillation at $\omega \approx m\Omega_{\rm ISCO}$ at early times and then asymptote to decaying sinusoids at late times. 
The complex frequency of the sinusoid displays the same qualitative behavior as the echo response functions; each echo decays less than the previous and has a slightly lower frequency, with consecutive early echoes differing more than consecutive late echoes. We explore these features in more detail in Sec.~\ref{sec:Gen}.

\subsection{Wormhole}

The echoes from specific ECO spacetimes can also be placed within the reflecting boundary formalism.
Consider for example a wormhole produced by identifying two Schwarzschild spacetimes of mass $M$ at an areal radius $r_0$.  In Appendix \ref{sec:wormdetails}, we show that an observer in one universe can describe the influence of the other universe on wave propagation by a reflecting boundary condition $\tilde \psi \propto \Rb(\omega)e^{i\omega(x-x_0)}+e^{-i\omega (x-x_0)}$ as $x\to x_0$, where
\begin{align}
\Rb(\omega)=\RBH(\omega)e^{-2i\omega x_0} \,.
\label{eq:wormbc}
\end{align}
The free propagation phase $e^{-2i\omega x_0}$ appearing in the reflectivity accounts for the additional delay as the waves propagate to the potential peak in the other universe and back again.

\begin{figure}[t]
\includegraphics[width = 1 \columnwidth]{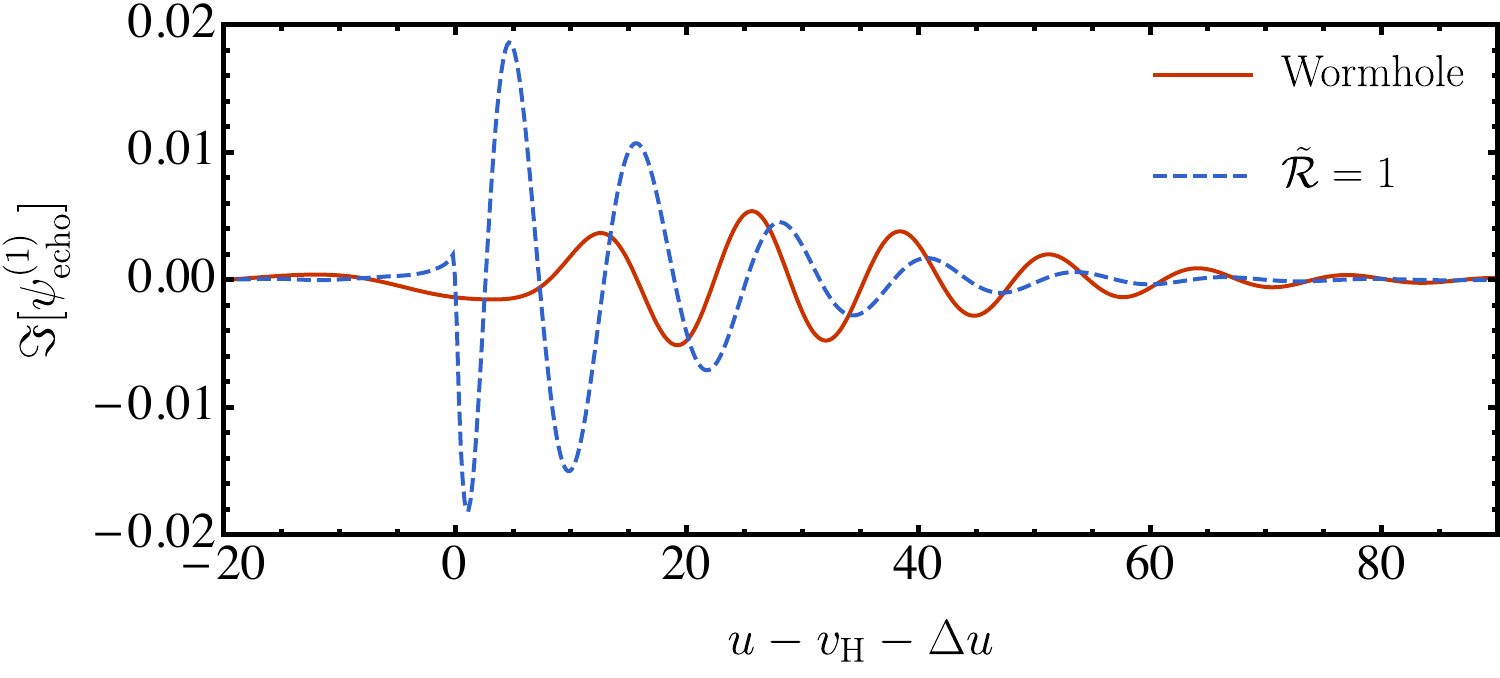} \\
\includegraphics[width = 1 \columnwidth]{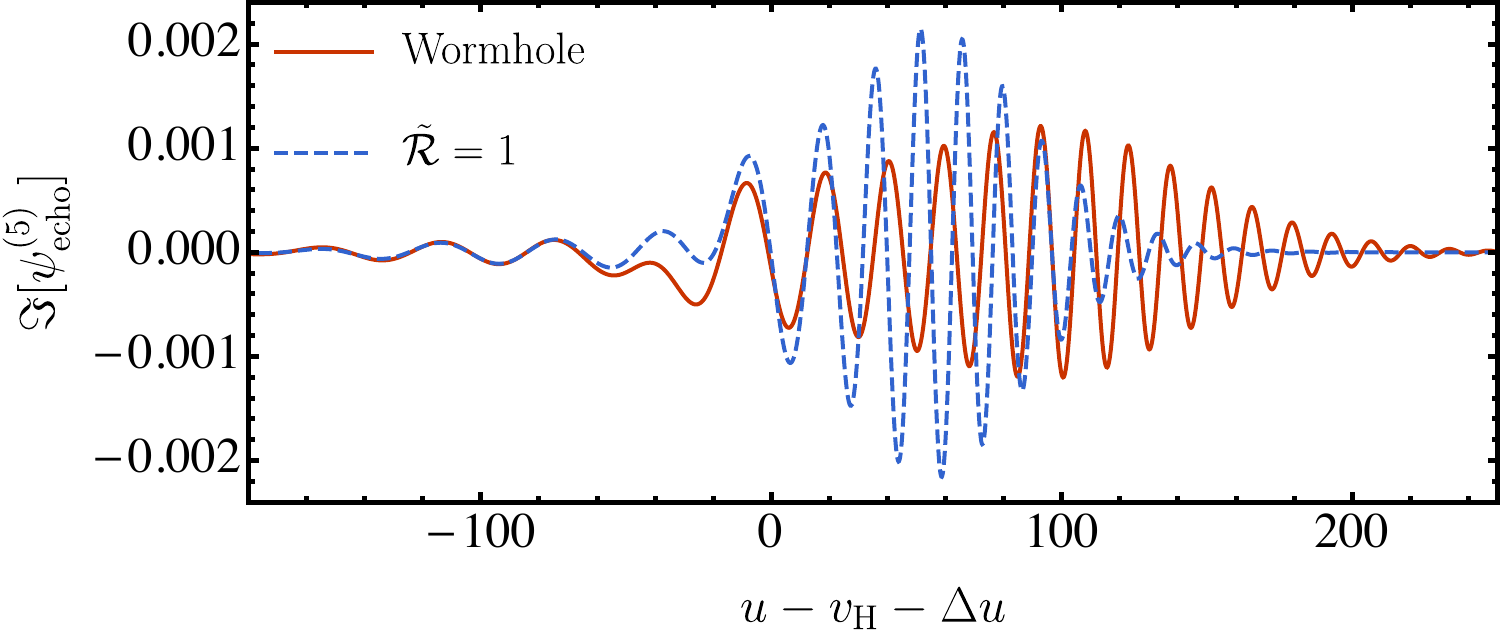}
\caption{
The imaginary part of the $(\ell,m)=(2,2)$ time domain echoes  excited by a test charge following the ISCO plunge orbit in a wormhole spacetime, as compared with the echoes of the $\Rb=1$ reflecting boundary. We plot the first echo (top) and fifth echo (bottom). Each wormhole echo is shifted by $\Delta u = 4 n |x_0|$, while each constant reflectivity echo is shifted by $\Delta u = 2 n |x_0|$.
}
\label{fig:eccom}
\end{figure}

Echoes in the wormhole spacetime are simply related to frequency independent $\Rb=1$ echoes. Namely the $n$th echo in the wormhole spacetime is the $2n$th echo of the $\Rb=1$ case, as can be seen from Eq.~\eqref{eq:Kn}. 
Therefore, the wormhole echoes exhibit the same patterns as the frequency-independent echoes. A comparison of the first echoes and the fifth echoes produced by a test charge following the ISCO plunge orbit is shown in Fig.~\ref{fig:eccom}.

\subsection{Echo interference}

\begin{figure}[t]
\includegraphics[width = 1 \columnwidth]{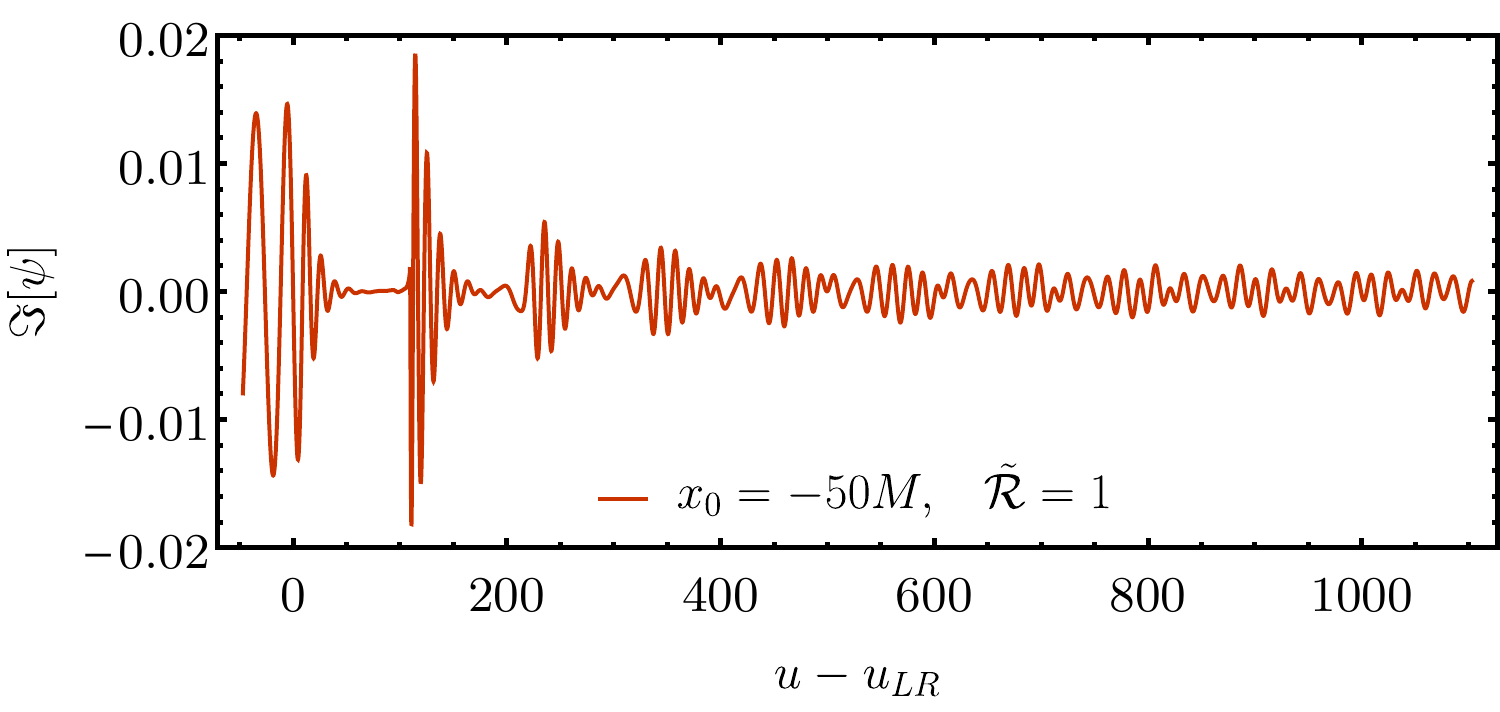} \\
\includegraphics[width = 1 \columnwidth]{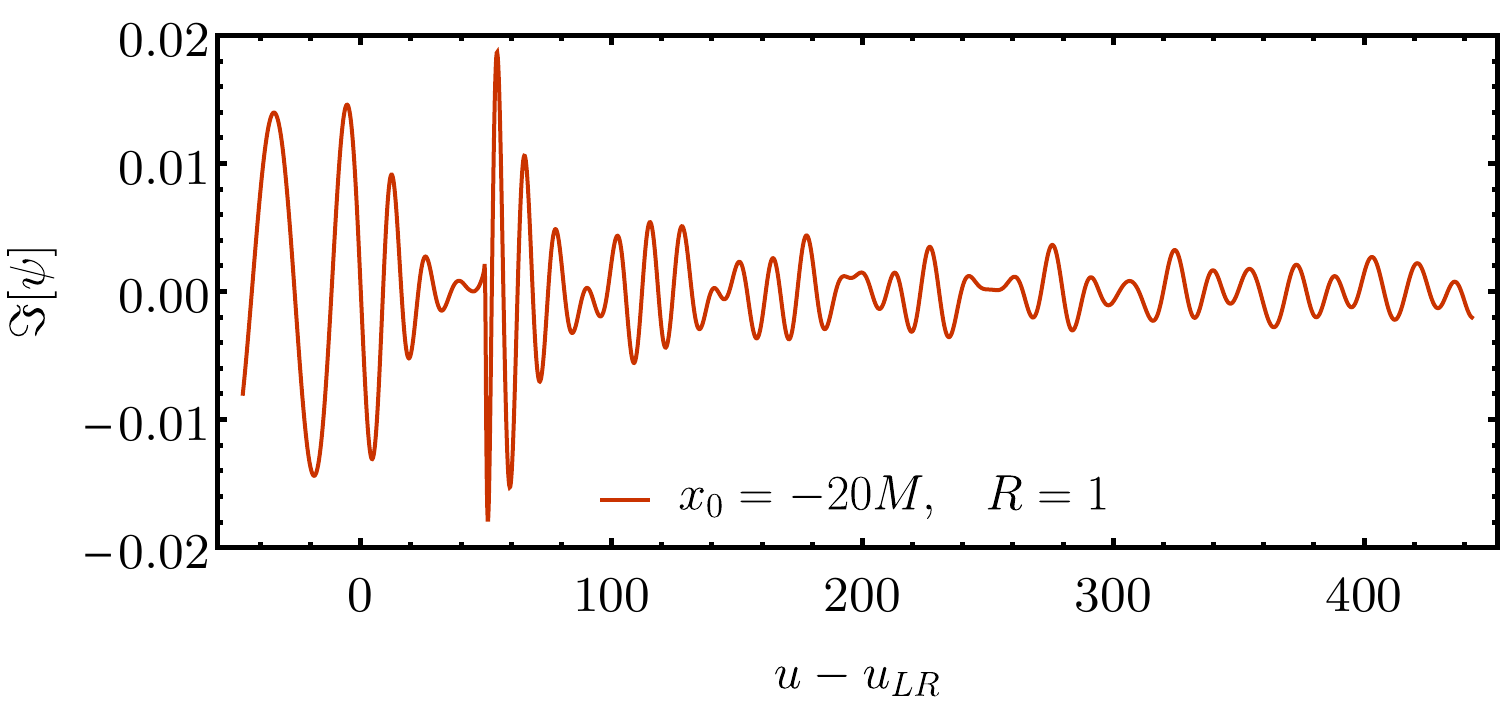} \\
\caption{
The imaginary part of the $(\ell,m)=(2,2)$ total  waveform $\psi^\infty$ excited by test charge following the ISCO plunge orbit. We show results for an ECO with $\Rb=1$ and $x_0 =-50 M$ (top),
and an ECO with $\Rb=1$ and $x_0 =-20 M$ (bottom). We shift the time axis by the retarded time that the charge crosses the spherical photon orbit, $u_{\rm LR}$.
}
\label{fig:Esum1}
\end{figure}

Having explored the individual echo pulses, we now examine the full echo waveform.
When the spacing between echoes is large compared to the duration of each echo, the echoes do not interfere and the total waveform appears as a sum of echo pulses. Figure \ref{fig:Esum1} shows the waveform $\psi^\infty(u)$ generated by the ISCO plunge orbit in the case $\Rb = 1$, truncating the echo sum at $n=11$. We illustrate the $\ell = 2$ waveform for two locations $x_0$ of the boundary.

The top panel shows the total waveform for $x_0=-50 M$.
The first part of the waveform is the BH waveform $\psi_{\rm BH}^\infty$, which initially oscillates at roughly a frequency of $m\Omega_{\rm ISCO}$ and transitions to ringing at the BH QNM frequencies. 
The transition occurs around a retarded time $u_{\rm LR}$, when the particle crosses the light ring.
Roughly $|2x_0|$ later, there are three to four distinct echo pulses, each spaced by roughly $|2x_0|$. 
As we observed earlier, the later echoes decay more slowly and do not appear distinct because they have a long enough duration to interfere with each other. 
The bottom panel shows the case $x_0=-20 M$, where there are only two distinct pulses before the echoes begin to interfere.

\begin{figure}[t]
\includegraphics[width = 1 \columnwidth]{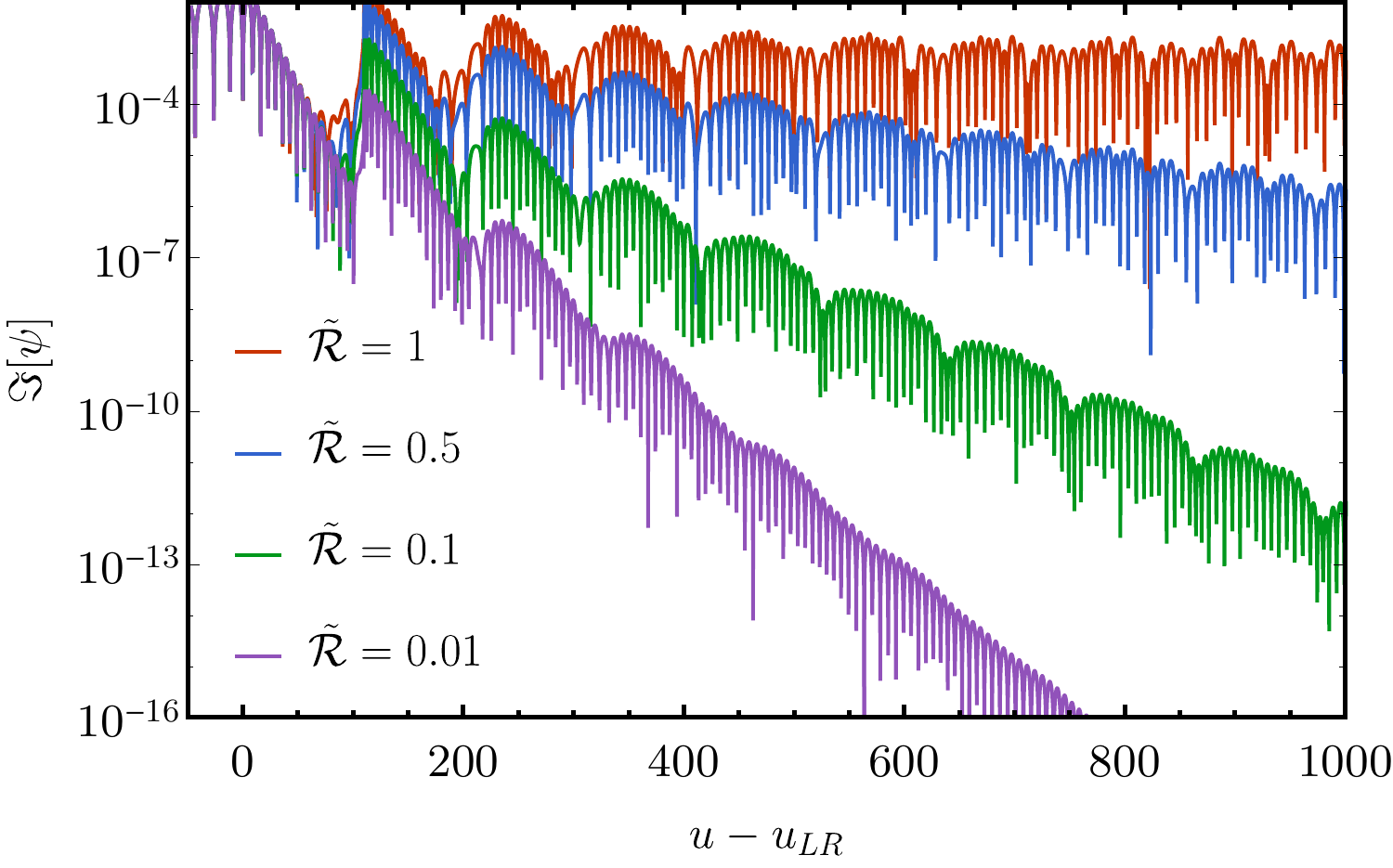}
\caption{The imaginary part of the $(\ell,m)=(2,2)$ total waveform $\psi^\infty$ excited by a test charge following the ISCO plunge orbit.
We show results for ECOs with $x_0 =-50 M$ and several different choices of a frequency independent $\Rb$.}
\label{fig:EsumManyRb}
\end{figure}

We show additional examples in Fig.~\ref{fig:EsumManyRb}, using our ISCO plunge waveform. In this figure, the ECO surface is located at $x_0=-50 M$ and $\Rb$ ranges from $0.01$ to $1$. While only three to four distinct echoes are visible at large $\Rb$, for $
\Rb = 0.1$ we can see many pulses in the rapidly decaying waveform.

\begin{figure}[t]
\includegraphics[width = 1 \columnwidth]{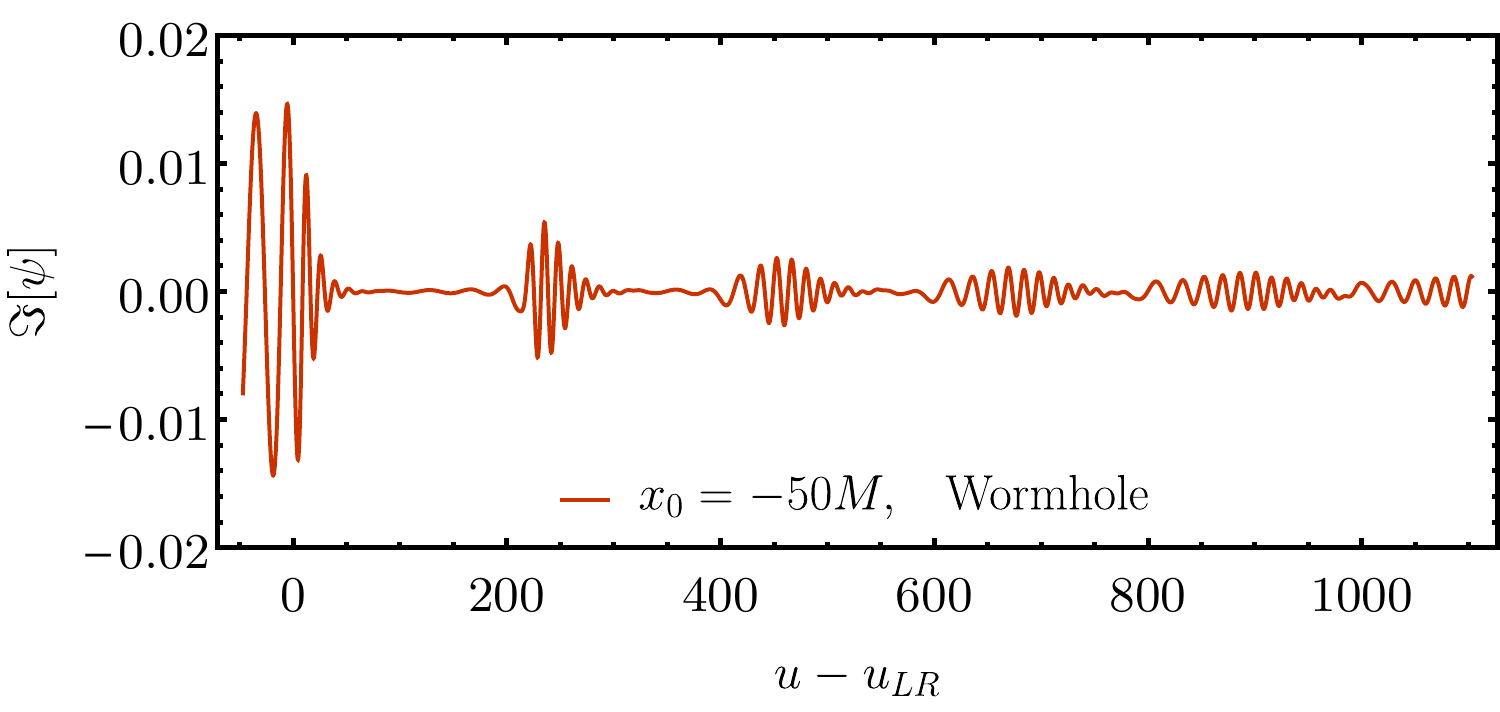} \\
\includegraphics[width = 1 \columnwidth]{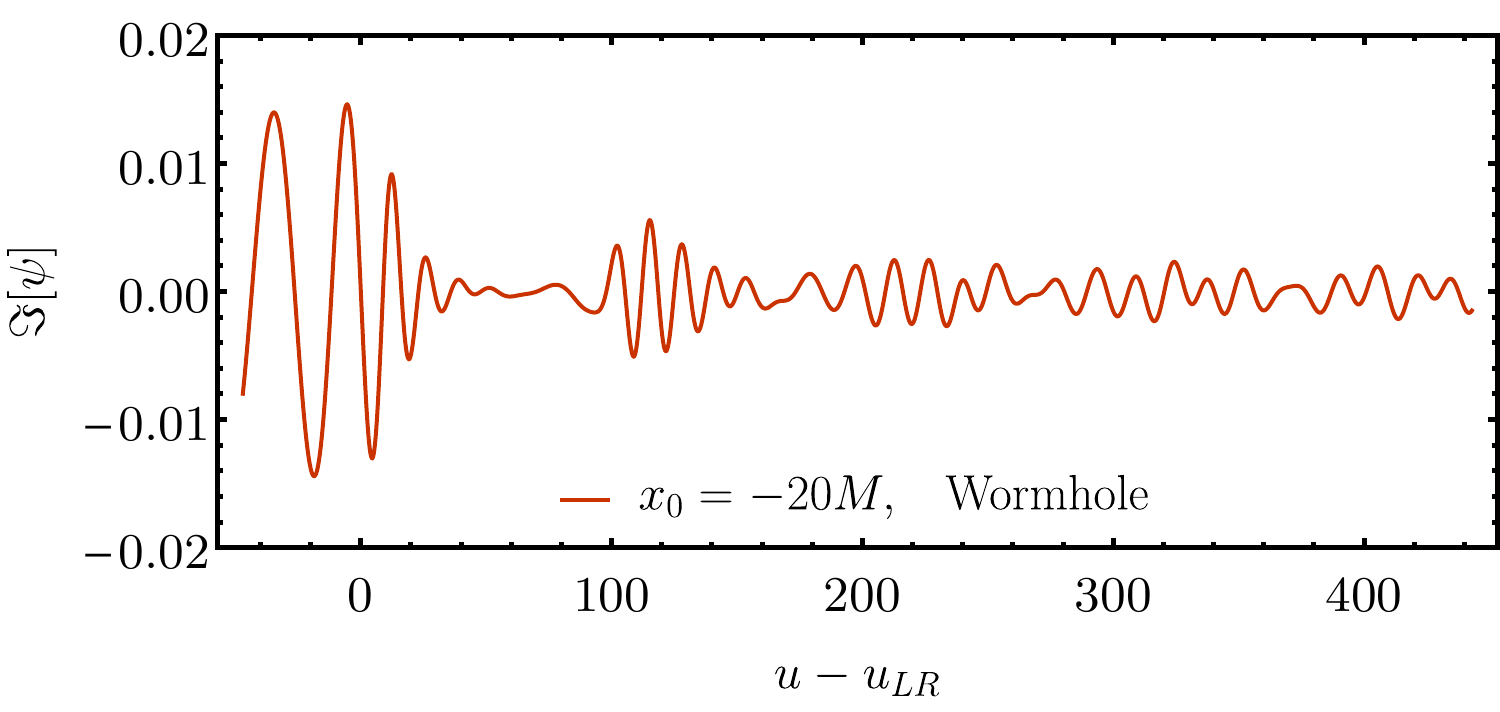} 
\caption{
The imaginary part of the $(\ell,m)=(2,2)$ total waveform $\psi^\infty$ excited by a test charge following the ISCO plunge orbit. We show results for a wormhole with $x_0=-50M$ (top) and $x_0=-20M$ (bottom).
}
\label{fig:EsumWormhole}
\end{figure}

The observation also holds for wormhole waveforms, which we show in Fig.~\ref{fig:EsumWormhole}.
The doubled propagation time as compared to the $\Rb = 1$ case produces a longer spacing between echoes.
As such, the early wormhole echoes are more distinct than early $\Rb =1$ echoes. 

\begin{figure}[t]
\includegraphics[width = 1 \columnwidth]{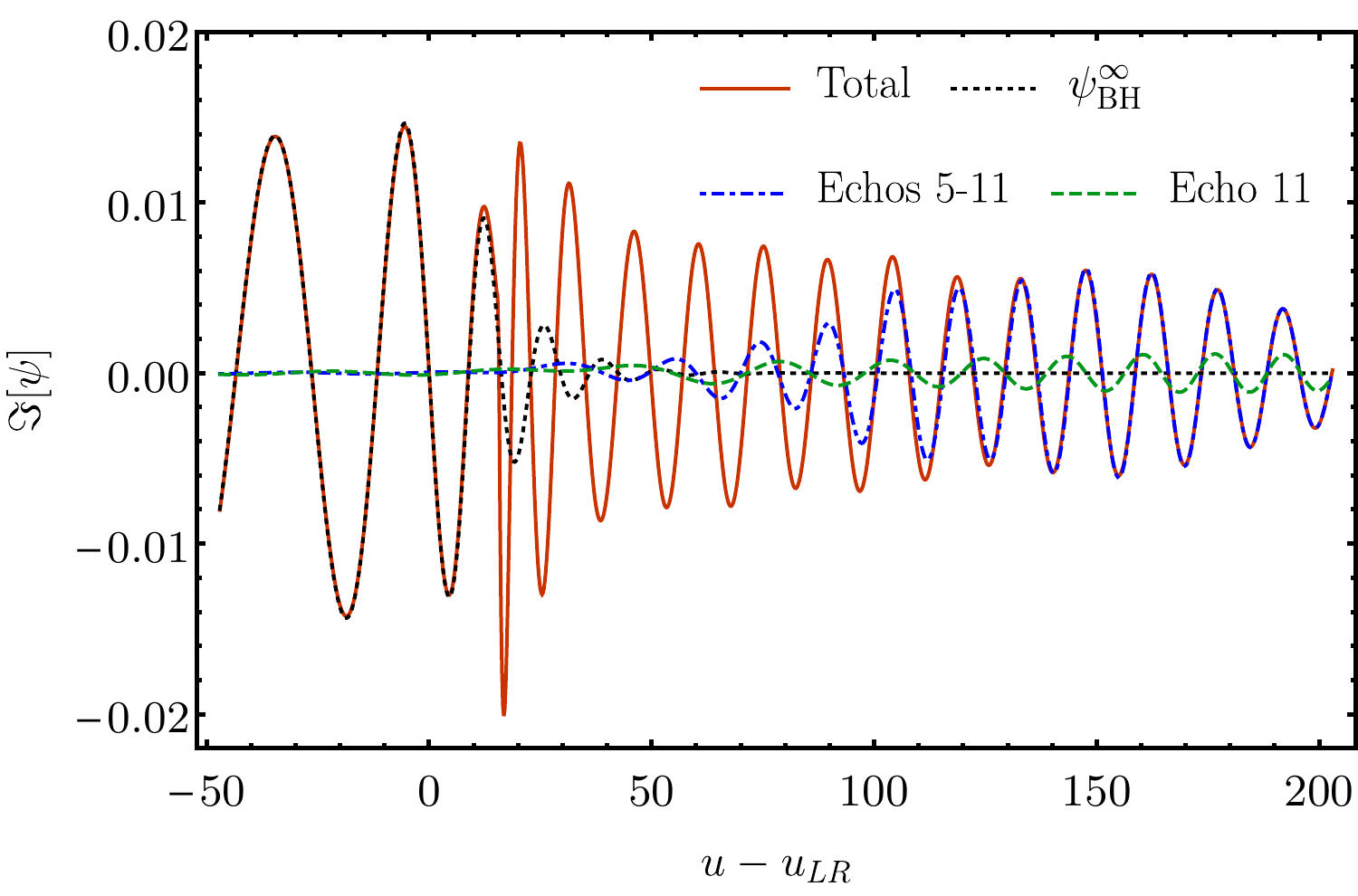}
\caption{
The imaginary part of the $(\ell,m)=(2,2)$, time domain, total waveform excited by a test charge following the ISCO plunge orbit. We show results from an ECO with $\Rb=1$ and $x_0 =-3M$. The total waveform is obtained by summing the black hole waveform $\psi^\infty_{\rm BH}$ and a finite number of echoes. Each curve contains a different numbers of echoes.
}
\label{fig:SingleECOModeTime}
\end{figure}

Meanwhile, when the spacing between the echoes is small compared to the echo duration, there can be no distinct pulses. 
Instead, the waveform resembles a single decaying sinusoid at a frequency different than the BH frequency.
Figure \ref{fig:SingleECOModeTime} shows an occurrence of this for $\Rb =1$, $x_0=-3M$ and the ISCO plunge orbit.
In this case, the total waveform, appearing as the red solid curve, initially agrees with the BH waveform $\psi^\infty_{\rm BH}$, appearing as the black dotted curve, but then transitions to a decaying sinusoid. 
Note that this case pushes the limits of our approximation that the waves propagate freely near $x_0$; for $x_0 = -3M$, $r_0 \approx 2.08 M$ and $V(r_0)$ is approximately $25\%$ its peak value.

This decaying sinusoid is in fact the coherent superposition of the late echoes, a fact that we illustrate by plotting the last seven echoes appearing in the echo sum in purple. 
This coherent superposition occurs because  the later echoes all have nearly the same frequency. 
Finally note that the missing echoes from the truncated sum are not negligible compared to the total waveform, a fact we illustrate by also plotting the last echo appearing in the sum in green. 
In Sec.~\ref{sec:SingleMode} we study this example in the frequency domain, and we find that this is an example of the excitation of a single resonant mode of the ECO spacetime as described by our reflecting boundary condition.

\section{Excitation of ECO Modes}
\label{sec:ECOExcitation}

The presence of the reflecting boundary condition drastically changes the spectrum of the spacetime.
The result is a different set of resonant frequencies, those of the ECO spacetime. 
In this section we explore how our model treats these modes, and how they relate to the echoes discussed in Sec.~\ref{sec:Echoes}.

\subsection{New Modes}
\label{sec:ECOModes}

\begin{figure}[t]
\includegraphics[width = 1 \columnwidth]{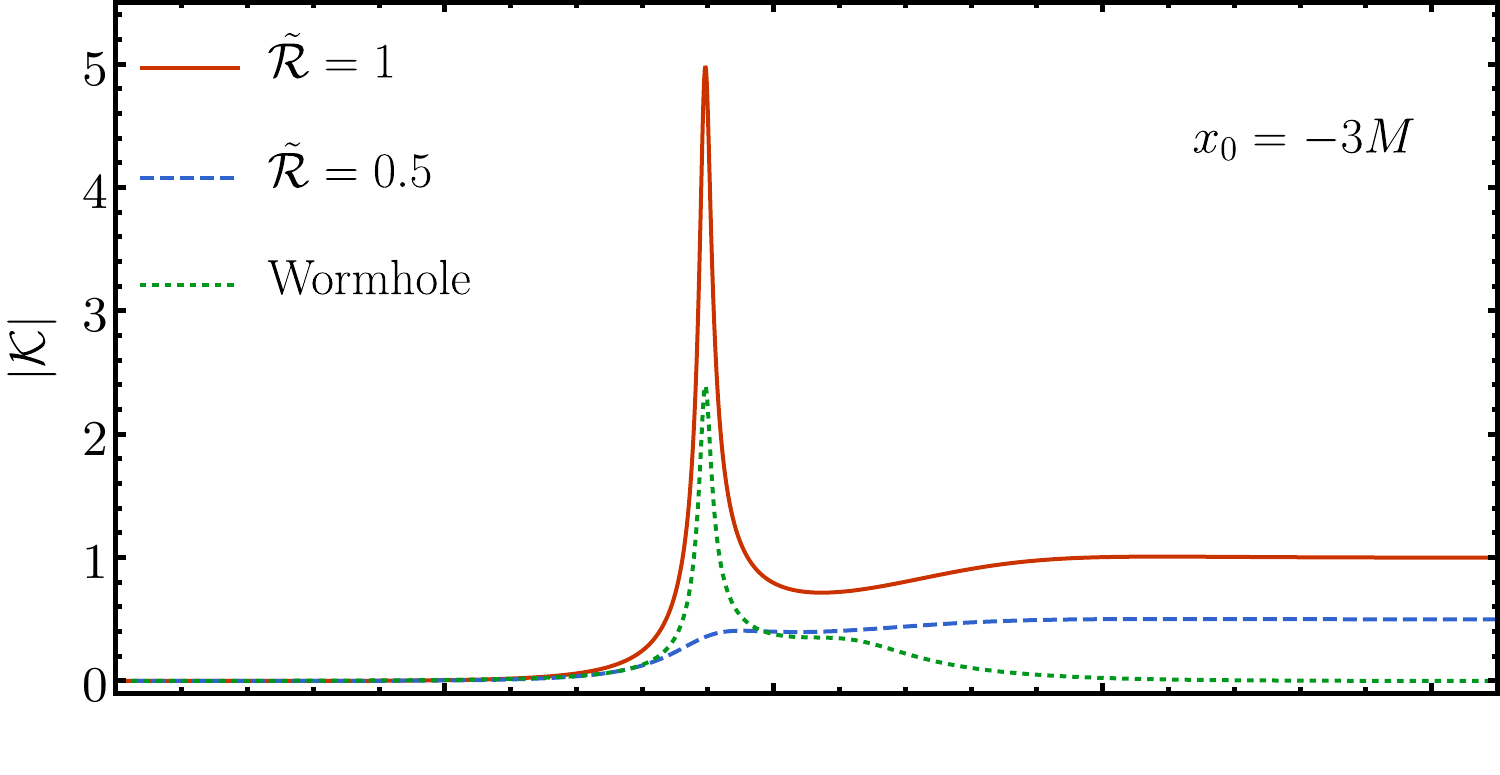}\\
\vspace{-14.5 pt}
\includegraphics[width = 1 \columnwidth]{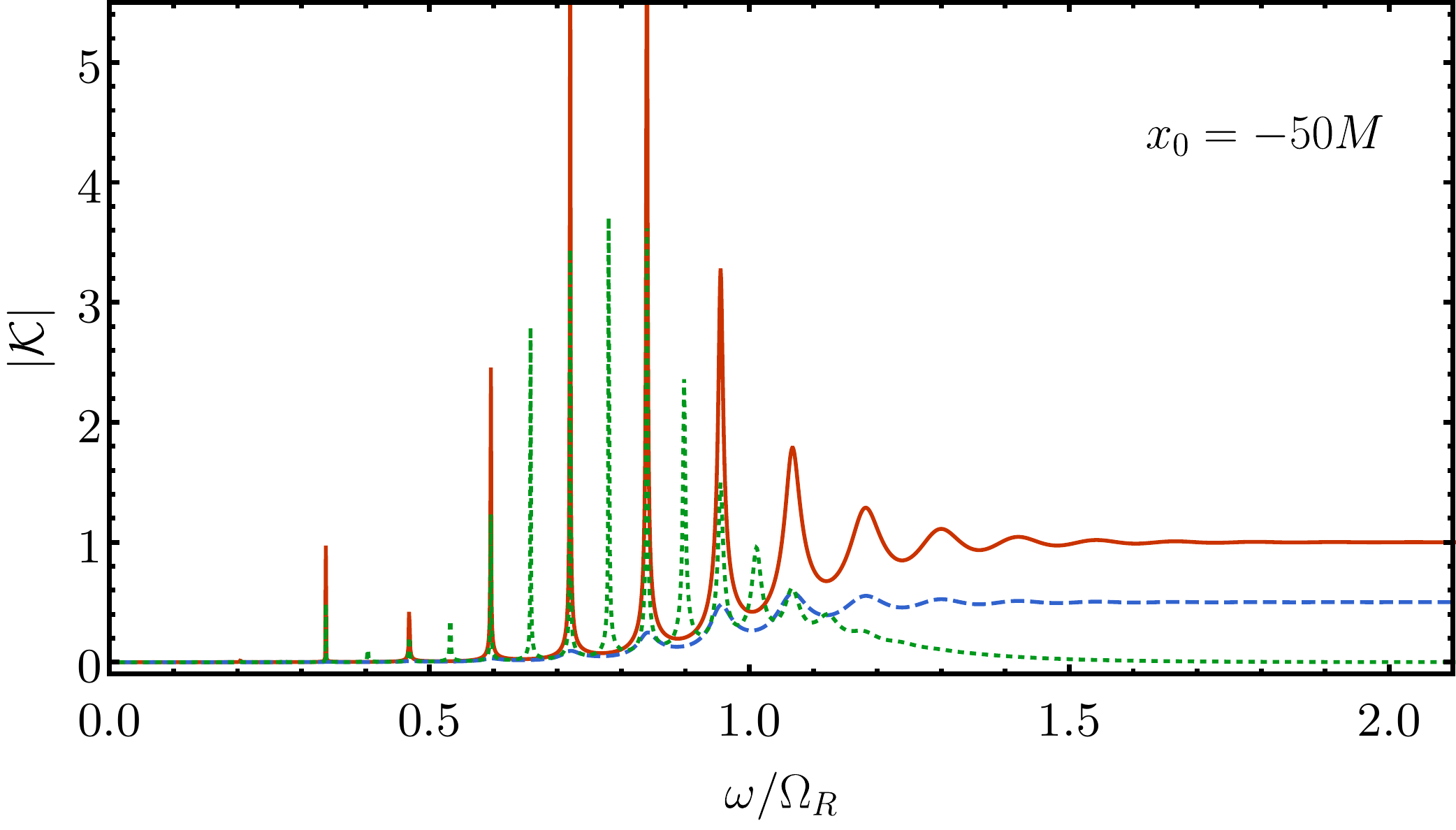}
\caption{
Top: The $\ell=2$ echo transfer function $|\K(\omega)|$ for $x_0=-3M$ and several choices of $\Rb$. Note that $|\K|$ is a symmetric function of $\omega$. Bottom: The same plot for $x_0=-50M$.
}
\label{fig:NewModes}
\end{figure}

The QNM resonances are the complex poles of the Green's function.
From Eq.~\eqref{eq:gBH}, we see that for a BH, they occur when $W_{\rm BH} = 0$. 
The BH QNMs are not poles of the ECO Green's function. 
As is seen from Eq.~\eqref{eq:Gref}, the first and second terms both have poles at the QNM frequencies, but these cancel in the full expression.

The modes of the ECO spacetime come from the poles of the response function $\K(\omega)$ appearing in the Green's function,
\begin{align}
\K = \frac{\TBH \Rb e^{-2i\omega x_0}} {1 - \RBH \Rb e^{-2i \omega x_0} } \,. \notag
\end{align}
These modes obey both the reflecting boundary condition at $x_0$ as well as the outgoing wave condition at $\mathcal I^+$.
Figure \ref{fig:NewModes} shows the $|\K|$ for $\Rb =1$, $\Rb =0.5$, and for the wormhole spacetime, each for two values of $x_0$: $x_0 = -3 M$ and $x_0 = -50 M$.
In the figure, each peak of $|\K|$ represents a resonance of the transfer function\footnote{
A peak of the transfer function $\K$ on the real axis is a resonance in the sense that amplification occurs at this frequency. To show that a complex pole of the Green's function is responsible for this peak, one must examine $\K$ in the complex $\omega$ plane.
}.

Observe that in all our cases there are no new modes at large frequencies $\omega \gg \Omega_R$.
This behavior can be understood analytically. Recall that at large frequencies $\RBH\to 0$ and $\TBH \to 1$. 
This means that
\begin{align}
&\K(\omega)\to \Rb(\omega)e^{-2i\omega x_0},& &\omega \to \infty \,,
\end{align}
and the additional resonances are exactly the poles of $\Rb$.

For $x_0=-3M$, Fig.~\ref{fig:NewModes} clearly displays a single new mode at a frequency close to the fundamental QNM of a BH, for both $\Rb = 1$ and the wormhole.
In the case $\Rb=0.5$, there is a small peak in $|\mathcal K|$ at about the same frequency, although it is less visible. 
 
For $x_0=-50M$ and constant $\Rb$, there is a set of new modes with a frequency spacing of $2\pi/(2|x_0|)$. 
For the wormhole, there is a set of new modes and with a spacing of $2\pi/(4|x_0|)$.
This frequency spacing corresponds to approximately the light travel time $T$ from the potential peak to the boundary and back.
For an optical cavity, this spacing is known as the free spectral range of the cavity, 
\begin{align}
\omega_{\rm FSR} = \frac{2\pi}{T} \,.
\end{align}

To understand the resonances, we can use techniques from similar problems involving optical cavities.
The zeros of the denominator of Eq.~\eqref{eq:EchoTransfer} contribute a set of resonances $\omega_n$ given by
\begin{align}
1=\RBH(\omega_n)\Rb(\omega_n)e^{-2i\omega_n x_0} \,.
\end{align}
Consider first the case that $\Rb(\omega)$ is frequency independent. In this case, there are two frequency scales in the problem; the scale $\delta \omega_{\rm BH} \approx \RBH(\omega)/\partial_{\omega}\RBH(\omega)$ on which the reflectivity changes and the scale $\omega_{\rm FSR}$ on which the exponent of the exponential changes.
When the frequency dependence of the $\RBH$ is weak, i.e.~$\omega_{\rm FSR}/\delta \omega_{\rm BH} \ll 1$, then to leading order in $\omega_{\rm FSR}/\delta \omega_{\rm BH} $
\begin{align}
\omega_n = n\omega_{\rm FSR} + i \frac{\omega_{\rm FSR}}{2\pi} \ln(\Rb\RBH)+\mathcal{O}\left(\frac{\omega_{\rm FSR}}{\delta \omega_{\rm BH} }\right), \label{eq:wapprox}
\end{align}
where $\RBH$ is evaluated at $n\omega_{\rm FSR}$.
We see that the new modes are spaced by $\omega_{\rm FSR}$ in agreement with Fig.~\ref{fig:NewModes}, and they decay provided $|\Rb|<1$.

More generally, when $\Rb(\omega)$ has frequency dependence we can often separate it into factors with fast and slow frequency dependence,
\begin{align}
\Rb(\omega) e^{-2i\omega x_0} = \hat {\mathcal R}(\omega)e^{i\omega T} \,,
\end{align}
where $\hat{\mathcal R}(\omega)$ varies appreciably over a characteristic range of frequencies $\delta \omega$ which is large compared to $2\pi/T$. Again, $T$ 
is approximately the round trip travel time between the potential peak and the major features in the true potential characterizing the ECO. For the wormhole, $\delta \omega = \delta \omega_{\rm BH}$ and $T=-4x_0$ is the light travel time. Provided both $\omega_{\rm FSR}/\delta \omega_{\rm BH}  \ll 1$ and $\omega_{\rm FSR}/\delta \omega \ll 1$, working to leading order, we again arrive at Eq.~\eqref{eq:wapprox} where $\omega_{\rm FSR} = 2 \pi /T$ and we must allow for $\mathcal O(\omega_{\rm FSR}/\delta \omega)$ errors.

Notice also that the \rm{ECO} resonances for $\Rb = 0.5$ are broader than the $\Rb=1$ resonances, while the width of the wormhole resonances is similar to the $\Rb=1$ resonances. 
This also follows from Eq.~\eqref{eq:wapprox} since the width of the resonances is controlled by the decay rate of the new modes, which is proportional to $\omega_{\rm FSR}\ln(\Rb\RBH)$. In the low frequency regime that the new modes appear at, $\Rb\approx 1$ for the wormhole and we expect the width to be similar to the $\Rb=1$ case.

\subsection{Single Mode Excitation}
\label{sec:SingleMode}

We return to Fig.~\ref{fig:SingleECOModeTime}, where for $\Rb = 1$ and $x_0 = -3 M$ the echo waveform appears as a single decaying sinusoid which differs from the QNMs of the BH.
This behavior can be interpreted as the excitation of a single resonant mode of $\K$ by the plunge.
This is clearest in the frequency domain.

\begin{figure}[t]
\includegraphics[width = 1 \columnwidth]{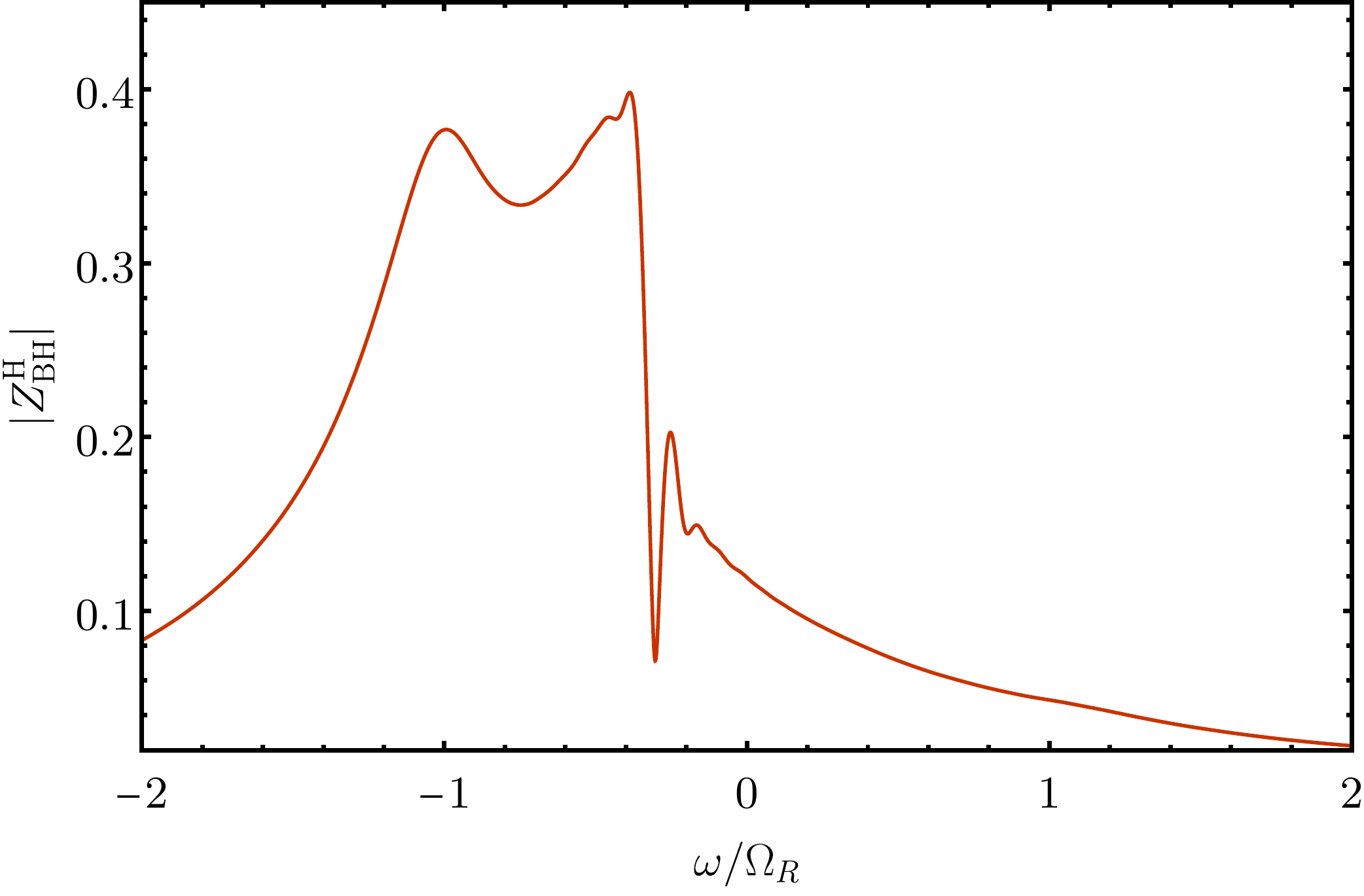}
\caption{
The modulus of the $(\ell,m)=(2,2)$ horizon waveform generated by a test charge following the ISCO plunge orbit.
}
\label{fig:HorizonFT}
\end{figure}

\begin{figure}[t]
\includegraphics[width = 1 \columnwidth]{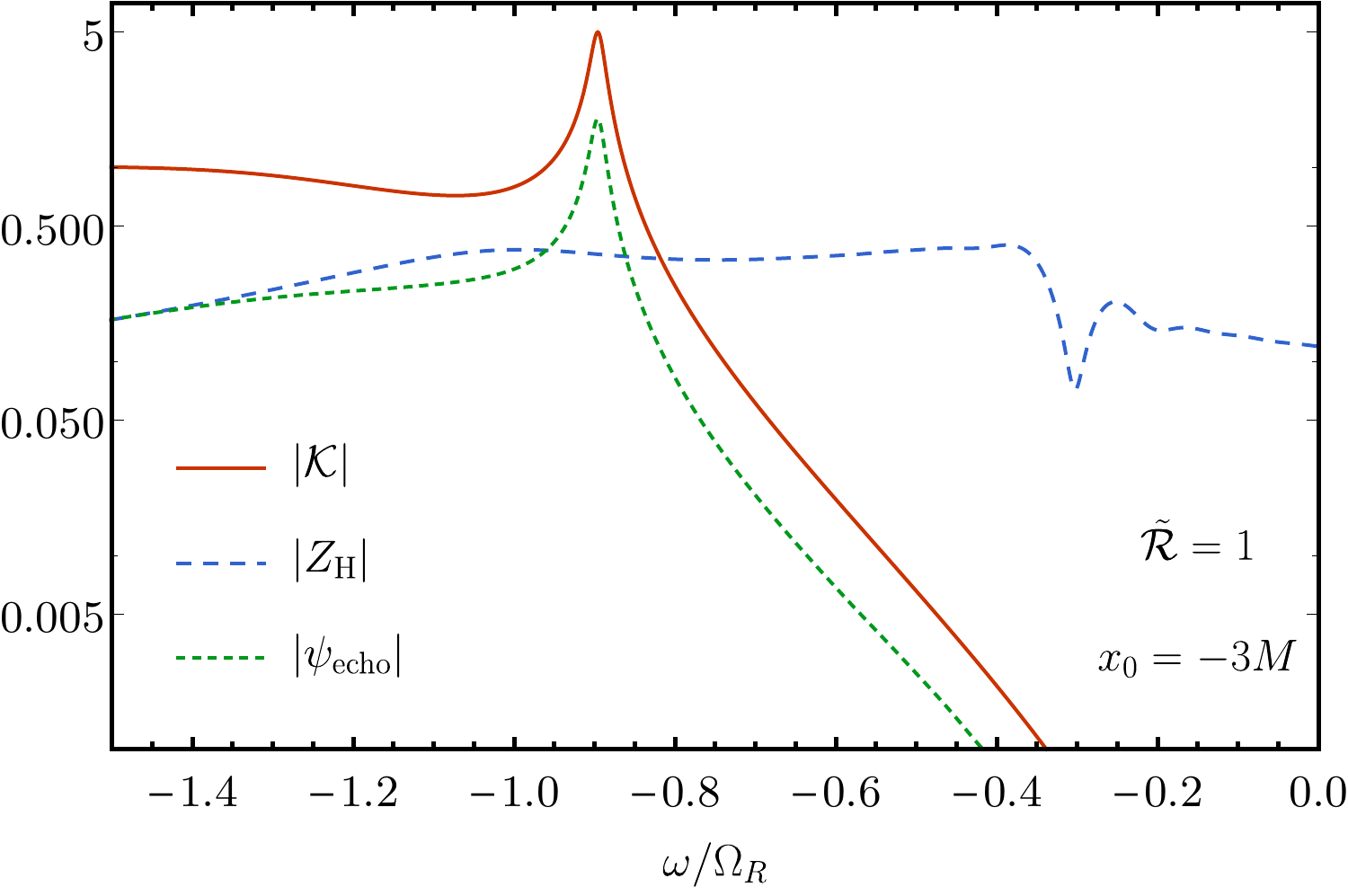}
\caption{
Single mode Excitation. The $(\ell,m)=(2,2)$ response function $|\K|$, the horizon waveform $\ZhBH$ , and the echo sum $\psiecho$ for $\Rb=1$ and $x_0=-3M$. The waveforms are generated by a test charge following the ISCO plunge orbit
}
\label{fig:SingleECOModeFrequency}
\end{figure}

The excitation of the modes is encoded in the product $\Zecho=\K \ZhBH$. Figure \ref{fig:HorizonFT} displays the horizon waveform $\ZhBH$. For this orbit, most of the power is at negative frequencies and there are strong peaks near orbital frequency $\omega =-m\Omega_{\rm ISCO}$ and fundamental BH QNM frequency $\omega =-\Omega_R$. Furthermore, $\ZhBH$ goes to zero at high frequencies. 

The echo waveform $\Zecho$ is shown in Fig.~\ref{fig:SingleECOModeFrequency} for the case $\Rb=1$, $x_0=-3M$.
Note that $\Zecho$ inherits the resonance from $\K$ . 
This resonant frequency is similar to the fundamental BH QNM, but has a much slower decay, as can be noted by the slenderness of the peak compared to the peak in the horizon  amplitude at the same frequency.

\subsection{Echoes from Interference of Modes}
\label{sec:EchoInt}

\begin{figure}[t]
\includegraphics[width = 1 \columnwidth]{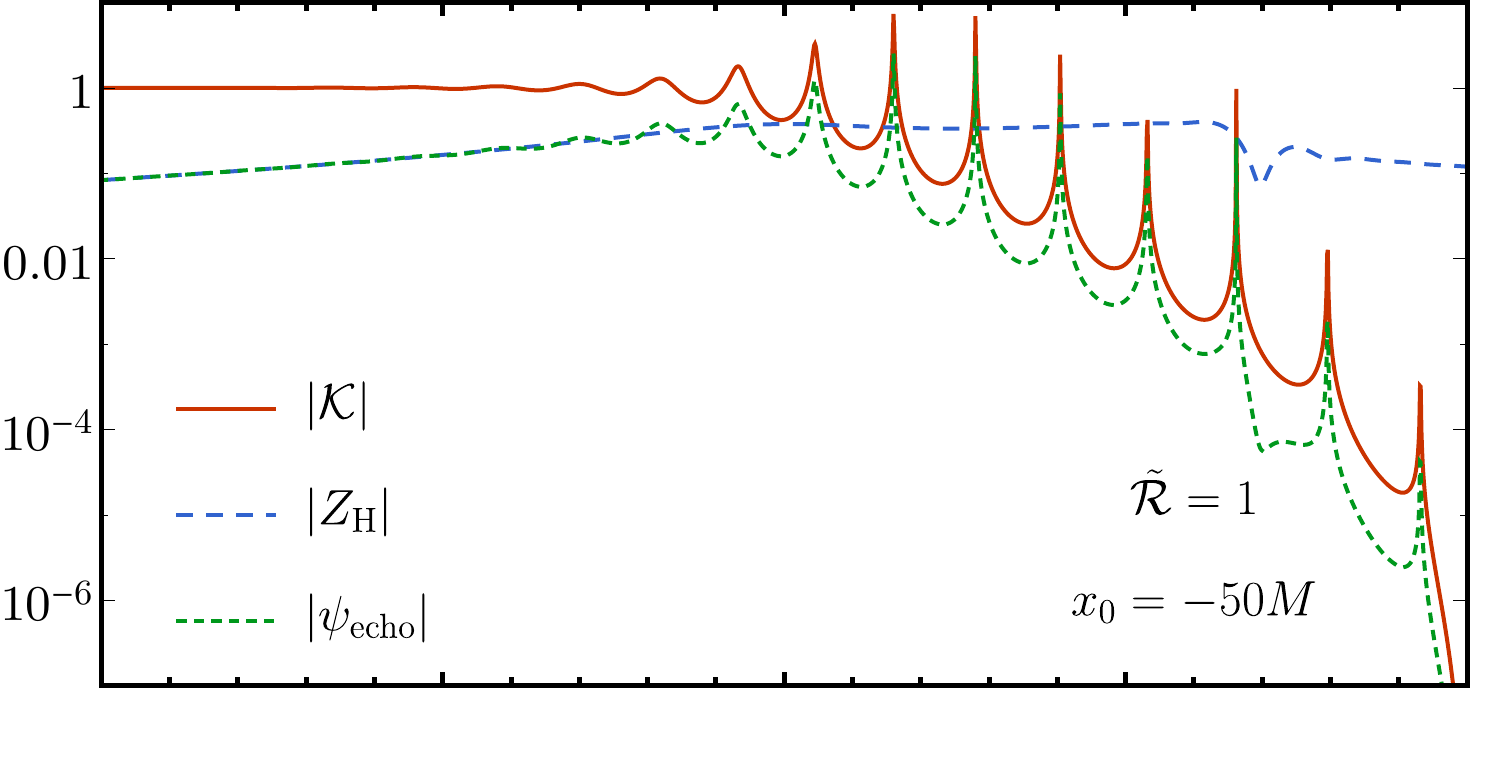}\\
\vspace{-14.75pt}
\includegraphics[width = 1 \columnwidth]{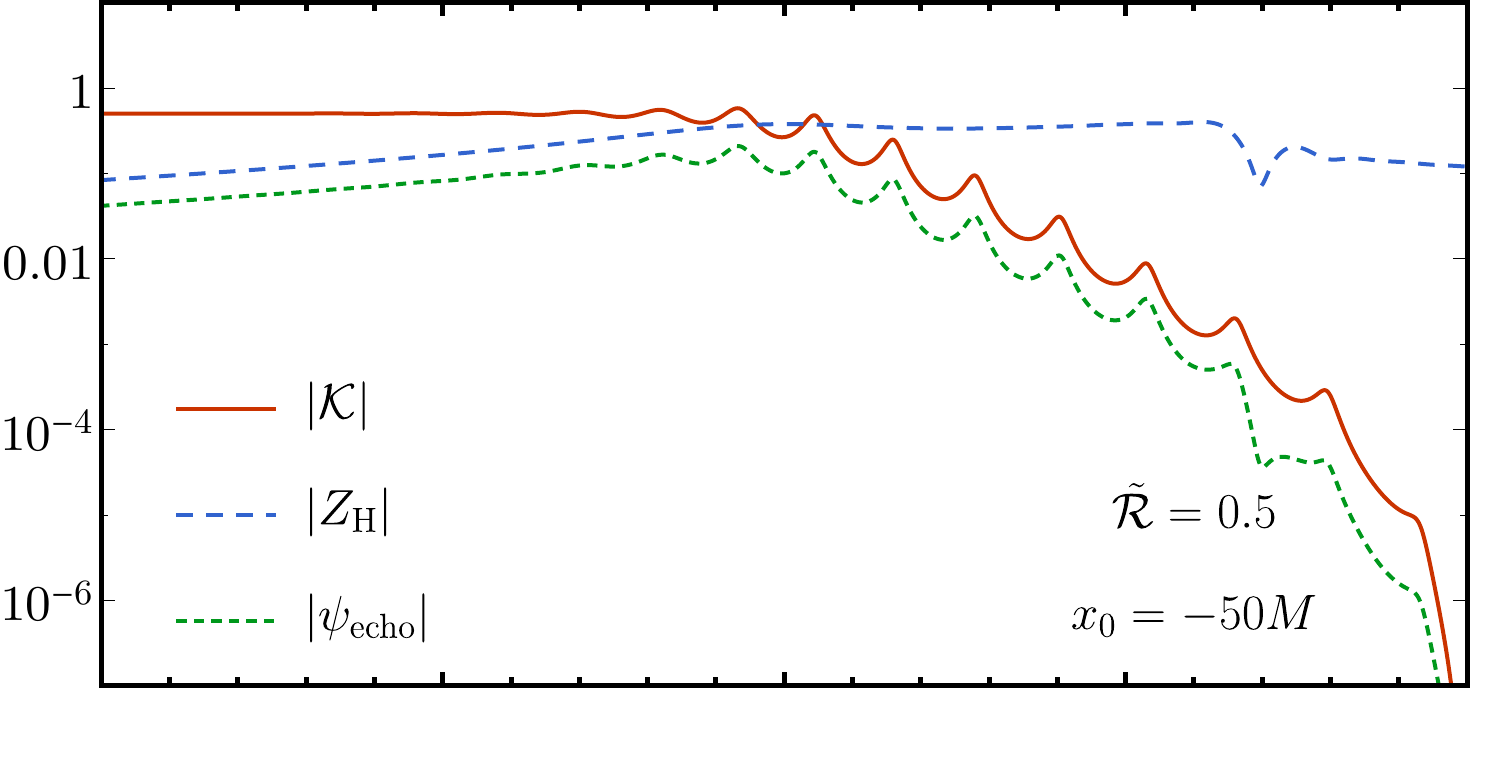}\\
\vspace{-14.75pt}
\includegraphics[width = 1\columnwidth]{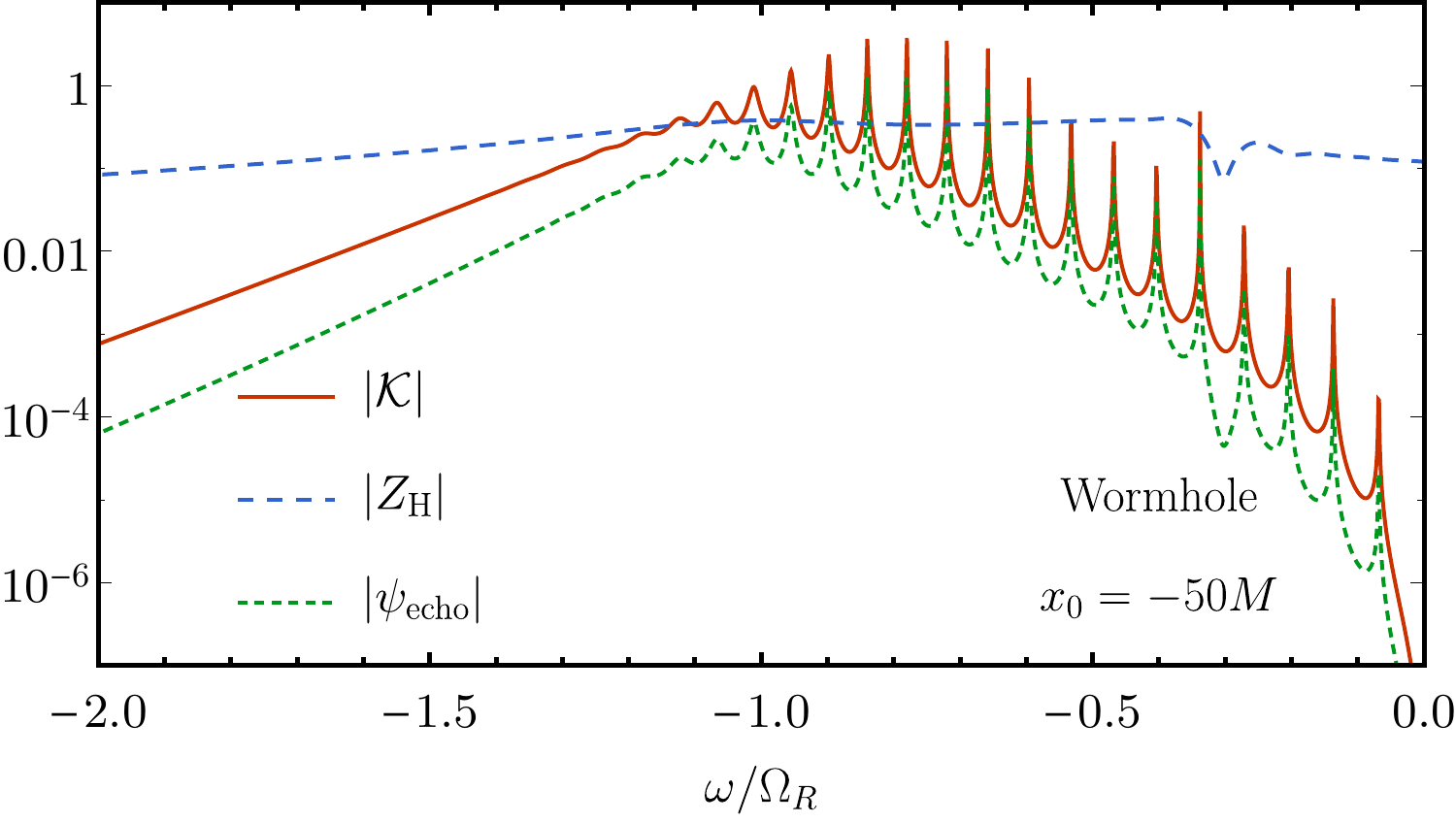}
\caption{
Multi-mode excitation. We fix $x_0=-50M$, a case where Fig.~\ref{fig:EsumManyRb} shows that the time domain waveform contains echoes for a range of $\Rb$. We show the $(\ell,m)=(2,2)$ response function $|\K|$, the horizon waveform $\ZhBH$, and the echo sum $\psiecho$. The waveforms ares generated by a test charge following the ISCO plunge orbit. The top panel corresponds to $\Rb=1$, the middle panel to $\Rb=0.5$, and the lower panel is the wormhole waveform.
}
\label{fig:MultiECOModeFrequency}
\end{figure}

Recall that for large values of $x_0$, the total waveform appears as a sum of distinct echo pulses. This scenario also can be understood in terms of the additional resonances of the ECO spacetime. 
Figure \ref{fig:MultiECOModeFrequency} shows  the frequency domain echo amplitude $\Zecho$  for three choices of $\Rb$, all with $x_0=-50M$: $\Rb=1$ appears in the top panel, $\Rb =0.5$ appears in the middle panel, and the wormhole appears in the bottom panel. 
The horizon amplitude  is substantial at all of the resonances of $\K$, which have spacing $\omega_{\rm FSR}$. 
The result is that all of the resonances appear in the $\Zecho$ in all three cases.

In fact, this is what we expect a sum of echo pulses to look like in the frequency domain. Suppose that in the time domain a function $f(t)$ is a sum of delta function pulses spaced by $T=2\pi/\Delta\omega$ beginning at time $t=0$, with each pulse $\gamma$ times smaller than than the previous,
\begin{align}
\label{eq:DDcomb}
f(t)=\sum_{n=0}^\infty \gamma^n \delta\left(t-nT\right) \,.
\end{align}
Then in the frequency domain $\tilde f(\omega)$ is an infinite sum of equally spaced, equally excited resonances (see Appendix \ref{sec:DDcomb} for a derivation)
\begin{align}
\tilde f(\omega)&=\frac{i\Delta \omega}{2\pi}\sum_{n=-\infty}^\infty\frac{1}{\omega-\omega_n} \,,\nonumber \\
&\omega_n=n\Delta \omega+i\frac{\Delta \omega}{2\pi}\ln \gamma \,. 
\label{eq:DDcombFT}
\end{align}
Before the echoes begin to blend together, but after the initial BH waveform decays, the waveforms $\psi^{\infty}(u)$ shown in Figs.~\ref{fig:Esum1}, \ref{fig:EsumManyRb} and \ref{fig:EsumWormhole} are loosely of the form of $f(u)$ if we view each pulse as a delta function and choose $ T = 2|x_0|$ (or $T = 4 |x_0|$ for the wormhole case). Therefore it is not surprising that $\Zecho(\omega)$ resembles $\tilde f(\omega)$ at low frequencies, where it is more reasonable to approximate each pulse appearing in the plots by a delta function.

\section{General Features of echoes}
\label{sec:Gen}

We turn now to some additional applications of our formalism for reprocessing black hole waveforms into waveforms from ECOs.
After reviewing some general features of echoes in our model, we develop a simple template that broadly reproduces the echoes seen by distant observers.
We also discuss the energy content of these echoes.

\subsection{General Features of echoes}
\label{sec:GFEA}

The horizon waveform $\psi^{\rm H}_{\rm BH}$ has some generic features which should hold for many sources.
Much like the inspiral, merger, and ringdown signal emitted from a compact binary, there are three phases to $\psi^{\rm H}_{\rm BH}$.
These phases are easily identifiable for the horizon waveform generated by the ISCO plunge, shown in the top panel of Fig.~\ref{fig:Echotime}.
At early times, when the small body is approximately on the ISCO orbit, the waveform frequency is approximately proportional to the ISCO orbital frequency, $\omega =m\Omega_{\rm ISCO}$. 
The waveform peaks around when the small body crosses the horizon at $v_{\rm H}$, and there is also a discontinuity in the derivative of $\psi^{\rm H}_{\rm BH}$ when the particle crosses the horizon (or $x_0$, in our large $|x_0|$ approximation).
At late times, after the particle has crossed the horizon, the waveform is dominated by a decaying sinusoid at the fundamental BH QNM frequency. 
These features are also seen in the frequency domain waveform shown in Fig.~\ref{fig:HorizonFT} and discussed in Sec.~\ref{sec:SingleMode}.

\begin{figure}[t]
\includegraphics[width =1 \columnwidth]{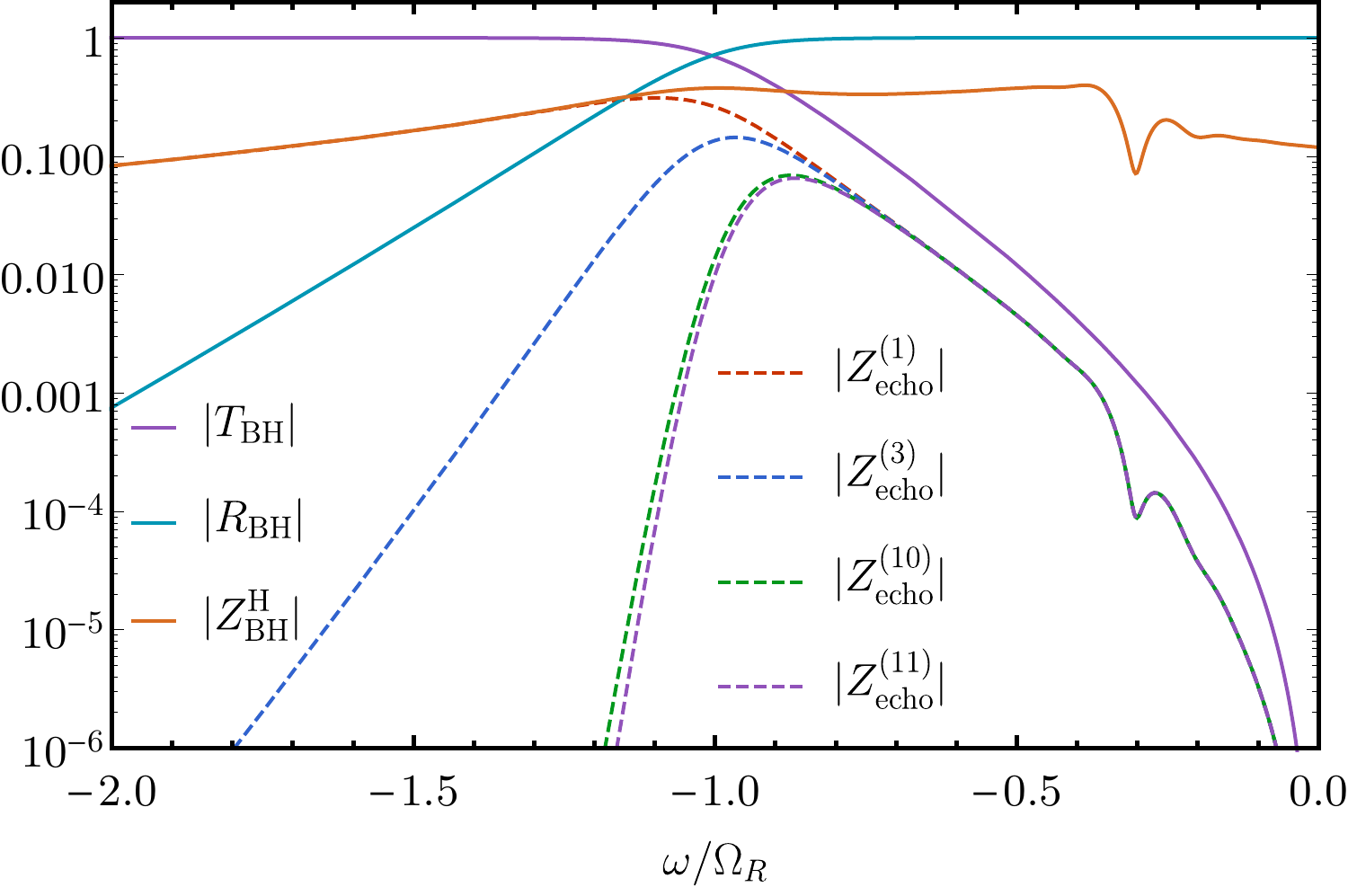}
\caption{The modulus of the $(\ell, m) =(2,2)$ horizon waveform $Z_{\rm BH}^{\rm H}$ and select $\Rb =1$ echoes $Z_{\rm echo}^{(n)}$ generated by a test charge following the ISCO plunge orbit. Also shown are $\RBH$ ad $\TBH$.
}
\label{fig:Genechos}
\end{figure}

The ringdown has a larger effect on the shape of the first few echoes than the earlier parts of $\psi^{\rm H}_{\rm BH}$, because the fundamental QNM frequency is transmitted more easily through the potential barrier.
Meanwhile, the horizon waveform at early times, which is generally at lower frequencies associated with the inspiral orbital timescale, mostly reflects off of the inside of the potential barrier and contributes less to the first echo. The later echoes, having already lost power at frequencies near $\omega =\Omega_R$ from each earlier scatter off of the potential barrier, depend more intricately on the details of the horizon waveform at early times.

We illustrate this in Fig.~\ref{fig:Genechos}, which examines echoes from the ISCO plunge for constant reflectivity $\Rb$.
Figure \ref{fig:Genechos} shows the frequency domain horizon waveform $\ZhBH$ as well as three echoes $\Zecho^{(n)}$, where 
\begin{align}
\label{eq:nZecho}
Z_{\rm echo}^{(n)}=\K^{(n)}\ZhBH
\end{align}
are the Fourier conjugates of the $n$th echoes $\psi_{\rm echo}^{(n)}(u)$.
The first echo inherits the peak of $\ZhBH$ near $\Omega_R$, but the peak near $m\Omega_{\rm ISCO}$ is removed by $\TBH$.
The third echo similarly retains a peak near $\omega =-\Omega_R$, although shifted to a slightly lower frequency compared to the first, and is significantly narrower.
By the tenth eleventh echoes, the differences between successive echoes has become small, and the echoes retain a suppressed peak near (but to the right of) $\omega = - \Omega_R$. 
Overall, we see that because of the low frequency suppression in all the echoes, the ringdown portion of the horizon waveform is most important for determining the shape of the first several echoes.

\subsection{Template for echoes}

The observation that the ringdown of the horizon waveform $\psi^{\rm H}_{\rm BH}$ is the most important factor for determining the shape of the echoes leads to a simple idea for a template for the echoes. 
Construct a template  $Z_{\rm T}^{\rm H}$ for the horizon waveform $\ZhBH$ consisting of only a ringdown at the fundamental QNM frequency. Then construct a template $Z_{\rm T}$  for the echoes  $Z_{\rm echo}$ and a template $Z_{\rm T}^{(n)}$ for each echo $Z_{\rm echo}^{(n)}$ using the transfer functions
\begin{align}
 &Z_{\rm T}=\K Z_{\rm T}^{\rm H}\,, &
 &Z_{\rm T}^{(n)}=\K^{(n)}Z_{\rm T}^{\rm H} \,.
\end{align}

To model the ringdown of the horizon waveform, we take a superposition of decaying sinusoids that each are excited at a slightly different time.
In the time domain our template for the horizon waveform is
\begin{align}
\psi^{\rm H}_{\rm T}(t)&=(\psi_{\rm QNM}*h)(t) \nonumber \\
h(t)&=\frac{\beta}{\sqrt{2\pi}}\exp\left(\frac{-(t-t_s)^2}{2/\beta^2}\right) \nonumber \\
\psi_{\rm QNM}(t)&=\theta(t)\left(-i\alpha_+ e^{-i\Omega_+ t}-i\alpha_- e^{-i\Omega_- t}\right),
\end{align}
where we use $\psi_{\rm T}^{\rm H}$ to indicate the Fourier conjugate of $Z_{\rm T}^{\rm H}$. 
We weight each decaying sinusoid by the Gaussian $h(t)$. 
The template is parametrized by two complex amplitudes $\alpha_\pm$ for the sinusoids at the positive and negative QNM frequencies, $\Omega_\pm=\pm \Omega_R +i\Omega_I$, a central start time $t_s$, and a frequency width $\beta$. 
In the frequency domain, the template for the horizon waveform takes the even simpler form
\begin{align}
Z^{\rm H}_{\rm T}(\omega; \vec p)=e^{i\omega t_s}e^{-\omega^2/(2\beta^2)}\left(\frac{\alpha_+}{\omega-\Omega_+}+\frac{\alpha_-}{\omega-\Omega_-}\right),
\end{align}
where $\vec p=(\alpha_+,\alpha_-,t_s, \beta)$ are the template parameters.

To evaluate the template we investigate its ability to match both individual echoes and complete waveforms produced from a test charge following the ISCO plunge orbit, in the case of a constant $\Rb$.
To quantify the match, we define the overlap of two waveforms as
\begin{align}
\label{eq:overlap}
\rho^2(Z_1, Z_2)=\frac{|\braket{Z_1|Z_2}|^2}{\braket{Z_1|Z_1}\braket{Z_2|Z_2}} \,,
\end{align}
in terms of the inner product
\begin{align}
\braket{a|b}=\int_{-\infty}^\infty \frac{d\omega}{2\pi}\tilde a^*(\omega)\tilde b(\omega) \,.
\end{align}
The overlap satisfies $0 \leq \rho \leq 1$, with $\rho \approx 1$ indicating a good match.

\begin{figure}[t]
\includegraphics[width = 1 \columnwidth]{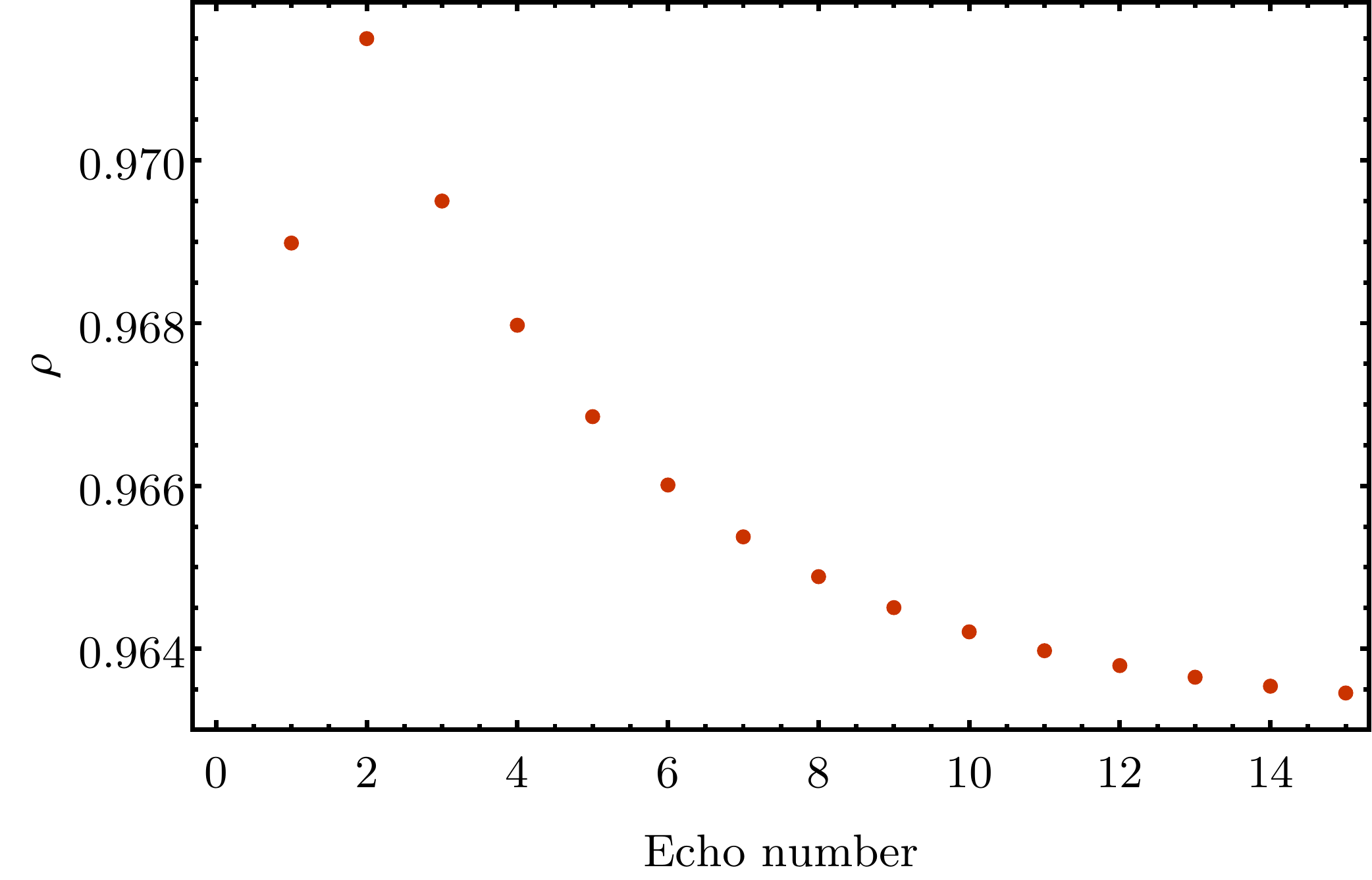}
\caption{ The overlap $\rho(Z_{\rm T}^{(n)}, Z_{\rm echo}^{(n)}; \vec p_1)$ for the $n$th individual echo  plotted versus echo number $n$. 
The parameters $\vec p_1$ are determined by maximizing the overlap for the first $n=1$ echo. We show results for $(\ell,m)=(2,2)$ and use a test charge following the ISCO plunge trajectory as a source for the $Z_{\rm echo}^{(n)}$.
}
\label{fig:rho2vsecho}
\end{figure}

For our first test of the model, we consider the overlap for the individual echoes, $\rho(Z_{\rm T}^{(n)},\Zecho^{(n)}; \vec p)$. 
Note that the overlap for the individual echoes is independent of $x_0$ and $\Rb$.
We set the template parameters $\vec p =\vec p_1$ by analytically maximizing the overlap over $\alpha_\pm$ \cite{Zimmerman:2011dx} at fixed nonlinear model parameters $t_s$ and $\beta$; we then numerically search for optimal parameters $t_s$ and $\beta$.
We compute the overlap for successive echoes using the same fixed $\vec p_1$.

\begin{figure}[t]
\includegraphics[width = 1 \columnwidth]{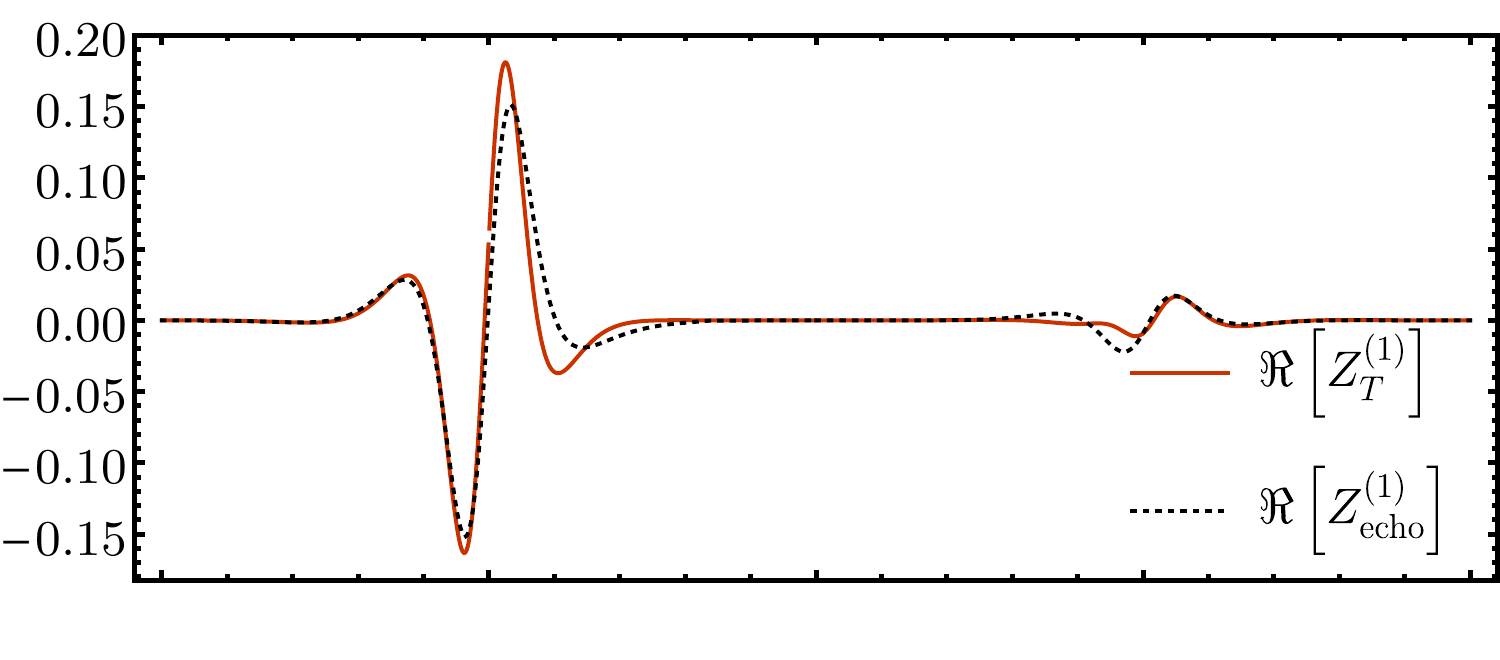}\\
\vspace{-14.5 pt}
\includegraphics[width = 1 \columnwidth]{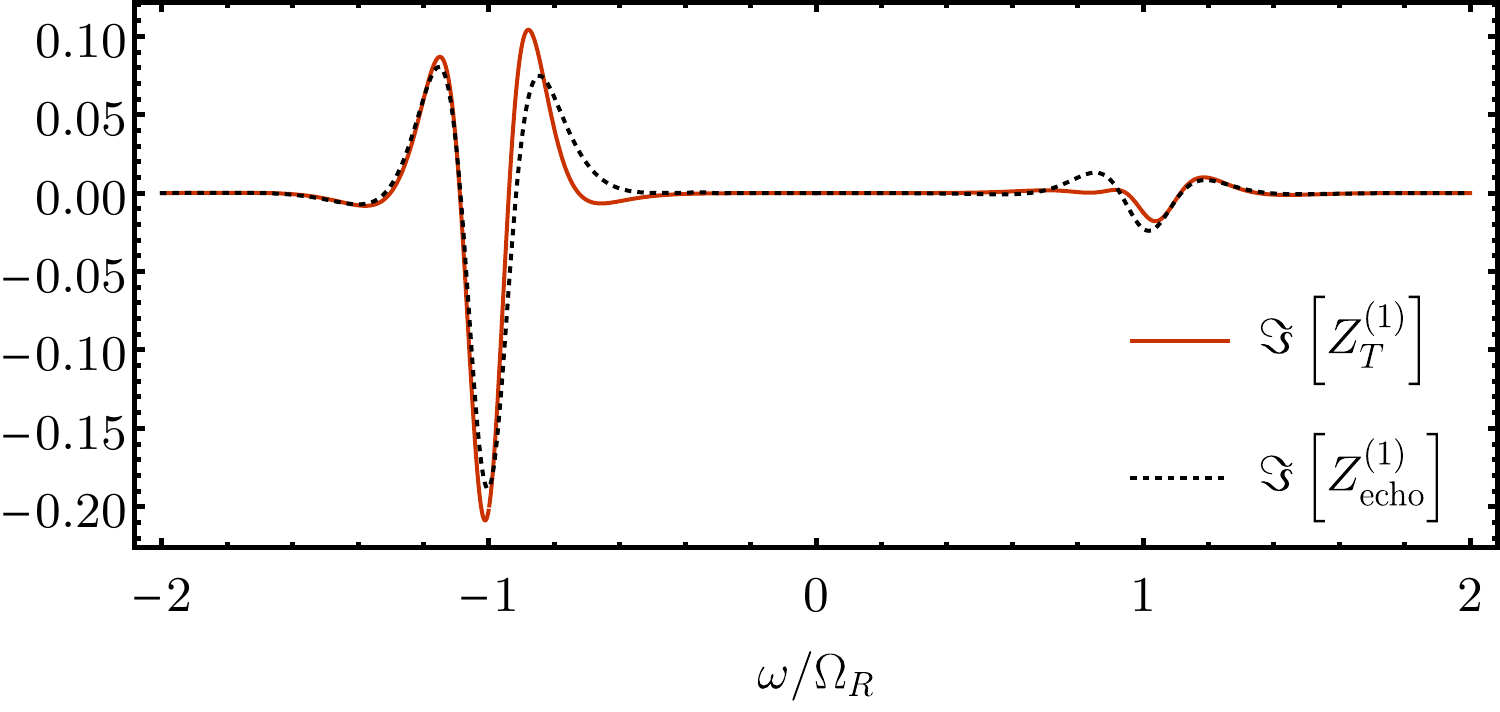}
\caption{ A comparison of the $(\ell, m) =(2,2)$ real (top) and imaginary (bottom) parts of the $n=1$ echo template $Z_{\rm T}^{(1)}$  and the first echo. The echo is generated by a test charge following the ISCO plunge orbit and the parameters for the template are determined by maximizing the overlap $\rho$ given by Eq.~\eqref{eq:overlap} between the template and the echo. The value of the overlap is $\rho=0.969$.
}
\label{fig:dirTemcompEchos}
\end{figure}

\begin{figure}[t]
\includegraphics[width = 1 \columnwidth]{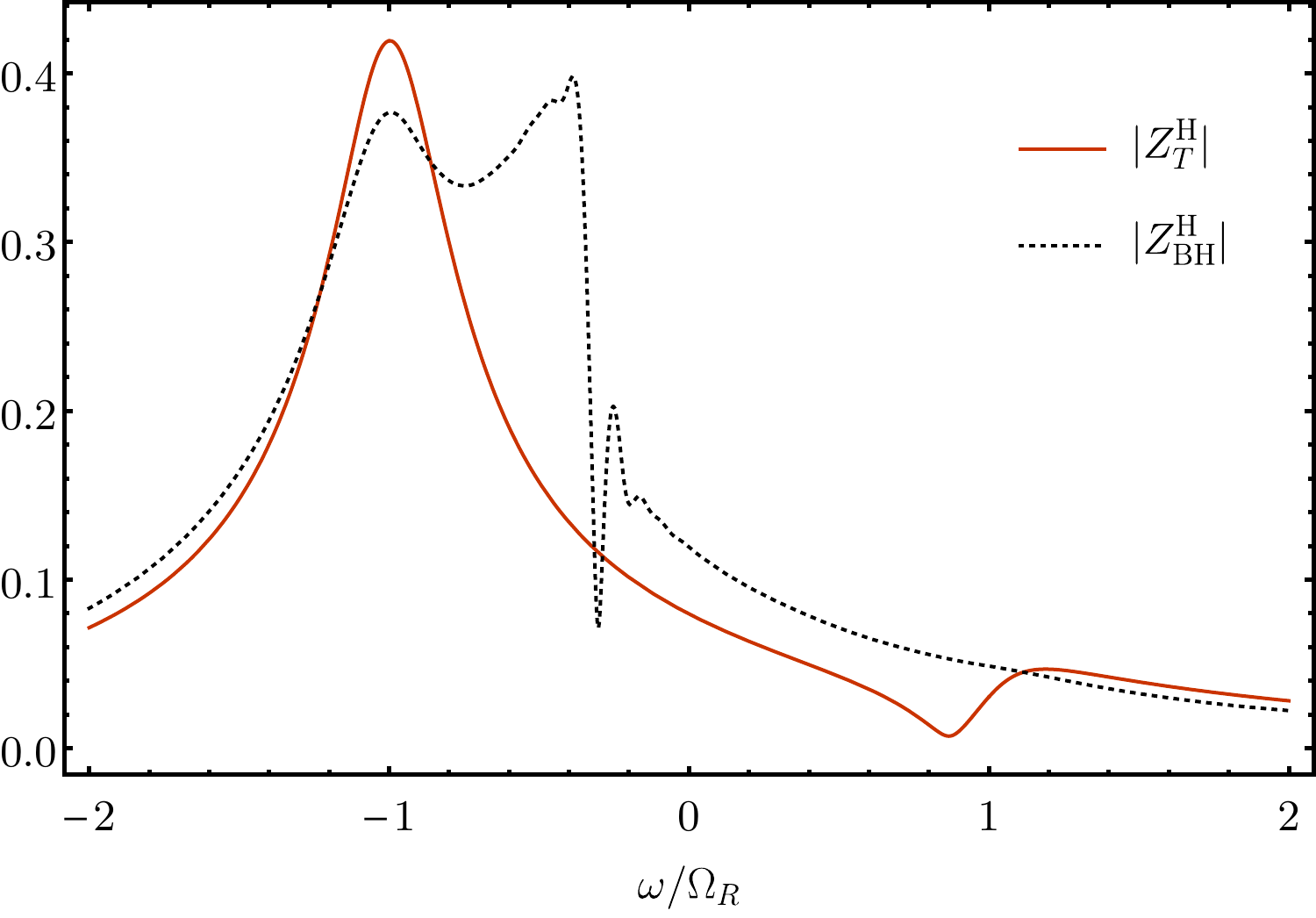}
\caption{A comparison of the modulus of the $(\ell, m) =(2,2)$ of the horizon waveform template $Z_{\rm T}^{\rm H}$ and numerically computed horizon waveform. The waveform is generated by a test charge following the ISCO plunge orbit and the parameters for the template are determined by maximizing the overlap $\rho$ between the first echo template and the numerically calculated first echo. The value of the overlap is $\rho=0.72$.
}
\label{fig:horizonmatch}
\end{figure}

In Fig.~\ref{fig:rho2vsecho}, 
we plot $\rho(Z_{\rm T}^{(n)},\Zecho^{(n)}; \vec p_1)$
versus $n$ for the first twenty echoes. 
We see that the overlap is approximately between $0.96$ and $0.97$ and asymptotes to a constant as the echo number $n$ grows. 
We show a direct comparison of the template and the first echo in Fig.~\ref{fig:dirTemcompEchos} to give an example of the type of match produced by an overlap in this range\footnote{
Note that our procedure does not completely fix the parameters $\alpha_\pm$ since the normalized overlap is invariant under shifts $Z_{\rm T}^{(n)}\to a Z_{\rm T}^{(n)}$ for any complex constant $a$. To completely fix the parameters for Figs.~\ref{fig:dirTemcompEchos} and \ref{fig:horizonmatch}, we also impose the constraints $\braket{Z_{\rm T}^{(n)}|Z_{\rm T}^{(n)}}=\braket{Z_{\rm echo}^{(n)}|Z_{\rm echo}^{(n)}}$ and ${\rm ph}(\braket{Z_{\rm T}^{(n)}|Z_{\rm echo}^{(n)}})=0$. This is equivalent to minimizing the least squares differences between the waveforms while holding $\braket{Z_{\rm T}^{(n)}|Z_{\rm T}^{(n)}}$ constant.
}. 
Importantly, this analysis shows that the first echo can be used to generate values of the template parameters that produce reasonably good overlaps for later echoes. 

It is insightful to compare these overlaps to the corresponding overlap
$\rho(Z_{\rm T}^{\rm H},Z_{\rm BH}^{\rm H}; \vec p_1)$ between the horizon waveform and its template at the same parameters $\vec p_1$. 
This overlap is $\rho = 0.72$, and it is smaller than the overlap for the individual echoes. 
A direct comparison of the horizon waveform and its template, shown in Fig.~\ref{fig:horizonmatch}, reveals that the template misses key features of the horizon waveform at low frequencies $|\omega| < \Omega_R$.  
We explain the enhanced performance of the template for the echoes compared to the horizon waveform as being due to the echo transfer functions $\K^{(n)}$, which filter out the low frequencies where the template performs poorly.

\begin{figure}[t]
\includegraphics[width = 1 \columnwidth]{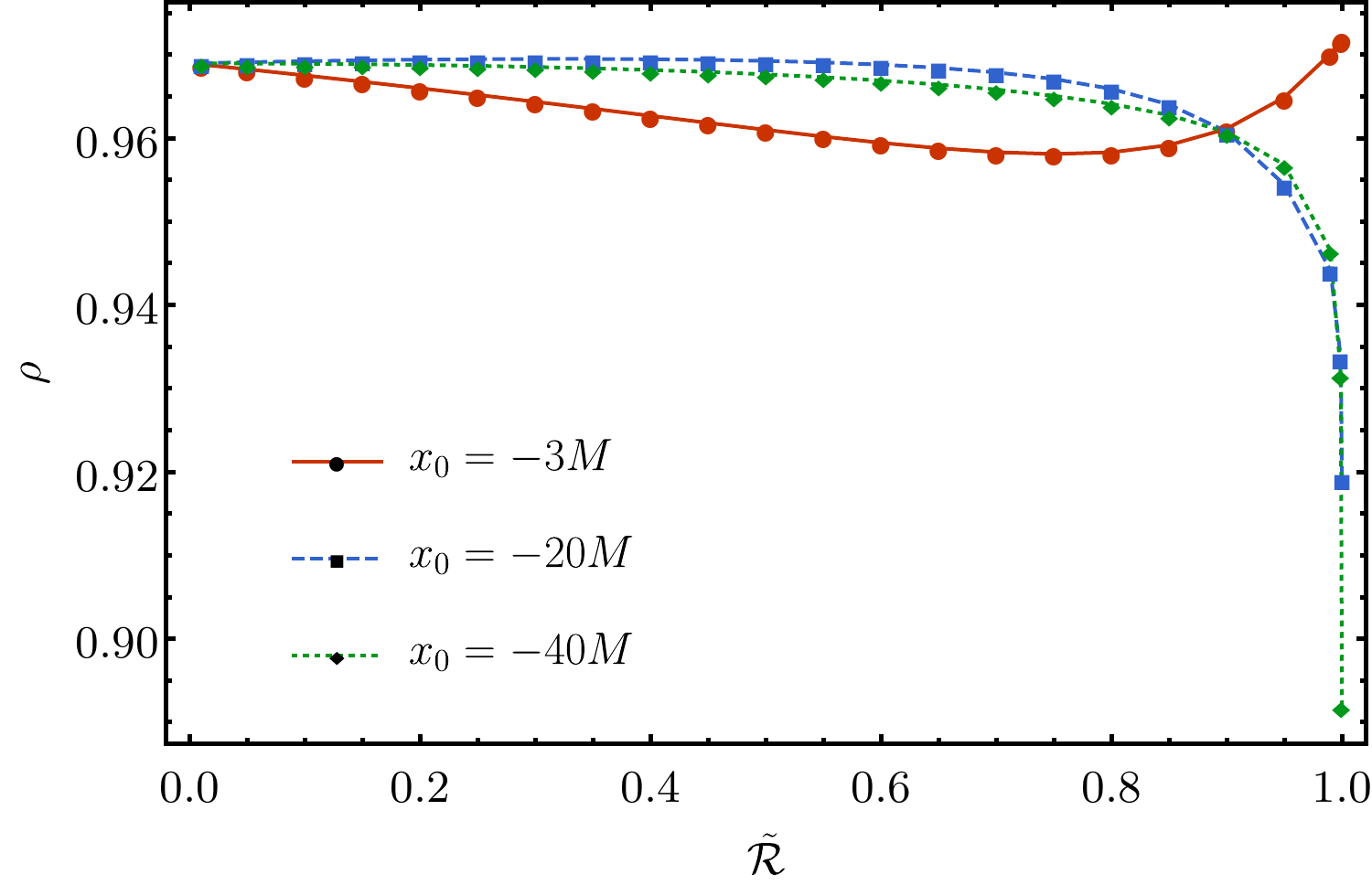}
\caption{ The overlap $\rho$ for the $(\ell,m) =(2,2)$ echo sum $Z_{\rm echo}$ for select values of $x_0$ and and $\Rb$. The waveform is generated by a test charge following the ISCO plunge orbit. The template parameters $\vec p$ are fixed in each case by maximizing the overlap for the corresponding parameters.
}
\label{fig:rho2Allecho}
\end{figure}

To investigate how the template models the full echo amplitude $\Zecho$, we investigate the overlap $\rho(Z_{\rm T},\Zecho;\vec p)$.
Note that this overlap does depend on $x_0$ and $\Rb$.
We fix $x_0$ and $\Rb$ and maximize over the template parameters $\vec p$. 
The results are shown in Fig.~\ref{fig:rho2Allecho} for $x_0= -3M, -20M$, and $-50M$ at several values of $\Rb$
ranging from $0.01$ to 1. 

We see that the overlap is generally greater than $0.96$ for $\Rb<0.99$. 
For $\Rb\geq 0.99$, the overlap for the larger values of $x_0$ drops significantly. 
The dramatic reduction in the overlap occurs because the amount of power (as determined by the power density $dP/d\omega=|\Zecho|^2$) in the echo waveform at low frequencies significantly increases as $\Rb\to 1$ when $x_0$ is large.
This power is contained in the narrow resonances appearing in Fig.~\ref{fig:MultiECOModeFrequency}. 
This degrades the overlap because the template is only designed to perform well for frequencies near the BH QNM frequency $\Omega_R$. 
For example when $x_0 =-50M$ and $\Rb=0.999$, less than $8\%$ of the power is at frequencies $|\omega | < 0.6\Omega_R$, while when $\Rb=1$, the number jumps to $35\%$.

\subsection{Energy in the echoes}

Our formalism also allows us to relate the energy in the ECO waveform to the energy in the BH waveforms on the horizon $\mathcal{H}^+$ and at asymptotic infinity $\mathcal{I}^+$. For very compact ECOs, we derive a simple relationship between the energy in the black hole waveform and the energy in the ECO waveform.

The stress energy tensor for the scalar field is $T_{\mu\nu}=\nabla_\mu \phi\nabla_\nu \phi-(1/2) g_{\mu \nu}\nabla^\rho\phi\nabla_\rho\phi$ and energy flow is governed by the energy flux vector $-T_{\mu\nu}(\partial/\partial t)^\nu$. 
Given a wave $\psi(v)$ that impinges on the horizon or a wave $\psi(u)$ that is incident on $\mathcal{I}^+$, the energy $\mathcal{E}[\psi]$ is the functional
\begin{align}
\mathcal{E}[\psi]&=\sum_{\ell m} E_{\ell m}[\psi], \\
E_{\ell m}[\psi]&=\int_{-\infty}^\infty d\tau |\dot\psi_{\ell m}(\tau)|^2 
=\int_{-\infty}^\infty \frac{d\omega}{2\pi} \omega^2|Z_{\ell m}(\omega)|^2,
\end{align}
where we have temporarily restored the harmonic indices. The last equality is an application of Parseval's theorem, and we have denoted $Z_{lm}$ as the Fourier conjugate of $\psi_{lm}$. 

The energy of the ECO waveform $E^\infty$ can be expressed in terms of the energy in the black hole waveform  $E_{\rm BH}^\infty=E[\psi_{\rm BH}^\infty]$, the energy in the echoes $E_{\rm echo}=E[\psi_{\rm echo}]$, and correlations between the echoes and the black hole waveform
\begin{align}
 E[\psi_{\rm BH}^\infty]&=E[\psi_{\rm BH}^\infty +\psi_{\rm echo}] \nonumber \\ 
 &= E_{\rm BH}^\infty + E_{\rm echo}
+2\Re
\left[\int_{-\infty}^\infty d\tau \, \dot \psi_{\rm BH}^\infty(\tau)\dot \psi_{\rm echo}(\tau)^*\right] \,.
\end{align}

In the limit that $x_0$ is much larger than the duration of each echo, the different echoes do not overlap, allowing us to neglect the correlations, so that
\begin{align}
 E^\infty \approx E_{\rm BH}^\infty + E_{\rm echo} \label{Eq:ennocor}.
\end{align}
An identical argument allows us to write the echo energy as an approximate sum of the energy in each echo. 
\begin{align}
E_{\rm echo}& \approx \sum_{n=1}^\infty E[\psi_{\rm echo}^{(n)}] 
= \sum_{n=1}^\infty \int \frac{d\omega}{2\pi}\omega^2|\Zecho^{(n)}|^2
\nonumber \\
&= \int\frac{d\omega}{2\pi}|\Rb \TBH|^2\sum_{m=0}^\infty |\Rb\RBH|^{2m}\omega^2|\ZhBH|^2
\nonumber \\
&= \int \frac{d\omega}{2\pi}\frac{|\Rb\TBH|^2}{1-|\Rb \RBH|^2}\omega^2 |\ZhBH|^2 \,,
\label{eq:KeyEnergy}
\end{align}
where we have used Eqs.~\eqref{eq:Kn} and \eqref{eq:nZecho}.

When $\Rb =1$, since $|\TBH|^2=1-|\RBH|^2$, the echo energy $E_{\rm echo}$ is precisely the energy $E_{\rm BH}^{\rm H}=E[\psi_{\rm BH}^{\rm H}]$ that would have gone down the horizon in the BH spacetime. 
When $|\Rb|<1$, there will be less energy in the echoes than the horizon waveform, falling to $0$ as $\Rb \to 0$. 
Finally, Eq.~\eqref{eq:KeyEnergy} predicts that for very compact ECOs, the relationship between the energy in the ECO waveform and BH waveforms on $\mathcal{H}^+$ and $\mathcal{I}^+$ is independent of $x_0$.

\begin{figure}[t]
\includegraphics[width = 1 \columnwidth]{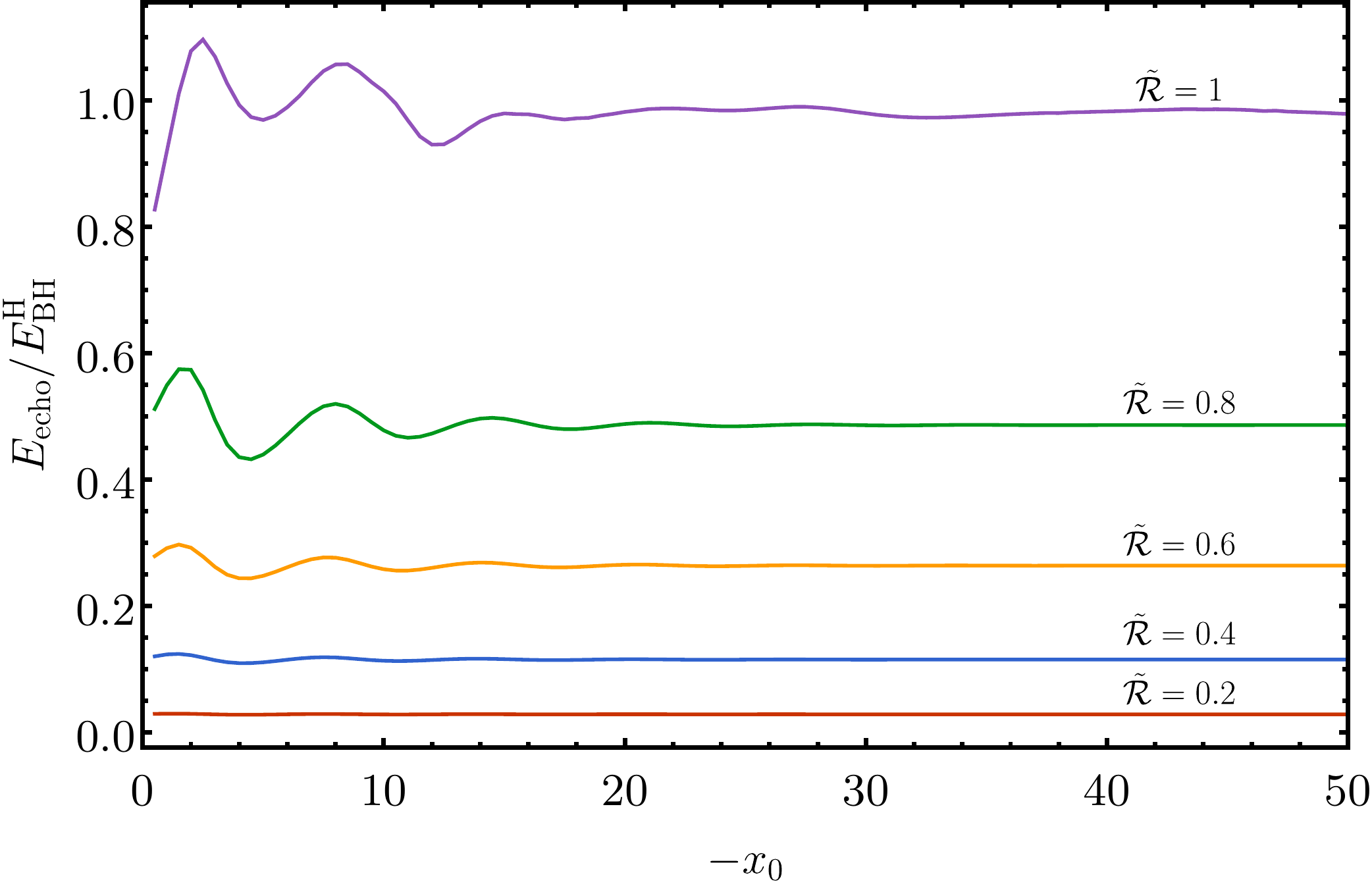}
\caption{
The energy $E_{\rm echo}$ in the $(\ell,m)=(2,2) $ component of the echo waveform compared to energy $E_{\rm BH}^{\rm H}$ in the horizon waveform for different values of $\Rb$ and $x_0$.  The waveforms come from a test charge following an ISCO plunge orbit.
}
\label{fig:Energy}
\end{figure}

Figure \ref{fig:Energy} shows $E_{\rm echo}/E_{\rm BH}^{\rm H}$ for $(\ell, m)=(2,2)$ waveforms from the ISCO plunge orbit for a variety of $\Rb$ and $x_0$. As expected, smaller values of $\Rb$ produce echoes containing less energy and the ratio becomes independent of $x_0$ as $x_0\to \infty$. 
For perfectly reflecting, extremely compact ECOs with $x_0>20M$, more than $97\%$ of the energy in the horizon waveform is radiated in the echoes.

\section{Conclusions}

In this work, we derive a relationship between the Green's functions for a massless scalar field in a BH spacetime and in the exterior region of ECOs.
This is accomplished by replacing the compact object with a reflecting boundary near the horizon of the BH. 
The exterior of any ECO can be modeled with a particular choice of boundary location and frequency dependent reflectivity.

We use the relationship between Green's functions to show that the ECO waveform seen by asymptotic observers is the same as that seen in the BH spacetime, plus additional emission from reflection off the boundary.
This additional emission can be computed by reprocessing the horizon waveform in the BH spacetime using a simple transfer function.
We find that the difference between the BH and ECO waveforms at infinity can be understood either as a superposition of echo pulses or a superposition of modes associated with poles in the ECO Green's function. 
Furthermore, we show how both the individual echoes and the new mode frequencies encode the information describing the ECO model; namely the boundary reflectivity and location.

Our formalism also explains how the BH QNMs imprint themselves in ECO waveforms: The ECO waveform has a main burst that rings down at the black hole QNM frequencies. 
In addition, the frequency content of the individual echo pulses is largely determined by the frequency content in the horizon waveform $\psi^{\rm H}_{\rm BH}$ near the BH QNM frequencies.
Despite the imprint of these frequencies on the ECO waveform, our formalism also shows that the BH QNM frequencies are not poles in the ECO Green's function. Rather, the piece of the Green's function responsible for producing the main burst and the piece responsible for the echoes both have poles at the BH QNM frequencies, which cancel in the full expression. 

We demonstrate how our formalism can be used to reprocess a black hole waveform into an ECO waveform by studying the echoes produced by a test charge spiralling in from the ISCO. We use our numerical results and analytic observations to design a simple template for the echoes that accurately reproduces our waveforms, with normalized overlaps $\rho > 0.95$ for most values of boundary location and reflectivity (taken here to be frequency independent). 

To determine the significance of our proposed template, future work will be required to extend the formalism to gravitational perturbations of Kerr. 
In addition to the added algebraic complexity, one will have to overcome the absence of Birkhoff's theorem in Kerr, as well as the lack of a simple scheme for parameterizing reflecting boundary conditions for gravitational perturbations \cite{Price:2017cjr} (see \cite{Nakano:2017fvh} for one possible prescription). 
Ideally, future work will also extend the formalism beyond test particle sources, so that comparable mass binaries can be treated. 
Nevertheless, our results indicate that a relatively simple template, combined with a prescription for reprocessing waveforms generated in black hole spacetimes, can be used to investigate the existence of ECOs and their echoes using gravitational wave observations.

\begin{acknowledgments}
We thank Vitor Cardoso, Baoyi Chen, Chad Galley, Davide Gerosa, Yiqiu Ma, David Nichols, Samaya Nissanke, Paolo Pani, Leo Stein, Saul Teukolsky, and Huan Yang for valuable discussions. 
We are grateful to Ofek Birnholtz, Vitor Cardoso, Gaurav Khanna, Hiroyuki Nakano, and Paolo Pani for providing feedback on a draft of this manuscript. The figures were made using the MaTeX package \cite{matex}. This research was supported at Caltech by NSF grant PHY-1404569, the Walter Burke Institute for Theoretical Physics, and the David and Barbara Groce startup fund.
\end{acknowledgments}

\appendix

\section{Calculation of the reflection and transmission coefficients}

\label{sec:RTcalc}

In this appendix we describe our calculation of the reflection and transmission coefficients $\mathcal R_{\rm BH}$ and $\mathcal T_{\rm BH}$, in both the time and frequency domains.

\subsection{Time Domain}

\begin{figure}[t]
\includegraphics[width = 1 \columnwidth]{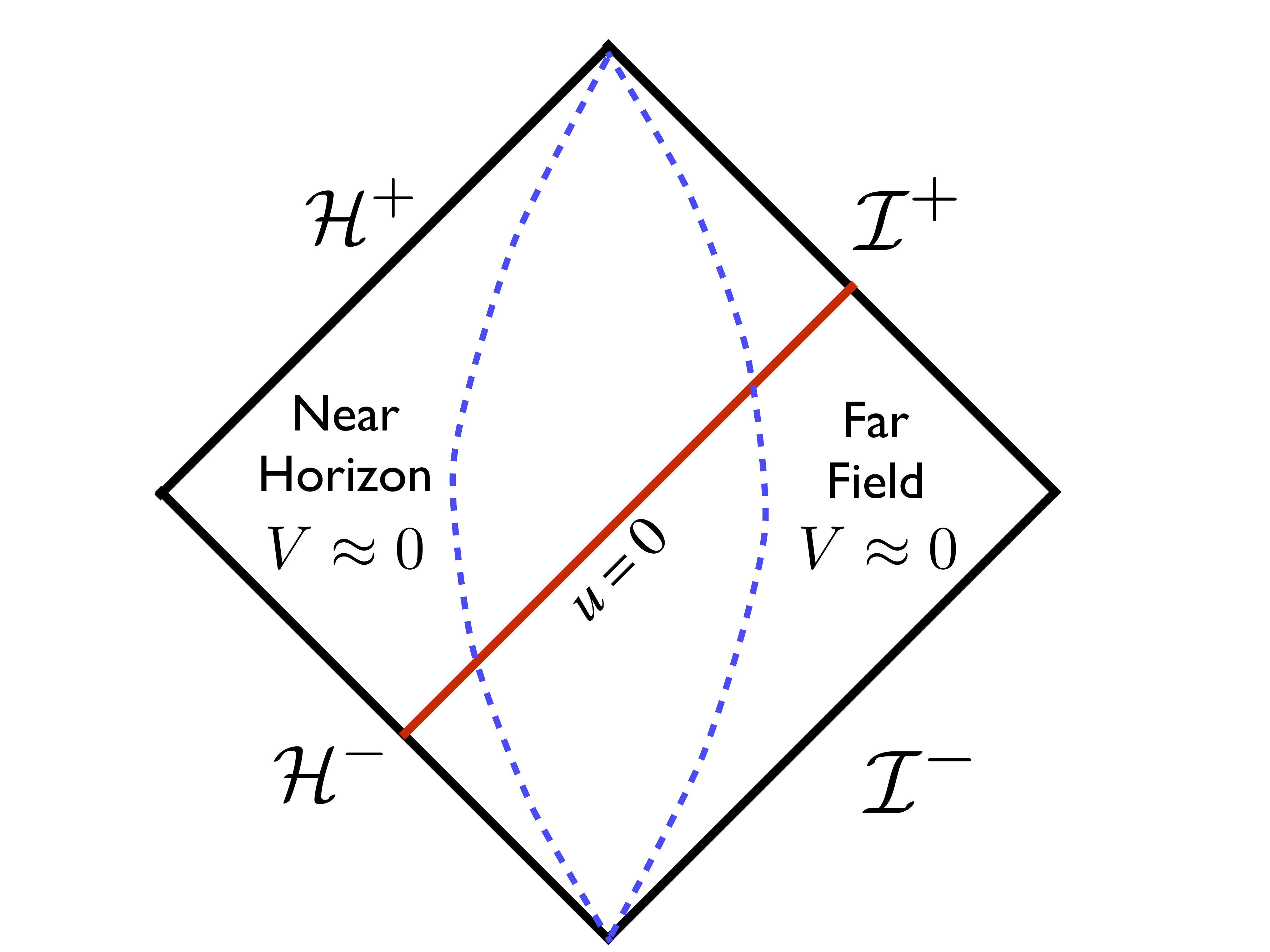}
\caption{
A Penrose diagram illustrating the relevant surfaces of the characteristic initial value definition of $\mathcal T_{\rm BH}$ and $\mathcal R_{\rm BH}$. 
Initial data, consisting of a delta function pulse at $u=0$ (red line), is posed on $\mathcal{H} ^-$ and $\mathcal{I}^-$. The transfer function $\mathcal T_{\rm BH}$ is extracted off of $\mathcal{I}^+$ and $\mathcal R_{\rm BH}$ is extracted off of $\mathcal{H} ^+$. The blue dashed lines approximately bound the near-horizon and far-field regions where $V\approx 0$. 
}
\label{fig:RTDom}
\end{figure}

The scattering coefficients $\RBH$ and $\TBH$ are defined from the frequency domain solution $\psiup$ to Eq.~\eqref{eq:RW}. 
An equivalent time-domain definition is found in terms of a solution $\psi$ to the characteristic initial value problem 
\begin{align}
\frac{\partial^2 \psi}{\partial u \partial v}+\frac{fV}{4}\psi= 0 \label{eq:psiwave}
\end{align}
with characteristic initial data posed on the past horizon $\mathcal{H}^-$ and past null infinity $\mathcal{I}^-$ consisting of a delta function pulse
\begin{align}
&\left. \psi(u)\right |_{\mathcal{H}^-} =\delta (u),& &\left. \psi(v)\right |_{\mathcal{I}^-}=0
\end{align}
as shown in Fig.~\ref{fig:RTDom}.
Then $\mathcal T_{\rm BH}(u)$ is the field $\psi(u)|_{\mathcal{I}^+}$ evaluated at future null infinity and $\mathcal R_{\rm BH}(v)$ is the field $\left. \psi(v)\right |_{\mathcal{H}^+}$ evaluated on the future horizon.
This is seen as follows.

When $V=0$, the general solution to Eq.~\eqref{eq:psiwave} is a superposition of an outward traveling wave and an inward traveling wave, 
\begin{align}
\psi(v,u)=h(u)+k(v) \label{eq:wavegen},
\end{align}
where $h$ and $k$ are free functions.
The potential can be neglected, $V\approx 0$, in the near horizon region, roughly bounded by the left blue dashed line in Fig.~\ref{fig:RTDom}, and also in the far field region, roughly bounded by the right blued dashed line. We match the general solution Eq.~\eqref{eq:wavegen} to the boundary data in these regions to obtain 
\begin{align}
\psi(v,u)=\begin{cases}
\delta(u)+\left. \psi(v)\right |_{\mathcal{H}^+}, & x \to -\infty \,, \\
\left. \psi(u)\right |_{\mathcal{I}^+}, & x \to \infty \,.
\end{cases}
\end{align}
Notice the field on the horizon is not determined by the initial conditions in the near horizon matching region. 
Likewise the field at future null infinity is not determined by the initial conditions in the far-field matching region. 
Calculating these fields requires all of the initial data.

Rewriting the solution in $(t,x)$ coordinates and taking the Fourier transform with respect to $t$ yields
\begin{align}
\tilde \psi(\omega,x)
&=\begin{cases}
e^{i\omega x}+\ \tilde \psi(\omega) |_{\mathcal{H}^+}e^{-i\omega x}, & x \to -\infty \\ \tilde \psi(\omega) |_{\mathcal{I}^+}e^{i\omega x}, & x \to \infty.
\end{cases}
\end{align}
Comparing this with frequency domain definition Eq. ~\eqref{eq:RTAmplitudes} of $\mathcal R_{\rm BH}$ and $\mathcal T_{\rm BH}$, we identify 
\begin{align}
\tilde \psi(\omega)|_{\mathcal{H}^+} &= \RBH(\omega) \,, \\ \tilde \psi(\omega)|_{\mathcal{I}^+} &= \TBH(\omega) \,,
\end{align}
establishing the equivalence of the two definitions.

For our numerical calculations, it is important to realize that $\mathcal T_{\rm BH}(u)$ only has support for $u\geq 0$ and $\mathcal T_{\rm BH}(0)=\delta(0)$. The first fact follows from $\psi(v,u)=0$ for $u<0$, since for these times $\psi$ lies in the domain of dependence of the portion of initial data which is equal to zero. 
The second conclusion follows from the high frequency behavior $\TBH\to 1$ as $\omega/\Omega_R \to \pm\infty$ \cite{Caponthesis}. This implies that $\TBH = 1 + f(\omega)$, where $f\to 0$ as $\omega \to \pm \infty$. Taking the Fourier transform of both sides gives the delta function at $u=0$. 

We use our characteristic code for homogeneous solutions to wave equations detailed in Sec.~\ref{sec:charcode} to solve this characteristic initial value problem.
Namely, we pose the initial data on the future part of a null cone described by $v=v_0$ and $u=0$ and choose $-v_0$ large enough that the delta function pulse $\delta(u)$ is deep in the near horizon region. 
We use a discrete approximation for the delta function in the initial data  
\begin{align}
\delta(u) =\begin{cases}
\displaystyle
\frac{1}{2(2h)}, & u =0 \\
0, & \text{otherwise}
\end{cases}.
\end{align}
where our numerical grid is spaced by $2h$.
We extract $\mathcal R_{\rm BH}$ off of the ray $u=u_E$ in our computational domain that is closest to $\mathcal{H}^+$. 
Similarly we extract $\mathcal T_{\rm BH}$ off of the ray $v=v_E$ in our computational domain that is closest to $\mathcal{I}^+$.

We performed convergence checks on our choice of stepsize $h$, initial data ray location $v_0$ and the location of the extraction rays $v_E$ and $u_E$. 
We used $h =0.025M$. 
We verified that the same numerical approximation of the $\delta(u)$ that we used in our initial data appears in $\mathcal T_{\rm BH}$. 
For calculations in the paper that rely on $\mathcal T_{\rm BH}$, we insert the $\delta$ function analytically and only use the smooth part of $\mathcal T_{\rm BH}$ from our code.
To obtain the smooth part of $\mathcal T_{\rm BH}(u)$ near zero we extrapolated this data backwards in time a single time step.

\subsection{Frequency Domain}

For computations that required accurate frequency domain representations of $\RBH$ and $\TBH$, we also computed  $\RBH$ and $\TBH$ directly in the frequency domain. This also provided an independent check of our time domain methods. 

At a fixed frequency, the homogeneous wave equation \eqref{eq:RW} together with one of the two boundary conditions in Eq.~\eqref{eq:psiup} forms a boundary value problem for  $\psiup(\omega, x)$. The coefficients $B_{\rm out}$ and $B_{\rm in}$ necessary to compute  $\RBH$ and $\TBH$ are determined from the solution and its derivative near the opposite boundary by comparing to the remaining boundary condition. 

We numerically integrated outward from the horizon, using an analytic third-order expansion of $\tilde \psi$ to match the boundary condition there.
We extracted the field at a large radius $r = 1000M$, matching to an asymptotic expansion of $\tilde \psi$ including terms up to third order in $1/r$.

\section{Point Particle Waveforms}
\label{sec:Numerics}

In this appendix we provide Green's functions solutions for the scalar field $\psi_{\rm BH}$ in the BH spacetime, specialized to point particle sources for observers at future null infinity $\mathcal{I}^+$ and the future horizon $\mathcal{H}^+$.

\subsection{Green's Function solution}
\label{sec:GFsol}

The boundary conditions for $\psi_{\rm BH}$ in Eq. \eqref{eq:ZBHdef} select the retarded solution to the Klein-Gordon equation
\begin{align}
&\psi_{\rm BH}(x,t)=\int_{-\infty}^{\infty}dt'\int_{-\infty}^{\infty}dx'S(x,t)g_{\rm BH}(x,x',t-t'), \nonumber \\
& S(x,t)=-rf(r)\rho_{\ell m}(x,t),
\end{align}
constructed from the retarded (biscalar) Green's function $g_{\rm BH}(x,x',\tau)$ and the spherical harmonic components of the scalar charge density\footnote{Note that $S$, $\psi_{\rm BH}$, $g_{\rm BH}$ and all variants of them which appear in this appendix have $(\ell,m)$ indices which we suppress for brevity.}. The retarded Green's function obeys $g_{\rm BH}(x,x',t-t')=0$ when $t-t'<|x-x'|$ and the differential equation
\begin{align}
&\frac{\partial^2g_{\rm BH}}{\partial x^2}-\frac{\partial^2g_{\rm BH}}{\partial t^2}-f(r)V(r)g_{\rm BH}=\delta(t-t')\delta(x-x'). \label{eq:gBHwave}
\end{align}

We are interested in the waveforms on either the BH horizon or at asymptotic infinity.
This leads us to consider the asymptotic Green's functions 
\begin{align}
g_{\rm BH} \sim
\begin{cases}
g_{\rm H}(x', v-v'), & \text{as } x \to -\infty,\, v\text{ fixed}\,,\\
g_\infty(x', u - u'), & \text{as } x \to \infty,\, u \text{ fixed}\,,
\end{cases}
\end{align}
which describe the response on the horizon and at infinity, respectively.

We also need the appropriate source functions, specialized to ingoing coordinates $(v,x)$ and outgoing coordinates $(u,x)$.
The scalar charge density of a point particle of scalar charge $q$, following the trajectory $x^\mu_p(\tau)$ is 
\begin{align}
\rho(x^\mu)=q\int d\tau \frac{\delta^{(4)}(x^\mu-x^\mu_p(\tau))}{\sqrt{-g}},
\end{align}
Resolving into spherical harmonics $\rho=\sum\lm\rho\lm Y\lm$, re-parameterizing by advanced time, and writing the result in ingoing coordinates leads to 
\begin{align}
&S(x,v)=\hat S_{\rm in}(v)\delta (x-x_p),\nonumber \\
& \hat S_{\rm in}(v)=\frac{-q Y_{\ell m}^*(\theta_p,\phi_p)}{r_p (dv_p/d\tau)}\,,
\end{align}
where the trajectory is evaluated at $v$. Similarly, if we re-parameterize by the retarded time, and write the result in outgoing coordinates, the source is
\begin{align}
&S(x,u)=\hat S_{\rm out}(u)\delta (x-x_p),\nonumber \\
& \hat S_{\rm out}(u)=\frac{-qY_{\ell m}^*(\theta_p,\phi_p)}{r_p (du_p/d\tau)} \,,
\end{align}
where the trajectory is evaluated at the retarded time $u$.

With these definitions, the horizon waveform is
\begin{align}
\psi_{\rm BH}^{\rm H}(v) & =
\int_{-\infty}^\infty dx'\int_{-\infty}^\infty dv' S(x',v')g_{\rm H}(x',v-v') \nonumber \\
& = \int_{-\infty}^\infty dv'\hat S_{\rm in}(v')g_{\rm H}(x_p(v'),v-v'). \label{eq:horizonwave1}
\end{align}
For a particle that crosses the horizon at an advance time $v =v_{\rm H}$, this becomes, using the causal property of the retarded Green's function,
\begin{align}
\psi^{\rm H}_{\rm BH}(v)=\begin{cases}
\displaystyle
\int_{-\infty}^v dv'\hat S_{\rm in} (v')g_{\rm H}(x_p(v'),v-v'), & v< v_{\rm H}  \,, \\
\displaystyle
\int_{-\infty}^{v_{\rm H}} dv'\hat S_{\rm in} (v')g_{\rm H}(x_p(v'),v-v'), & v\geq v_{\rm H} \,. \label{eq:horizonwave}
\end{cases}
\end{align}
Meanwhile, the asymptotic waveform is given by
\begin{align}
\psi_{\rm BH}^{\infty}(u)& =
\int_{-\infty}^\infty dx'\int_{-\infty}^\infty du' S(x',u')g_{\infty}(x',u-u')
\nonumber \\
&=
\int_{-\infty}^u du'\hat S_{\rm out}(u')g_{\infty}(x_p(u'),u-u'). \label{eq:infinitywave}
\end{align}
where we have again used causality to truncate the upper limit of the integration to $u$.

In this paper, we extensively study the radiation produced by a test charge on the ISCO plunge orbit. Such a particle asymptotes to the ISCO radius $r=6M$ as $t\to - \infty$ and has a specific energy $E_{\rm ISCO}=2\sqrt{2}/3$ and a specific angular momentum of $L_{\rm ISCO}=\sqrt{12}M$. 
To calculate the waveforms $\psi_{\rm BH}^\infty$ and $\psi_{\rm BH}^{\rm H}$ we rely on Eqs.~\eqref{eq:horizonwave} and \eqref{eq:infinitywave} with analytic expressions for the trajectory found in \cite{Hadar:2009ip}, and a Green's function that we compute numerically using a characteristic code detailed in Sec.~\ref{sec:charcode}.

\subsection{Characteristic Initial Value Problem for the Green's function}

We obtain the retarded Green's function $g_{\rm BH}$ for the scalar field in the BH spacetime as the solution of a characteristic initial value problem.
In null coordinates $(v,u)$, Eq.~\eqref{eq:gBHwave} for $g_{\rm BH}(v,v',u,u')$ takes the form
\begin{align}
\frac{\partial ^2 g_{\rm BH}}{\partial v\partial u}+\frac{fV}{4}g_{\rm BH}=-\frac{1}{2}\delta(\Delta u)\delta(\Delta v) \label{eq:gwavenull}
\end{align}
where $\Delta u =u-u'$, $\Delta v =v-v'$. 
Causality motivates us to look for a distributional solution
\begin{align}
g_{\rm BH}(v,v',u,u')=\hat g(v,v',u,u')\theta(\Delta u)\theta(\Delta v), \label{eq:ansatz}
\end{align}
where $\hat g$ is a smooth function defined in the future light cone of the source point $(v',u')$. Substitution of the ansatz \eqref{eq:ansatz} in $\eqref{eq:gwavenull}$ yields
\begin{align}
&\delta (\Delta u)\delta(\Delta v)\hat g+\theta(\Delta u)\delta(\Delta v)\frac{\partial \hat g}{\partial u} +\theta(\Delta v)\delta (\Delta u)\frac{\partial \hat g}{\partial v} \nonumber \\
&+\theta(\Delta u)\theta(\Delta v)\left(\frac{\partial ^2 \hat g}{\partial v\partial u}+\frac{fV}{4}\hat g\right)=-\frac{1}{2}\delta (\Delta u)\delta(\Delta v)
\end{align}
We now equate terms of equal singularity strength.
The first term on the LHS balances the RHS if we demand $[\hat g]\equiv g(v',v',u',u')=-1/2$. The second term, which is only nonzero along $v=v'$, vanishes if we demand $\left.\partial_u \hat g\right |_{v=v'}=0$, which can be integrated to yield $\hat g|_{v=v'}=-1/2$. Likewise, setting the third term to zero yields $\hat g|_{u=u'}=-1/2$. Finally, the fourth term vanishes if $\hat g$ satisfies the homogeneous wave equation equation 
\begin{align}
\frac{\partial ^2 \hat g}{\partial v\partial u}+\frac{fV}{4}\hat g =0 
\label{eq:ghom}
\end{align}
in the forward light cone of source point.

Equation \eqref{eq:ghom}, together with the initial data $\hat g =-1/2$ posed on the future part of the null cone formed by the rays $u=u'$ and $v=v'$, is a characteristic initial value problem for $g_{\rm BH}$.
We solve this numerically using a characteristic code described in Sec.~\ref{sec:charcode}.

\subsection{Characteristic Code}
\label{sec:charcode}
\begin{figure}[t]
\includegraphics[width = 1 \columnwidth]{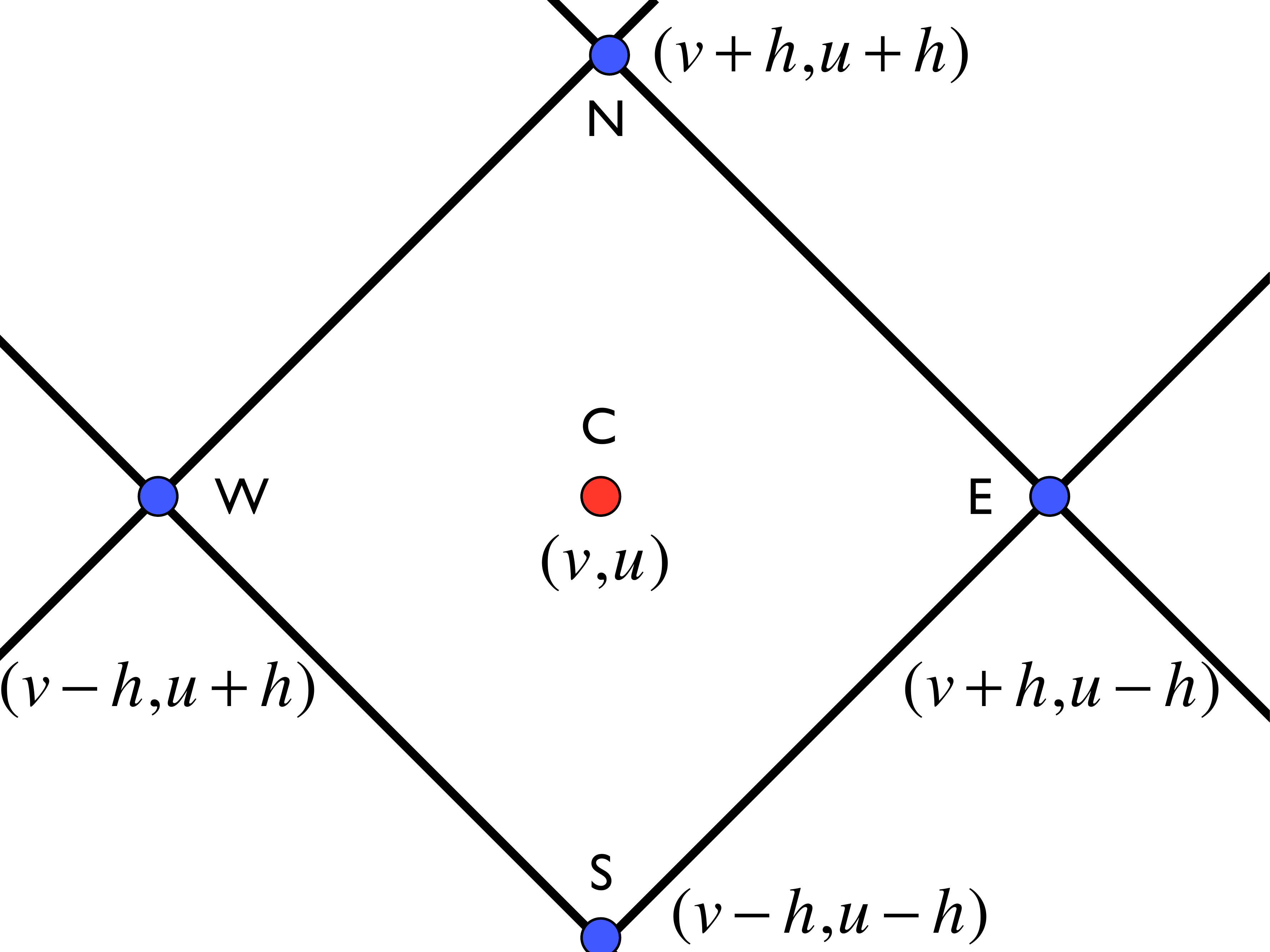}
\caption{
A generic computational cell in our characteristic evolution scheme.
}
\label{fig:ComCell}
\end{figure}

We numerically compute $\hat g_{\rm BH}$ using  a finite-difference characteristic code based on the method of Price and Lousto \cite{Lousto:1997wf}. 
For this, we fix a source point $(v',u')$ and solve the homogeneous wave equation \eqref{eq:ghom} obeyed by $\hat g(v,u)$. 
We discretize the field point coordinates $(v,u)$  onto a rectangular grid with nodes spaced by $2h$. 

A standard computational cell centered on the point $C=(v,u)$ is shown in Fig.~\ref{fig:ComCell}.
Referring to the figure, given the data $\psi_S$, $\psi_W$, and $\psi_E$ on the bottom three corners of a computational cell, the value on top corner $\psi_N$ can be obtained with the stepping algorithm
\begin{align}
\label{eq:step}
&\psi_N=-\psi_S+(1+2W_Ch^2)(\psi_E+\psi_W) \\
&W_C=\left .-\frac{fV}{4}\right |_{C}. 
\end{align}
This algorithm can be derived by integrating the homogeneous wave equation \eqref{eq:ghom} over a computational cell with $\mathcal{O}(h^4)$ accuracy.
Our code inputs initial data on the future part of the light cone formed by the rays $v=v'$ and $u=u'$ and 
is second order convergent. 
We generate values for $\psi$ on the remaining nodes of the grid using the stepping algorithm \eqref{eq:step}.

To obtain $g_\infty(x',\Delta u)$, we further fix $\Delta u$ and use our characteristic code to obtain $g_{\rm BH}$ as a function of field point radius $r$. 
Using the fact that the field has an expansion in powers of $1/r$, we then extrapolate the field to future null infinity using Richardson extrapolation.

To extract $g_{\rm H}(x',\Delta v)$, we  use our characteristic code to obtain $g_{\rm BH}$ evaluated on the ray $u= \text{constant}$ that is closest to the horizon in our computational domain. 
For early advanced time $\Delta v$, this ray is buried deep in the near horizon region, and we approximate $g_{\rm H}(x',\Delta v)$ as $g_{\rm BH}$ evaluated on this ray. 
We check that this scheme converges as we move the extraction ray $u= \text{constant}$ towards $\mathcal{H}^+$.

We perform these calculations for radii between $r'-2M =1.7\times 10^{-5}$ and $r' = r_{\rm ISCO} =6M$ with $\Delta x' =1$. 
We then interpolate between these values to obtain $g_{\rm H}(x',\Delta v)$ and $g_{\infty}(x',\Delta u)$ that we use in the calculations presented in this paper.

\subsection{Windowing and Frequency Domain Waveforms}
\label{sec:window}

Waveforms from physically relevant orbits are finite in duration. The waveforms produced by the exact ISCO plunge orbit are not; at late times, the waveforms ringdown to zero, but at arbitrarily early times they have an oscillation at $\omega = m\Omega_{\rm ISCO}$ due to the test charge orbiting on the ISCO. 

Hence, for all calculations in this paper we consider the echoes produced by a windowed horizon waveform. More precisely we apply a one-sided version of the Planck-Taper \cite{McKechan:2010kp}  window function to the exact ISCO plunge horizon waveforms:
\begin{align}
\sigma_T(t, t_1, n) =\begin{cases}
0, & t\leq t_1 \\ 
\displaystyle
\frac{1}{1+e^{z}}, & t_1 < t <t_2\\
1, & t\geq t_2 \\ 
\end{cases},
\end{align}
where $t_1$ is free parameter indicating when the window starts, $t_2=t_1+ 2a \pi/\Omega_{\rm ISCO}$ with a $a$ free parameter, and z is a function that goes from $\infty$ at $t_1$ to $-\infty$ at $t_2$,
\begin{align}
z=\frac{t_2-t_1}{t-t_1}+\frac{t_2-t_1}{t-t_2}.
\end{align}
We choose parameters that leave 3 oscillations at early times near $\omega \approx m\Omega_{\rm ISCO}$ and smoothly turn on over the course of two oscillations.

We obtain the horizon waveform $\ZhBH$ in the frequency domain by numerically performing the inverse Fourier transform of the time domain waveform $\psi_{\rm BH}^{\rm H}$.

\section{Wormhole Reflectivity}
\label{sec:wormdetails}

In this appendix, we compute $\Rb(\omega)$ for a wormhole \cite{Cardoso:2016rao} describing two Schwarzschild spacetimes of mass $M$ identified with a thin shell of exotic stress-energy at an areal radius $r_0$ corresponding to a tortoise coordinate location of $x_0$. 
Note that the value of $\Rb$ depends on our phase convention, and we use that of Eq.~\eqref{eq:refBC}, which is invariant under shifts of the origin of the tortoise coordinate $x$. 

To begin, define a new tortoise coordinate $y$ covering the entire wormhole spacetime, 
\begin{align}
\frac{dr}{dy}=\begin{cases}
\displaystyle
\left(1-\frac{2M}{r}\right), & y > 0 \\
\displaystyle
-\left(1-\frac{2M}{r}\right), & y <0,
\end{cases}
\end{align}
with a different origin $y(r_0)=0$ than
is used for the coordinate $x$. 
Scalar waves propagating in the wormhole spacetime are described by the scalar wave equation on the domain $-\infty<y<\infty$, with a non-differentiable, but continuous potential $V(y)$ at $y=r_0$. 
The reflectivity $\Rb$ is determined by matching the  solution obeying the outgoing boundary condition in the left half of the universe to a solution in the right half.

We accomplish this using the homogeneous solution $\psiup$, although with a different phase normalization than in Eq.~\eqref{eq:psiup} due to the shift in the origin $y$,
\begin{align}
\psiup(y) & \sim 
e^{i\omega y} \,, & y \to \infty \label{eq:psiupapp}
\end{align}
For compact wormholes $r_0\to 2M$ the potential $V\approx 0$ near the location $x= 0$ and we have
\begin{align}
 \label{eq:psiupzero}
\psiup(y) & \sim 
C_{\rm out}(\omega) e^{i \omega y} + C_{\rm in}(\omega) e^{-i\omega y}\,, & y \to 0
\end{align} 
From these we define $\RW=C_{\rm in}/C_{\rm out}$
to denote the reflection coefficient using the phase convention \eqref{eq:psiupapp}. 

In the left half of the universe, $\psiup(-y)$ is the solution describing waves that are completely outgoing at null infinity. 
Near the matching radius $y=0$, we have by definition
\begin{align}
\psiup(-y)\propto e^{-i\omega y}+\RW e^{i\omega y}.
\end{align}
This matches to the form of desired boundary condition for waves in the right half, $\psi \propto e^{-i\omega y}+\Rb e^{i\omega y}$, if we choose $\Rb=\RW$. 

Finally, we express this result in terms of the BH scattering coefficients, which use the phase convention of Eq.~\eqref{eq:psiup}. 
The scattering coefficients defined by Eq.~\eqref{eq:psiupapp} are related to those of Eq.~\eqref{eq:psiup} through a simple shift of the origin of $y$.
This means that
\begin{align}
\RW = e^{-2i\omega x_0}\RBH
\end{align}
We see then that the wormhole can be treated using a reflecting boundary at $x_0$ with
\begin{align}
\Rb(\omega)=\RBH(\omega)e^{-2i\omega x_0}\,.
\end{align}
We use this simple result to explore the echoes in wormhole spacetimes.

\section{Fourier Transform of Decaying Sequence of Pulses}
\label{sec:DDcomb}

In this appendix, we derive the Fourier transform of the $f(t)$ given in Eq.~\eqref{eq:DDcomb}
\begin{align}
\label{eq:DDcombApp}
&f(t)=\sum_{n=0}^\infty \gamma^n \delta\left(t-nT\right) \,,
\end{align}
which involves some nontrivial manipulations to arrive at the form in Eq.~\eqref{eq:DDcombFT}.
Namely, directly evaluating the Fourier transform with the delta functions gives
\begin{align}
\tilde f(\omega)=\int_{-\infty}^{\infty}dt f(t) e^{i\omega t}=\sum_{n=0}^\infty \gamma^n e^{i\omega nT}
\end{align}
The derivation of the two different forms is related to the fact that one can write the Fourier transform $\tilde c(\omega)$ of a Dirac comb with period $T=2\pi/\Delta\omega$
\begin{align}
c(t)=\sum_{n=-\infty}^{\infty}\delta(t-nT)
\end{align}
in two different ways. On one hand, directly integrating over the $\delta$ functions gives
\begin{align}
\tilde c(\omega)=\sum_{n=-\infty}^{\infty}e^{i\omega n T}.
\end{align}
On the other hand, the Dirac comb is a periodic function with a period $T$ and can be expanded as a Fourier series
\begin{align}
&c(t)=\sum_{n=-\infty}^{\infty}c_n e^{-i\Delta \omega n t}, \nonumber \\
&c_n=\frac{1}{T}\int_{-T/2}^{T/2}dte^{i\Delta \omega nt}c(t) =\frac{1}{T} \,.
\end{align}
Comparing this to the expression for the inverse Fourier transform $c(t)=1/(2\pi)\int d\omega e^{-i\omega t}\tilde c(\omega)$ leads to the alternate form of $\tilde c(\omega)$
\begin{align}
\tilde c(\omega)=\Delta \omega \sum_{n=-\infty}^{\infty}\delta(\omega- n \Delta \omega )
\label{eq:cFT}
\end{align}
We use this result to derive Eq.~\eqref{eq:DDcombFT} for $\tilde f$. 

First note that $f$ has a simple relationship to the Dirac comb
\begin{align}
f(t)&=\sum_{n=0}^{\infty}\gamma^n \delta(t-nT)
\nonumber \\ &
=e^{(t/T) \ln\gamma}\sum_{n=0}^{\infty} \delta(t-nT)
=b(t)c(t) \,, \\
b(t)& \equiv \theta(t)e^{(t/T) \ln\gamma} \,,
\end{align}
where $\theta(t)$ is the unit step function.
Then the convolution property of the Fourier transform implies that 
\begin{align}
\tilde f(\omega)=\int_{-\infty}^{\infty}\frac{d\omega'}{2\pi}\tilde b(\omega -\omega')\tilde c(\omega').
\label{eq:convprop}
\end{align}
The Fourier transform $\tilde b(\omega)$ is
\begin{align}
\tilde b(\omega)=\int_{0}^{\infty}dt \, e^{i \omega t + (t/T) \ln\gamma}=-\frac{1}{i\omega+\ln\gamma/T} \,,
\label{eq:bFT}
\end{align}
where we have used the fact that $\ln\gamma <0$ for $0<\gamma \leq 1$.
Substituting Eq.~\eqref{eq:cFT} and Eq.~\eqref{eq:bFT} into Eq.~\eqref{eq:convprop} and integrating over the Dirac comb then yields Eq.~\eqref{eq:DDcombFT}.

\bibliography{main.bbl}

\begin{thebibliography}{57}%
\makeatletter
\providecommand \@ifxundefined [1]{%
 \@ifx{#1\undefined}
}%
\providecommand \@ifnum [1]{%
 \ifnum #1\expandafter \@firstoftwo
 \else \expandafter \@secondoftwo
 \fi
}%
\providecommand \@ifx [1]{%
 \ifx #1\expandafter \@firstoftwo
 \else \expandafter \@secondoftwo
 \fi
}%
\providecommand \natexlab [1]{#1}%
\providecommand \enquote  [1]{``#1''}%
\providecommand \bibnamefont  [1]{#1}%
\providecommand \bibfnamefont [1]{#1}%
\providecommand \citenamefont [1]{#1}%
\providecommand \href@noop [0]{\@secondoftwo}%
\providecommand \href [0]{\begingroup \@sanitize@url \@href}%
\providecommand \@href[1]{\@@startlink{#1}\@@href}%
\providecommand \@@href[1]{\endgroup#1\@@endlink}%
\providecommand \@sanitize@url [0]{\catcode `\\12\catcode `\$12\catcode
  `\&12\catcode `\#12\catcode `\^12\catcode `\_12\catcode `\%12\relax}%
\providecommand \@@startlink[1]{}%
\providecommand \@@endlink[0]{}%
\providecommand \url  [0]{\begingroup\@sanitize@url \@url }%
\providecommand \@url [1]{\endgroup\@href {#1}{\urlprefix }}%
\providecommand \urlprefix  [0]{URL }%
\providecommand \Eprint [0]{\href }%
\providecommand \doibase [0]{http://dx.doi.org/}%
\providecommand \selectlanguage [0]{\@gobble}%
\providecommand \bibinfo  [0]{\@secondoftwo}%
\providecommand \bibfield  [0]{\@secondoftwo}%
\providecommand \translation [1]{[#1]}%
\providecommand \BibitemOpen [0]{}%
\providecommand \bibitemStop [0]{}%
\providecommand \bibitemNoStop [0]{.\EOS\space}%
\providecommand \EOS [0]{\spacefactor3000\relax}%
\providecommand \BibitemShut  [1]{\csname bibitem#1\endcsname}%
\let\auto@bib@innerbib\@empty
\bibitem [{\citenamefont {Thorne}(1972)}]{Thorne:1972ji}%
  \BibitemOpen
  \bibfield  {author} {\bibinfo {author} {\bibfnamefont {K.~S.}\ \bibnamefont
  {Thorne}},\ }\bibfield  {title} {\enquote {\bibinfo {title} {{NONSPHERICAL
  GRAVITATIONAL COLLAPSE: A SHORT REVIEW}},}\ }\href@noop {} {\  (\bibinfo
  {year} {1972})}\BibitemShut {NoStop}%
\bibitem [{\citenamefont {Abbott}\ \emph
  {et~al.}(2016{\natexlab{a}})\citenamefont {Abbott} \emph
  {et~al.}}]{Abbott:2016blz}%
  \BibitemOpen
  \bibfield  {author} {\bibinfo {author} {\bibfnamefont {B.~P.}\ \bibnamefont
  {Abbott}} \emph {et~al.} (\bibinfo {collaboration} {Virgo, LIGO
  Scientific}),\ }\bibfield  {title} {\enquote {\bibinfo {title} {{Observation
  of Gravitational Waves from a Binary Black Hole Merger}},}\ }\href {\doibase
  10.1103/PhysRevLett.116.061102} {\bibfield  {journal} {\bibinfo  {journal}
  {Phys. Rev. Lett.}\ }\textbf {\bibinfo {volume} {116}},\ \bibinfo {pages}
  {061102} (\bibinfo {year} {2016}{\natexlab{a}})},\ \Eprint
  {http://arxiv.org/abs/1602.03837} {arXiv:1602.03837 [gr-qc]} \BibitemShut
  {NoStop}%
\bibitem [{\citenamefont {Abbott}\ \emph
  {et~al.}(2016{\natexlab{b}})\citenamefont {Abbott} \emph
  {et~al.}}]{Abbott:2016nmj}%
  \BibitemOpen
  \bibfield  {author} {\bibinfo {author} {\bibfnamefont {B.~P.}\ \bibnamefont
  {Abbott}} \emph {et~al.} (\bibinfo {collaboration} {Virgo, LIGO
  Scientific}),\ }\bibfield  {title} {\enquote {\bibinfo {title} {{GW151226:
  Observation of Gravitational Waves from a 22-Solar-Mass Binary Black Hole
  Coalescence}},}\ }\href {\doibase 10.1103/PhysRevLett.116.241103} {\bibfield
  {journal} {\bibinfo  {journal} {Phys. Rev. Lett.}\ }\textbf {\bibinfo
  {volume} {116}},\ \bibinfo {pages} {241103} (\bibinfo {year}
  {2016}{\natexlab{b}})},\ \Eprint {http://arxiv.org/abs/1606.04855}
  {arXiv:1606.04855 [gr-qc]} \BibitemShut {NoStop}%
\bibitem [{\citenamefont {Abbott}\ \emph
  {et~al.}(2016{\natexlab{c}})\citenamefont {Abbott} \emph
  {et~al.}}]{TheLIGOScientific:2016pea}%
  \BibitemOpen
  \bibfield  {author} {\bibinfo {author} {\bibfnamefont {B.~P.}\ \bibnamefont
  {Abbott}} \emph {et~al.} (\bibinfo {collaboration} {Virgo, LIGO
  Scientific}),\ }\bibfield  {title} {\enquote {\bibinfo {title} {{Binary Black
  Hole Mergers in the first Advanced LIGO Observing Run}},}\ }\href {\doibase
  10.1103/PhysRevX.6.041015} {\bibfield  {journal} {\bibinfo  {journal} {Phys.
  Rev.}\ }\textbf {\bibinfo {volume} {X6}},\ \bibinfo {pages} {041015}
  (\bibinfo {year} {2016}{\natexlab{c}})},\ \Eprint
  {http://arxiv.org/abs/1606.04856} {arXiv:1606.04856 [gr-qc]} \BibitemShut
  {NoStop}%
\bibitem [{\citenamefont {Abbott}\ \emph {et~al.}(2017)\citenamefont {Abbott}
  \emph {et~al.}}]{LIGO:GW170104}%
  \BibitemOpen
  \bibfield  {author} {\bibinfo {author} {\bibfnamefont {B.~P.}\ \bibnamefont
  {Abbott}} \emph {et~al.} (\bibinfo {collaboration} {LIGO Scientific and Virgo
  Collaboration}),\ }\bibfield  {title} {\enquote {\bibinfo {title} {{GW170104:
  Observation of a 50-Solar-Mass Binary Black Hole Coalescence at Redshift
  0.2}},}\ }\href {\doibase 10.1103/PhysRevLett.118.221101} {\bibfield
  {journal} {\bibinfo  {journal} {Phys. Rev. Lett.}\ }\textbf {\bibinfo
  {volume} {118}},\ \bibinfo {pages} {221101} (\bibinfo {year}
  {2017})}\BibitemShut {NoStop}%
\bibitem [{\citenamefont {Falcke}\ \emph {et~al.}(2000)\citenamefont {Falcke},
  \citenamefont {Melia},\ and\ \citenamefont {Agol}}]{Falcke:1999pj}%
  \BibitemOpen
  \bibfield  {author} {\bibinfo {author} {\bibfnamefont {Heino}\ \bibnamefont
  {Falcke}}, \bibinfo {author} {\bibfnamefont {Fulvio}\ \bibnamefont {Melia}},
  \ and\ \bibinfo {author} {\bibfnamefont {Eric}\ \bibnamefont {Agol}},\
  }\bibfield  {title} {\enquote {\bibinfo {title} {{Viewing the shadow of the
  black hole at the galactic center}},}\ }\href {\doibase 10.1086/312423}
  {\bibfield  {journal} {\bibinfo  {journal} {Astrophys. J.}\ }\textbf
  {\bibinfo {volume} {528}},\ \bibinfo {pages} {L13} (\bibinfo {year}
  {2000})},\ \Eprint {http://arxiv.org/abs/astro-ph/9912263}
  {arXiv:astro-ph/9912263 [astro-ph]} \BibitemShut {NoStop}%
\bibitem [{\citenamefont {Johnson}\ \emph {et~al.}(2015)\citenamefont {Johnson}
  \emph {et~al.}}]{Johnson:2015iwg}%
  \BibitemOpen
  \bibfield  {author} {\bibinfo {author} {\bibfnamefont {Michael~D.}\
  \bibnamefont {Johnson}} \emph {et~al.},\ }\bibfield  {title} {\enquote
  {\bibinfo {title} {{Resolved Magnetic-Field Structure and Variability Near
  the Event Horizon of Sagittarius A*}},}\ }\href {\doibase
  10.1126/science.aac7087} {\bibfield  {journal} {\bibinfo  {journal}
  {Science}\ }\textbf {\bibinfo {volume} {350}},\ \bibinfo {pages} {1242--1245}
  (\bibinfo {year} {2015})},\ \Eprint {http://arxiv.org/abs/1512.01220}
  {arXiv:1512.01220 [astro-ph.HE]} \BibitemShut {NoStop}%
\bibitem [{\citenamefont {Abbott}\ \emph
  {et~al.}(2016{\natexlab{d}})\citenamefont {Abbott} \emph
  {et~al.}}]{TheLIGOScientific:2016src}%
  \BibitemOpen
  \bibfield  {author} {\bibinfo {author} {\bibfnamefont {B.~P.}\ \bibnamefont
  {Abbott}} \emph {et~al.} (\bibinfo {collaboration} {Virgo, LIGO
  Scientific}),\ }\bibfield  {title} {\enquote {\bibinfo {title} {{Tests of
  general relativity with GW150914}},}\ }\href {\doibase
  10.1103/PhysRevLett.116.221101} {\bibfield  {journal} {\bibinfo  {journal}
  {Phys. Rev. Lett.}\ }\textbf {\bibinfo {volume} {116}},\ \bibinfo {pages}
  {221101} (\bibinfo {year} {2016}{\natexlab{d}})},\ \Eprint
  {http://arxiv.org/abs/1602.03841} {arXiv:1602.03841 [gr-qc]} \BibitemShut
  {NoStop}%
\bibitem [{\citenamefont {Yagi}\ and\ \citenamefont
  {Stein}(2016)}]{Yagi:2016jml}%
  \BibitemOpen
  \bibfield  {author} {\bibinfo {author} {\bibfnamefont {Kent}\ \bibnamefont
  {Yagi}}\ and\ \bibinfo {author} {\bibfnamefont {Leo~C.}\ \bibnamefont
  {Stein}},\ }\bibfield  {title} {\enquote {\bibinfo {title} {{Black Hole Based
  Tests of General Relativity}},}\ }\href {\doibase
  10.1088/0264-9381/33/5/054001} {\bibfield  {journal} {\bibinfo  {journal}
  {Class. Quant. Grav.}\ }\textbf {\bibinfo {volume} {33}},\ \bibinfo {pages}
  {054001} (\bibinfo {year} {2016})},\ \Eprint
  {http://arxiv.org/abs/1602.02413} {arXiv:1602.02413 [gr-qc]} \BibitemShut
  {NoStop}%
\bibitem [{\citenamefont {Yunes}\ \emph {et~al.}(2016)\citenamefont {Yunes},
  \citenamefont {Yagi},\ and\ \citenamefont {Pretorius}}]{Yunes:2016jcc}%
  \BibitemOpen
  \bibfield  {author} {\bibinfo {author} {\bibfnamefont {Nicolas}\ \bibnamefont
  {Yunes}}, \bibinfo {author} {\bibfnamefont {Kent}\ \bibnamefont {Yagi}}, \
  and\ \bibinfo {author} {\bibfnamefont {Frans}\ \bibnamefont {Pretorius}},\
  }\bibfield  {title} {\enquote {\bibinfo {title} {{Theoretical Physics
  Implications of the Binary Black-Hole Mergers GW150914 and GW151226}},}\
  }\href {\doibase 10.1103/PhysRevD.94.084002} {\bibfield  {journal} {\bibinfo
  {journal} {Phys. Rev.}\ }\textbf {\bibinfo {volume} {D94}},\ \bibinfo {pages}
  {084002} (\bibinfo {year} {2016})},\ \Eprint
  {http://arxiv.org/abs/1603.08955} {arXiv:1603.08955 [gr-qc]} \BibitemShut
  {NoStop}%
\bibitem [{\citenamefont {Eckart}\ \emph {et~al.}(2017)\citenamefont {Eckart},
  \citenamefont {Hüttemann}, \citenamefont {Kiefer}, \citenamefont {Britzen},
  \citenamefont {Zajaček}, \citenamefont {Lämmerzahl}, \citenamefont
  {Stöckler}, \citenamefont {Valencia-S}, \citenamefont {Karas},\ and\
  \citenamefont {García-Marín}}]{Eckart:2017bhq}%
  \BibitemOpen
  \bibfield  {author} {\bibinfo {author} {\bibfnamefont {Andreas}\ \bibnamefont
  {Eckart}}, \bibinfo {author} {\bibfnamefont {Andreas}\ \bibnamefont
  {Hüttemann}}, \bibinfo {author} {\bibfnamefont {Claus}\ \bibnamefont
  {Kiefer}}, \bibinfo {author} {\bibfnamefont {Silke}\ \bibnamefont {Britzen}},
  \bibinfo {author} {\bibfnamefont {Michal}\ \bibnamefont {Zajaček}}, \bibinfo
  {author} {\bibfnamefont {Claus}\ \bibnamefont {Lämmerzahl}}, \bibinfo
  {author} {\bibfnamefont {Manfred}\ \bibnamefont {Stöckler}}, \bibinfo
  {author} {\bibfnamefont {Monica}\ \bibnamefont {Valencia-S}}, \bibinfo
  {author} {\bibfnamefont {Vladimir}\ \bibnamefont {Karas}}, \ and\ \bibinfo
  {author} {\bibfnamefont {Macarena}\ \bibnamefont {García-Marín}},\
  }\bibfield  {title} {\enquote {\bibinfo {title} {{The Milky Way’s
  Supermassive Black Hole: How Good a Case Is It?}}}\ }\href {\doibase
  10.1007/s10701-017-0079-2} {\bibfield  {journal} {\bibinfo  {journal} {Found.
  Phys.}\ }\textbf {\bibinfo {volume} {47}},\ \bibinfo {pages} {553--624}
  (\bibinfo {year} {2017})},\ \Eprint {http://arxiv.org/abs/1703.09118}
  {arXiv:1703.09118 [astro-ph.HE]} \BibitemShut {NoStop}%
\bibitem [{\citenamefont {Abramowicz}\ \emph {et~al.}(2002)\citenamefont
  {Abramowicz}, \citenamefont {Kluzniak},\ and\ \citenamefont
  {Lasota}}]{Abramowicz:2002vt}%
  \BibitemOpen
  \bibfield  {author} {\bibinfo {author} {\bibfnamefont {Marek~A.}\
  \bibnamefont {Abramowicz}}, \bibinfo {author} {\bibfnamefont {Wlodek}\
  \bibnamefont {Kluzniak}}, \ and\ \bibinfo {author} {\bibfnamefont
  {Jean-Pierre}\ \bibnamefont {Lasota}},\ }\bibfield  {title} {\enquote
  {\bibinfo {title} {{No observational proof of the black hole
  event-horizon}},}\ }\href {\doibase 10.1051/0004-6361:20021645} {\bibfield
  {journal} {\bibinfo  {journal} {Astron. Astrophys.}\ }\textbf {\bibinfo
  {volume} {396}},\ \bibinfo {pages} {L31--L34} (\bibinfo {year} {2002})},\
  \Eprint {http://arxiv.org/abs/astro-ph/0207270} {arXiv:astro-ph/0207270
  [astro-ph]} \BibitemShut {NoStop}%
\bibitem [{\citenamefont {Cardoso}\ and\ \citenamefont
  {Pani}(2017)}]{CardosoReview}%
  \BibitemOpen
  \bibfield  {author} {\bibinfo {author} {\bibfnamefont {Vitor}\ \bibnamefont
  {Cardoso}}\ and\ \bibinfo {author} {\bibfnamefont {Paolo}\ \bibnamefont
  {Pani}},\ }\href@noop {} {\enquote {\bibinfo {title} {The observational
  evidence for black holes},}\ } (\bibinfo {year} {2017}),\ \bibinfo {note}
  {unpublished}\BibitemShut {NoStop}%
\bibitem [{\citenamefont {Unruh}\ and\ \citenamefont
  {Wald}(2017)}]{Unruh:2017uaw}%
  \BibitemOpen
  \bibfield  {author} {\bibinfo {author} {\bibfnamefont {William~G.}\
  \bibnamefont {Unruh}}\ and\ \bibinfo {author} {\bibfnamefont {Robert~M.}\
  \bibnamefont {Wald}},\ }\bibfield  {title} {\enquote {\bibinfo {title}
  {{Information Loss}},}\ }\href@noop {} {\  (\bibinfo {year} {2017})},\
  \Eprint {http://arxiv.org/abs/1703.02140} {arXiv:1703.02140 [hep-th]}
  \BibitemShut {NoStop}%
\bibitem [{\citenamefont {Mazur}\ and\ \citenamefont
  {Mottola}(2001)}]{Mazur:2001fv}%
  \BibitemOpen
  \bibfield  {author} {\bibinfo {author} {\bibfnamefont {Pawel~O.}\
  \bibnamefont {Mazur}}\ and\ \bibinfo {author} {\bibfnamefont {Emil}\
  \bibnamefont {Mottola}},\ }\bibfield  {title} {\enquote {\bibinfo {title}
  {{Gravitational condensate stars: An alternative to black holes}},}\
  }\href@noop {} {\  (\bibinfo {year} {2001})},\ \Eprint
  {http://arxiv.org/abs/gr-qc/0109035} {arXiv:gr-qc/0109035 [gr-qc]}
  \BibitemShut {NoStop}%
\bibitem [{\citenamefont {Schunck}\ and\ \citenamefont
  {Mielke}(2003)}]{Schunck:2003kk}%
  \BibitemOpen
  \bibfield  {author} {\bibinfo {author} {\bibfnamefont {F.~E.}\ \bibnamefont
  {Schunck}}\ and\ \bibinfo {author} {\bibfnamefont {E.~W.}\ \bibnamefont
  {Mielke}},\ }\bibfield  {title} {\enquote {\bibinfo {title} {{General
  relativistic boson stars}},}\ }\href {\doibase 10.1088/0264-9381/20/20/201}
  {\bibfield  {journal} {\bibinfo  {journal} {Class. Quant. Grav.}\ }\textbf
  {\bibinfo {volume} {20}},\ \bibinfo {pages} {R301--R356} (\bibinfo {year}
  {2003})},\ \Eprint {http://arxiv.org/abs/0801.0307} {arXiv:0801.0307
  [astro-ph]} \BibitemShut {NoStop}%
\bibitem [{\citenamefont {Morris}\ \emph {et~al.}(1988)\citenamefont {Morris},
  \citenamefont {Thorne},\ and\ \citenamefont {Yurtsever}}]{Morris:1988tu}%
  \BibitemOpen
  \bibfield  {author} {\bibinfo {author} {\bibfnamefont {M.~S.}\ \bibnamefont
  {Morris}}, \bibinfo {author} {\bibfnamefont {K.~S.}\ \bibnamefont {Thorne}},
  \ and\ \bibinfo {author} {\bibfnamefont {U.}~\bibnamefont {Yurtsever}},\
  }\bibfield  {title} {\enquote {\bibinfo {title} {{Wormholes, Time Machines,
  and the Weak Energy Condition}},}\ }\href {\doibase
  10.1103/PhysRevLett.61.1446} {\bibfield  {journal} {\bibinfo  {journal}
  {Phys. Rev. Lett.}\ }\textbf {\bibinfo {volume} {61}},\ \bibinfo {pages}
  {1446--1449} (\bibinfo {year} {1988})}\BibitemShut {NoStop}%
\bibitem [{\citenamefont {Mathur}(2005)}]{Mathur:2005zp}%
  \BibitemOpen
  \bibfield  {author} {\bibinfo {author} {\bibfnamefont {Samir~D.}\
  \bibnamefont {Mathur}},\ }\bibfield  {title} {\enquote {\bibinfo {title}
  {{The Fuzzball proposal for black holes: An Elementary review}},}\ }\bibfield
   {booktitle} {\emph {\bibinfo {booktitle} {{The quantum structure of
  space-time and the geometric nature of fundamental interactions. Proceedings,
  4th Meeting, RTN2004, Kolymbari, Crete, Greece, September 5-10, 2004}}},\
  }\href {\doibase 10.1002/prop.200410203} {\bibfield  {journal} {\bibinfo
  {journal} {Fortsch. Phys.}\ }\textbf {\bibinfo {volume} {53}},\ \bibinfo
  {pages} {793--827} (\bibinfo {year} {2005})},\ \Eprint
  {http://arxiv.org/abs/hep-th/0502050} {arXiv:hep-th/0502050 [hep-th]}
  \BibitemShut {NoStop}%
\bibitem [{\citenamefont {Barcelo}\ \emph {et~al.}(2011)\citenamefont
  {Barcelo}, \citenamefont {Garay},\ and\ \citenamefont
  {Jannes}}]{Barcelo:2010vc}%
  \BibitemOpen
  \bibfield  {author} {\bibinfo {author} {\bibfnamefont {C.}~\bibnamefont
  {Barcelo}}, \bibinfo {author} {\bibfnamefont {L.~J.}\ \bibnamefont {Garay}},
  \ and\ \bibinfo {author} {\bibfnamefont {G.}~\bibnamefont {Jannes}},\
  }\bibfield  {title} {\enquote {\bibinfo {title} {{Quantum Non-Gravity and
  Stellar Collapse}},}\ }\href {\doibase 10.1007/s10701-011-9577-9} {\bibfield
  {journal} {\bibinfo  {journal} {Found. Phys.}\ }\textbf {\bibinfo {volume}
  {41}},\ \bibinfo {pages} {1532--1541} (\bibinfo {year} {2011})},\ \Eprint
  {http://arxiv.org/abs/1002.4651} {arXiv:1002.4651 [gr-qc]} \BibitemShut
  {NoStop}%
\bibitem [{\citenamefont {Barcelo}\ \emph {et~al.}(2015)\citenamefont
  {Barcelo}, \citenamefont {Carballo-Rubio}, \citenamefont {Garay},\ and\
  \citenamefont {Jannes}}]{Barcelo:2014cla}%
  \BibitemOpen
  \bibfield  {author} {\bibinfo {author} {\bibfnamefont {Carlo}\ \bibnamefont
  {Barcelo}}, \bibinfo {author} {\bibfnamefont {Raúl}\ \bibnamefont
  {Carballo-Rubio}}, \bibinfo {author} {\bibfnamefont {Luis~J.}\ \bibnamefont
  {Garay}}, \ and\ \bibinfo {author} {\bibfnamefont {Gil}\ \bibnamefont
  {Jannes}},\ }\bibfield  {title} {\enquote {\bibinfo {title} {{The lifetime
  problem of evaporating black holes: mutiny or resignation}},}\ }\href
  {\doibase 10.1088/0264-9381/32/3/035012} {\bibfield  {journal} {\bibinfo
  {journal} {Class. Quant. Grav.}\ }\textbf {\bibinfo {volume} {32}},\ \bibinfo
  {pages} {035012} (\bibinfo {year} {2015})},\ \Eprint
  {http://arxiv.org/abs/1409.1501} {arXiv:1409.1501 [gr-qc]} \BibitemShut
  {NoStop}%
\bibitem [{\citenamefont {Holdom}\ and\ \citenamefont
  {Ren}(2017)}]{Holdom:2016nek}%
  \BibitemOpen
  \bibfield  {author} {\bibinfo {author} {\bibfnamefont {Bob}\ \bibnamefont
  {Holdom}}\ and\ \bibinfo {author} {\bibfnamefont {Jing}\ \bibnamefont
  {Ren}},\ }\bibfield  {title} {\enquote {\bibinfo {title} {{Not quite a black
  hole}},}\ }\href {\doibase 10.1103/PhysRevD.95.084034} {\bibfield  {journal}
  {\bibinfo  {journal} {Phys. Rev.}\ }\textbf {\bibinfo {volume} {D95}},\
  \bibinfo {pages} {084034} (\bibinfo {year} {2017})},\ \Eprint
  {http://arxiv.org/abs/1612.04889} {arXiv:1612.04889 [gr-qc]} \BibitemShut
  {NoStop}%
\bibitem [{\citenamefont {Maggio}\ \emph {et~al.}(2017)\citenamefont {Maggio},
  \citenamefont {Pani},\ and\ \citenamefont {Ferrari}}]{Maggio:2017ivp}%
  \BibitemOpen
  \bibfield  {author} {\bibinfo {author} {\bibfnamefont {Elisa}\ \bibnamefont
  {Maggio}}, \bibinfo {author} {\bibfnamefont {Paolo}\ \bibnamefont {Pani}}, \
  and\ \bibinfo {author} {\bibfnamefont {Valeria}\ \bibnamefont {Ferrari}},\
  }\bibfield  {title} {\enquote {\bibinfo {title} {{Exotic Compact Objects and
  How to Quench their Ergoregion Instability}},}\ }\href@noop {} {\  (\bibinfo
  {year} {2017})},\ \Eprint {http://arxiv.org/abs/1703.03696} {arXiv:1703.03696
  [gr-qc]} \BibitemShut {NoStop}%
\bibitem [{\citenamefont {Hod}(2017)}]{Hod:2017cga}%
  \BibitemOpen
  \bibfield  {author} {\bibinfo {author} {\bibfnamefont {Shahar}\ \bibnamefont
  {Hod}},\ }\bibfield  {title} {\enquote {\bibinfo {title} {{Onset of
  superradiant instabilities in rotating spacetimes of exotic compact
  objects}},}\ }\href@noop {} {\  (\bibinfo {year} {2017})},\ \Eprint
  {http://arxiv.org/abs/1704.05856} {arXiv:1704.05856 [hep-th]} \BibitemShut
  {NoStop}%
\bibitem [{\citenamefont {Cardoso}\ \emph {et~al.}(2014)\citenamefont
  {Cardoso}, \citenamefont {Crispino}, \citenamefont {Macedo}, \citenamefont
  {Okawa},\ and\ \citenamefont {Pani}}]{Cardoso:2014sna}%
  \BibitemOpen
  \bibfield  {author} {\bibinfo {author} {\bibfnamefont {Vitor}\ \bibnamefont
  {Cardoso}}, \bibinfo {author} {\bibfnamefont {Luís C.~B.}\ \bibnamefont
  {Crispino}}, \bibinfo {author} {\bibfnamefont {Caio F.~B.}\ \bibnamefont
  {Macedo}}, \bibinfo {author} {\bibfnamefont {Hirotada}\ \bibnamefont
  {Okawa}}, \ and\ \bibinfo {author} {\bibfnamefont {Paolo}\ \bibnamefont
  {Pani}},\ }\bibfield  {title} {\enquote {\bibinfo {title} {{Light rings as
  observational evidence for event horizons: long-lived modes, ergoregions and
  nonlinear instabilities of ultracompact objects}},}\ }\href {\doibase
  10.1103/PhysRevD.90.044069} {\bibfield  {journal} {\bibinfo  {journal} {Phys.
  Rev.}\ }\textbf {\bibinfo {volume} {D90}},\ \bibinfo {pages} {044069}
  (\bibinfo {year} {2014})},\ \Eprint {http://arxiv.org/abs/1406.5510}
  {arXiv:1406.5510 [gr-qc]} \BibitemShut {NoStop}%
\bibitem [{\citenamefont {Cardoso}\ \emph {et~al.}(2017)\citenamefont
  {Cardoso}, \citenamefont {Franzin}, \citenamefont {Maselli}, \citenamefont
  {Pani},\ and\ \citenamefont {Raposo}}]{Cardoso:2017cfl}%
  \BibitemOpen
  \bibfield  {author} {\bibinfo {author} {\bibfnamefont {Vitor}\ \bibnamefont
  {Cardoso}}, \bibinfo {author} {\bibfnamefont {Edgardo}\ \bibnamefont
  {Franzin}}, \bibinfo {author} {\bibfnamefont {Andrea}\ \bibnamefont
  {Maselli}}, \bibinfo {author} {\bibfnamefont {Paolo}\ \bibnamefont {Pani}}, \
  and\ \bibinfo {author} {\bibfnamefont {Guilherme}\ \bibnamefont {Raposo}},\
  }\bibfield  {title} {\enquote {\bibinfo {title} {{Testing strong-field
  gravity with tidal Love numbers}},}\ }\href {\doibase
  10.1103/PhysRevD.95.084014} {\bibfield  {journal} {\bibinfo  {journal} {Phys.
  Rev.}\ }\textbf {\bibinfo {volume} {D95}},\ \bibinfo {pages} {084014}
  (\bibinfo {year} {2017})},\ \Eprint {http://arxiv.org/abs/1701.01116}
  {arXiv:1701.01116 [gr-qc]} \BibitemShut {NoStop}%
\bibitem [{\citenamefont {Maselli}\ \emph {et~al.}(2017)\citenamefont
  {Maselli}, \citenamefont {Pani}, \citenamefont {Cardoso}, \citenamefont
  {Abdelsalhin}, \citenamefont {Gualtieri},\ and\ \citenamefont
  {Ferrari}}]{Maselli:2017cmm}%
  \BibitemOpen
  \bibfield  {author} {\bibinfo {author} {\bibfnamefont {Andrea}\ \bibnamefont
  {Maselli}}, \bibinfo {author} {\bibfnamefont {Paolo}\ \bibnamefont {Pani}},
  \bibinfo {author} {\bibfnamefont {Vitor}\ \bibnamefont {Cardoso}}, \bibinfo
  {author} {\bibfnamefont {Tiziano}\ \bibnamefont {Abdelsalhin}}, \bibinfo
  {author} {\bibfnamefont {Leonardo}\ \bibnamefont {Gualtieri}}, \ and\
  \bibinfo {author} {\bibfnamefont {Valeria}\ \bibnamefont {Ferrari}},\
  }\bibfield  {title} {\enquote {\bibinfo {title} {{Probing Planckian
  corrections at the horizon scale with LISA binaries}},}\ }\href@noop {} {\
  (\bibinfo {year} {2017})},\ \Eprint {http://arxiv.org/abs/1703.10612}
  {arXiv:1703.10612 [gr-qc]} \BibitemShut {NoStop}%
\bibitem [{\citenamefont {Amaro-Seoane}\ \emph {et~al.}(2013)\citenamefont
  {Amaro-Seoane}, \citenamefont {Aoudia}, \citenamefont {Babak}, \citenamefont
  {Binetruy}, \citenamefont {Berti} \emph {et~al.}}]{AmaroSeoane:2012km}%
  \BibitemOpen
  \bibfield  {author} {\bibinfo {author} {\bibfnamefont {Pau}\ \bibnamefont
  {Amaro-Seoane}}, \bibinfo {author} {\bibfnamefont {Sofiane}\ \bibnamefont
  {Aoudia}}, \bibinfo {author} {\bibfnamefont {Stanislav}\ \bibnamefont
  {Babak}}, \bibinfo {author} {\bibfnamefont {Pierre}\ \bibnamefont
  {Binetruy}}, \bibinfo {author} {\bibfnamefont {Emanuele}\ \bibnamefont
  {Berti}},  \emph {et~al.},\ }\bibfield  {title} {\enquote {\bibinfo {title}
  {{eLISA/NGO: Astrophysics and cosmology in the gravitational-wave millihertz
  regime}},}\ }\href@noop {} {\bibfield  {journal} {\bibinfo  {journal} {GW
  Notes}\ }\textbf {\bibinfo {volume} {6}},\ \bibinfo {pages} {4--110}
  (\bibinfo {year} {2013})},\ \Eprint {http://arxiv.org/abs/1201.3621}
  {arXiv:1201.3621 [astro-ph.CO]} \BibitemShut {NoStop}%
\bibitem [{\citenamefont {Pani}\ \emph {et~al.}(2010)\citenamefont {Pani},
  \citenamefont {Berti}, \citenamefont {Cardoso}, \citenamefont {Chen},\ and\
  \citenamefont {Norte}}]{Pani:2010em}%
  \BibitemOpen
  \bibfield  {author} {\bibinfo {author} {\bibfnamefont {Paolo}\ \bibnamefont
  {Pani}}, \bibinfo {author} {\bibfnamefont {Emanuele}\ \bibnamefont {Berti}},
  \bibinfo {author} {\bibfnamefont {Vitor}\ \bibnamefont {Cardoso}}, \bibinfo
  {author} {\bibfnamefont {Yanbei}\ \bibnamefont {Chen}}, \ and\ \bibinfo
  {author} {\bibfnamefont {Richard}\ \bibnamefont {Norte}},\ }\bibfield
  {title} {\enquote {\bibinfo {title} {{Gravitational-wave signatures of the
  absence of an event horizon. II. Extreme mass ratio inspirals in the
  spacetime of a thin-shell gravastar}},}\ }\href {\doibase
  10.1103/PhysRevD.81.084011} {\bibfield  {journal} {\bibinfo  {journal} {Phys.
  Rev.}\ }\textbf {\bibinfo {volume} {D81}},\ \bibinfo {pages} {084011}
  (\bibinfo {year} {2010})},\ \Eprint {http://arxiv.org/abs/1001.3031}
  {arXiv:1001.3031 [gr-qc]} \BibitemShut {NoStop}%
\bibitem [{\citenamefont {Krishnendu}\ \emph {et~al.}(2017)\citenamefont
  {Krishnendu}, \citenamefont {Arun},\ and\ \citenamefont
  {Mishra}}]{Krishnendu:2017shb}%
  \BibitemOpen
  \bibfield  {author} {\bibinfo {author} {\bibfnamefont {N.~V.}\ \bibnamefont
  {Krishnendu}}, \bibinfo {author} {\bibfnamefont {K.~G.}\ \bibnamefont
  {Arun}}, \ and\ \bibinfo {author} {\bibfnamefont {Chandra~Kant}\ \bibnamefont
  {Mishra}},\ }\bibfield  {title} {\enquote {\bibinfo {title} {{Testing the
  binary black hole nature of a compact binary coalescence}},}\ }\href@noop {}
  {\  (\bibinfo {year} {2017})},\ \Eprint {http://arxiv.org/abs/1701.06318}
  {arXiv:1701.06318 [gr-qc]} \BibitemShut {NoStop}%
\bibitem [{\citenamefont {Nollert}(1999)}]{Nollert}%
  \BibitemOpen
  \bibfield  {author} {\bibinfo {author} {\bibfnamefont {Hans-Peter}\
  \bibnamefont {Nollert}},\ }\bibfield  {title} {\enquote {\bibinfo {title}
  {Quasinormal modes: the characteristic `sound' of black holes and neutron
  stars},}\ }\href {http://stacks.iop.org/0264-9381/16/i=12/a=201} {\bibfield
  {journal} {\bibinfo  {journal} {Classical and Quantum Gravity}\ }\textbf
  {\bibinfo {volume} {16}},\ \bibinfo {pages} {R159} (\bibinfo {year}
  {1999})}\BibitemShut {NoStop}%
\bibitem [{\citenamefont {{Berti}}\ \emph {et~al.}(2009)\citenamefont
  {{Berti}}, \citenamefont {{Cardoso}},\ and\ \citenamefont
  {{Starinets}}}]{Berti2009}%
  \BibitemOpen
  \bibfield  {author} {\bibinfo {author} {\bibfnamefont {E.}~\bibnamefont
  {{Berti}}}, \bibinfo {author} {\bibfnamefont {V.}~\bibnamefont {{Cardoso}}},
  \ and\ \bibinfo {author} {\bibfnamefont {A.~O.}\ \bibnamefont
  {{Starinets}}},\ }\bibfield  {title} {\enquote {\bibinfo {title} {{TOPICAL
  REVIEW: Quasinormal modes of black holes and black branes}},}\ }\href
  {\doibase 10.1088/0264-9381/26/16/163001} {\bibfield  {journal} {\bibinfo
  {journal} {Classical and Quantum Gravity}\ }\textbf {\bibinfo {volume}
  {26}},\ \bibinfo {pages} {163001} (\bibinfo {year} {2009})},\ \Eprint
  {http://arxiv.org/abs/0905.2975} {arXiv:0905.2975 [gr-qc]} \BibitemShut
  {NoStop}%
\bibitem [{\citenamefont {{Dreyer}}\ \emph {et~al.}(2004)\citenamefont
  {{Dreyer}}, \citenamefont {{Kelly}}, \citenamefont {{Krishnan}},
  \citenamefont {{Finn}}, \citenamefont {{Garrison}},\ and\ \citenamefont
  {{Lopez-Aleman}}}]{Dreyer2004}%
  \BibitemOpen
  \bibfield  {author} {\bibinfo {author} {\bibfnamefont {O.}~\bibnamefont
  {{Dreyer}}}, \bibinfo {author} {\bibfnamefont {B.}~\bibnamefont {{Kelly}}},
  \bibinfo {author} {\bibfnamefont {B.}~\bibnamefont {{Krishnan}}}, \bibinfo
  {author} {\bibfnamefont {L.~S.}\ \bibnamefont {{Finn}}}, \bibinfo {author}
  {\bibfnamefont {D.}~\bibnamefont {{Garrison}}}, \ and\ \bibinfo {author}
  {\bibfnamefont {R.}~\bibnamefont {{Lopez-Aleman}}},\ }\bibfield  {title}
  {\enquote {\bibinfo {title} {{Black-hole spectroscopy: testing general
  relativity through gravitational-wave observations}},}\ }\href {\doibase
  10.1088/0264-9381/21/4/003} {\bibfield  {journal} {\bibinfo  {journal}
  {Classical and Quantum Gravity}\ }\textbf {\bibinfo {volume} {21}},\ \bibinfo
  {pages} {787--803} (\bibinfo {year} {2004})},\ \Eprint
  {http://arxiv.org/abs/arXiv:gr-qc/0309007} {arXiv:gr-qc/0309007} \BibitemShut
  {NoStop}%
\bibitem [{\citenamefont {Berti}\ \emph {et~al.}(2006)\citenamefont {Berti},
  \citenamefont {Cardoso},\ and\ \citenamefont {Will}}]{Berti:2005ys}%
  \BibitemOpen
  \bibfield  {author} {\bibinfo {author} {\bibfnamefont {Emanuele}\
  \bibnamefont {Berti}}, \bibinfo {author} {\bibfnamefont {Vitor}\ \bibnamefont
  {Cardoso}}, \ and\ \bibinfo {author} {\bibfnamefont {Clifford~M.}\
  \bibnamefont {Will}},\ }\bibfield  {title} {\enquote {\bibinfo {title} {{On
  gravitational-wave spectroscopy of massive black holes with the space
  interferometer LISA}},}\ }\href {\doibase 10.1103/PhysRevD.73.064030}
  {\bibfield  {journal} {\bibinfo  {journal} {Phys. Rev.}\ }\textbf {\bibinfo
  {volume} {D73}},\ \bibinfo {pages} {064030} (\bibinfo {year} {2006})},\
  \Eprint {http://arxiv.org/abs/gr-qc/0512160} {arXiv:gr-qc/0512160 [gr-qc]}
  \BibitemShut {NoStop}%
\bibitem [{\citenamefont {Pani}(2013)}]{Pani:2013pma}%
  \BibitemOpen
  \bibfield  {author} {\bibinfo {author} {\bibfnamefont {Paolo}\ \bibnamefont
  {Pani}},\ }\bibfield  {title} {\enquote {\bibinfo {title} {{Advanced Methods
  in Black-Hole Perturbation Theory}},}\ }\bibfield  {booktitle} {\emph
  {\bibinfo {booktitle} {{Proceedings, Spring School on Numerical Relativity
  and High Energy Physics (NR/HEP2): Lisbon, Portugal, March 11-14, 2013}}},\
  }\href {\doibase 10.1142/S0217751X13400186} {\bibfield  {journal} {\bibinfo
  {journal} {Int. J. Mod. Phys.}\ }\textbf {\bibinfo {volume} {A28}},\ \bibinfo
  {pages} {1340018} (\bibinfo {year} {2013})},\ \Eprint
  {http://arxiv.org/abs/1305.6759} {arXiv:1305.6759 [gr-qc]} \BibitemShut
  {NoStop}%
\bibitem [{\citenamefont {Berti}\ \emph {et~al.}(2016)\citenamefont {Berti},
  \citenamefont {Sesana}, \citenamefont {Barausse}, \citenamefont {Cardoso},\
  and\ \citenamefont {Belczynski}}]{Berti:2016lat}%
  \BibitemOpen
  \bibfield  {author} {\bibinfo {author} {\bibfnamefont {Emanuele}\
  \bibnamefont {Berti}}, \bibinfo {author} {\bibfnamefont {Alberto}\
  \bibnamefont {Sesana}}, \bibinfo {author} {\bibfnamefont {Enrico}\
  \bibnamefont {Barausse}}, \bibinfo {author} {\bibfnamefont {Vitor}\
  \bibnamefont {Cardoso}}, \ and\ \bibinfo {author} {\bibfnamefont {Krzysztof}\
  \bibnamefont {Belczynski}},\ }\bibfield  {title} {\enquote {\bibinfo {title}
  {{Spectroscopy of Kerr black holes with Earth- and space-based
  interferometers}},}\ }\href {\doibase 10.1103/PhysRevLett.117.101102}
  {\bibfield  {journal} {\bibinfo  {journal} {Phys. Rev. Lett.}\ }\textbf
  {\bibinfo {volume} {117}},\ \bibinfo {pages} {101102} (\bibinfo {year}
  {2016})},\ \Eprint {http://arxiv.org/abs/1605.09286} {arXiv:1605.09286
  [gr-qc]} \BibitemShut {NoStop}%
\bibitem [{\citenamefont {Yang}\ \emph {et~al.}(2017)\citenamefont {Yang},
  \citenamefont {Yagi}, \citenamefont {Blackman}, \citenamefont {Lehner},
  \citenamefont {Paschalidis}, \citenamefont {Pretorius},\ and\ \citenamefont
  {Yunes}}]{Yang:2017zxs}%
  \BibitemOpen
  \bibfield  {author} {\bibinfo {author} {\bibfnamefont {Huan}\ \bibnamefont
  {Yang}}, \bibinfo {author} {\bibfnamefont {Kent}\ \bibnamefont {Yagi}},
  \bibinfo {author} {\bibfnamefont {Jonathan}\ \bibnamefont {Blackman}},
  \bibinfo {author} {\bibfnamefont {Luis}\ \bibnamefont {Lehner}}, \bibinfo
  {author} {\bibfnamefont {Vasileios}\ \bibnamefont {Paschalidis}}, \bibinfo
  {author} {\bibfnamefont {Frans}\ \bibnamefont {Pretorius}}, \ and\ \bibinfo
  {author} {\bibfnamefont {Nicolas}\ \bibnamefont {Yunes}},\ }\bibfield
  {title} {\enquote {\bibinfo {title} {{Black hole spectroscopy with coherent
  mode stacking}},}\ }\href {\doibase 10.1103/PhysRevLett.118.161101}
  {\bibfield  {journal} {\bibinfo  {journal} {Phys. Rev. Lett.}\ }\textbf
  {\bibinfo {volume} {118}},\ \bibinfo {pages} {161101} (\bibinfo {year}
  {2017})},\ \Eprint {http://arxiv.org/abs/1701.05808} {arXiv:1701.05808
  [gr-qc]} \BibitemShut {NoStop}%
\bibitem [{\citenamefont {Cardoso}\ \emph
  {et~al.}(2016{\natexlab{a}})\citenamefont {Cardoso}, \citenamefont
  {Franzin},\ and\ \citenamefont {Pani}}]{Cardoso:2016rao}%
  \BibitemOpen
  \bibfield  {author} {\bibinfo {author} {\bibfnamefont {Vitor}\ \bibnamefont
  {Cardoso}}, \bibinfo {author} {\bibfnamefont {Edgardo}\ \bibnamefont
  {Franzin}}, \ and\ \bibinfo {author} {\bibfnamefont {Paolo}\ \bibnamefont
  {Pani}},\ }\bibfield  {title} {\enquote {\bibinfo {title} {{Is the
  gravitational-wave ringdown a probe of the event horizon?}}}\ }\href
  {\doibase 10.1103/PhysRevLett.117.089902, 10.1103/PhysRevLett.116.171101}
  {\bibfield  {journal} {\bibinfo  {journal} {Phys. Rev. Lett.}\ }\textbf
  {\bibinfo {volume} {116}},\ \bibinfo {pages} {171101} (\bibinfo {year}
  {2016}{\natexlab{a}})},\ \bibinfo {note} {[Erratum: Phys. Rev.
  Lett.117,no.8,089902(2016)]},\ \Eprint {http://arxiv.org/abs/1602.07309}
  {arXiv:1602.07309 [gr-qc]} \BibitemShut {NoStop}%
\bibitem [{\citenamefont {Ching}\ \emph {et~al.}(1996)\citenamefont {Ching},
  \citenamefont {Leung}, \citenamefont {Suen},\ and\ \citenamefont
  {Young}}]{Ching:1995rt}%
  \BibitemOpen
  \bibfield  {author} {\bibinfo {author} {\bibfnamefont {E.~S.~C.}\
  \bibnamefont {Ching}}, \bibinfo {author} {\bibfnamefont {P.~T.}\ \bibnamefont
  {Leung}}, \bibinfo {author} {\bibfnamefont {W.~M.}\ \bibnamefont {Suen}}, \
  and\ \bibinfo {author} {\bibfnamefont {K.}~\bibnamefont {Young}},\ }\bibfield
   {title} {\enquote {\bibinfo {title} {{Wave propagation in gravitational
  systems: Completeness of quasinormal modes}},}\ }\href {\doibase
  10.1103/PhysRevD.54.3778} {\bibfield  {journal} {\bibinfo  {journal} {Phys.
  Rev.}\ }\textbf {\bibinfo {volume} {D54}},\ \bibinfo {pages} {3778--3791}
  (\bibinfo {year} {1996})},\ \Eprint {http://arxiv.org/abs/gr-qc/9507034}
  {arXiv:gr-qc/9507034 [gr-qc]} \BibitemShut {NoStop}%
\bibitem [{\citenamefont {Nollert}\ and\ \citenamefont
  {Price}(1999)}]{Nollert:1998ys}%
  \BibitemOpen
  \bibfield  {author} {\bibinfo {author} {\bibfnamefont {Hans-Peter}\
  \bibnamefont {Nollert}}\ and\ \bibinfo {author} {\bibfnamefont {Richard~H.}\
  \bibnamefont {Price}},\ }\bibfield  {title} {\enquote {\bibinfo {title}
  {{Quantifying excitations of quasinormal mode systems}},}\ }\href {\doibase
  10.1063/1.532698} {\bibfield  {journal} {\bibinfo  {journal} {J. Math.
  Phys.}\ }\textbf {\bibinfo {volume} {40}},\ \bibinfo {pages} {980--1010}
  (\bibinfo {year} {1999})},\ \Eprint {http://arxiv.org/abs/gr-qc/9810074}
  {arXiv:gr-qc/9810074 [gr-qc]} \BibitemShut {NoStop}%
\bibitem [{\citenamefont {Pani}\ \emph {et~al.}(2009)\citenamefont {Pani},
  \citenamefont {Berti}, \citenamefont {Cardoso}, \citenamefont {Chen},\ and\
  \citenamefont {Norte}}]{Pani:2009ss}%
  \BibitemOpen
  \bibfield  {author} {\bibinfo {author} {\bibfnamefont {Paolo}\ \bibnamefont
  {Pani}}, \bibinfo {author} {\bibfnamefont {Emanuele}\ \bibnamefont {Berti}},
  \bibinfo {author} {\bibfnamefont {Vitor}\ \bibnamefont {Cardoso}}, \bibinfo
  {author} {\bibfnamefont {Yanbei}\ \bibnamefont {Chen}}, \ and\ \bibinfo
  {author} {\bibfnamefont {Richard}\ \bibnamefont {Norte}},\ }\bibfield
  {title} {\enquote {\bibinfo {title} {{Gravitational wave signatures of the
  absence of an event horizon. I. Nonradial oscillations of a thin-shell
  gravastar}},}\ }\href {\doibase 10.1103/PhysRevD.80.124047} {\bibfield
  {journal} {\bibinfo  {journal} {Phys. Rev.}\ }\textbf {\bibinfo {volume}
  {D80}},\ \bibinfo {pages} {124047} (\bibinfo {year} {2009})},\ \Eprint
  {http://arxiv.org/abs/0909.0287} {arXiv:0909.0287 [gr-qc]} \BibitemShut
  {NoStop}%
\bibitem [{\citenamefont {Cardoso}\ \emph
  {et~al.}(2016{\natexlab{b}})\citenamefont {Cardoso}, \citenamefont {Hopper},
  \citenamefont {Macedo}, \citenamefont {Palenzuela},\ and\ \citenamefont
  {Pani}}]{Cardoso:2016oxy}%
  \BibitemOpen
  \bibfield  {author} {\bibinfo {author} {\bibfnamefont {Vitor}\ \bibnamefont
  {Cardoso}}, \bibinfo {author} {\bibfnamefont {Seth}\ \bibnamefont {Hopper}},
  \bibinfo {author} {\bibfnamefont {Caio F.~B.}\ \bibnamefont {Macedo}},
  \bibinfo {author} {\bibfnamefont {Carlos}\ \bibnamefont {Palenzuela}}, \ and\
  \bibinfo {author} {\bibfnamefont {Paolo}\ \bibnamefont {Pani}},\ }\bibfield
  {title} {\enquote {\bibinfo {title} {{Gravitational-wave signatures of exotic
  compact objects and of quantum corrections at the horizon scale}},}\ }\href
  {\doibase 10.1103/PhysRevD.94.084031} {\bibfield  {journal} {\bibinfo
  {journal} {Phys. Rev.}\ }\textbf {\bibinfo {volume} {D94}},\ \bibinfo {pages}
  {084031} (\bibinfo {year} {2016}{\natexlab{b}})},\ \Eprint
  {http://arxiv.org/abs/1608.08637} {arXiv:1608.08637 [gr-qc]} \BibitemShut
  {NoStop}%
\bibitem [{\citenamefont {Price}\ and\ \citenamefont
  {Khanna}(2017)}]{Price:2017cjr}%
  \BibitemOpen
  \bibfield  {author} {\bibinfo {author} {\bibfnamefont {Richard~H.}\
  \bibnamefont {Price}}\ and\ \bibinfo {author} {\bibfnamefont {Gaurav}\
  \bibnamefont {Khanna}},\ }\bibfield  {title} {\enquote {\bibinfo {title}
  {{Gravitational wave sources: reflections and echoes}},}\ }\href@noop {} {\
  (\bibinfo {year} {2017})},\ \Eprint {http://arxiv.org/abs/1702.04833}
  {arXiv:1702.04833 [gr-qc]} \BibitemShut {NoStop}%
\bibitem [{\citenamefont {Völkel}\ and\ \citenamefont
  {Kokkotas}(2017)}]{Volkel:2017kfj}%
  \BibitemOpen
  \bibfield  {author} {\bibinfo {author} {\bibfnamefont {Sebastian~H.}\
  \bibnamefont {Völkel}}\ and\ \bibinfo {author} {\bibfnamefont {Kostas~D.}\
  \bibnamefont {Kokkotas}},\ }\bibfield  {title} {\enquote {\bibinfo {title}
  {{Ultra Compact Stars: Reconstructing the Perturbation Potential}},}\
  }\href@noop {} {\  (\bibinfo {year} {2017})},\ \Eprint
  {http://arxiv.org/abs/1704.07517} {arXiv:1704.07517 [gr-qc]} \BibitemShut
  {NoStop}%
\bibitem [{\citenamefont {Abedi}\ \emph {et~al.}(2016)\citenamefont {Abedi},
  \citenamefont {Dykaar},\ and\ \citenamefont {Afshordi}}]{Abedi:2016hgu}%
  \BibitemOpen
  \bibfield  {author} {\bibinfo {author} {\bibfnamefont {Jahed}\ \bibnamefont
  {Abedi}}, \bibinfo {author} {\bibfnamefont {Hannah}\ \bibnamefont {Dykaar}},
  \ and\ \bibinfo {author} {\bibfnamefont {Niayesh}\ \bibnamefont {Afshordi}},\
  }\bibfield  {title} {\enquote {\bibinfo {title} {{Echoes from the Abyss:
  Evidence for Planck-scale structure at black hole horizons}},}\ }\href@noop
  {} {\  (\bibinfo {year} {2016})},\ \Eprint {http://arxiv.org/abs/1612.00266}
  {arXiv:1612.00266 [gr-qc]} \BibitemShut {NoStop}%
\bibitem [{\citenamefont {Abedi}\ \emph {et~al.}(2017)\citenamefont {Abedi},
  \citenamefont {Dykaar},\ and\ \citenamefont {Afshordi}}]{Abedi:2017isz}%
  \BibitemOpen
  \bibfield  {author} {\bibinfo {author} {\bibfnamefont {Jahed}\ \bibnamefont
  {Abedi}}, \bibinfo {author} {\bibfnamefont {Hannah}\ \bibnamefont {Dykaar}},
  \ and\ \bibinfo {author} {\bibfnamefont {Niayesh}\ \bibnamefont {Afshordi}},\
  }\bibfield  {title} {\enquote {\bibinfo {title} {{Echoes from the Abyss: The
  Holiday Edition!}}}\ }\href@noop {} {\  (\bibinfo {year} {2017})},\ \Eprint
  {http://arxiv.org/abs/1701.03485} {arXiv:1701.03485 [gr-qc]} \BibitemShut
  {NoStop}%
\bibitem [{\citenamefont {Ashton}\ \emph {et~al.}(2016)\citenamefont {Ashton},
  \citenamefont {Birnholtz}, \citenamefont {Cabero}, \citenamefont {Capano},
  \citenamefont {Dent}, \citenamefont {Krishnan}, \citenamefont {Meadors},
  \citenamefont {Nielsen}, \citenamefont {Nitz},\ and\ \citenamefont
  {Westerweck}}]{Ashton:2016xff}%
  \BibitemOpen
  \bibfield  {author} {\bibinfo {author} {\bibfnamefont {Gregory}\ \bibnamefont
  {Ashton}}, \bibinfo {author} {\bibfnamefont {Ofek}\ \bibnamefont
  {Birnholtz}}, \bibinfo {author} {\bibfnamefont {Miriam}\ \bibnamefont
  {Cabero}}, \bibinfo {author} {\bibfnamefont {Collin}\ \bibnamefont {Capano}},
  \bibinfo {author} {\bibfnamefont {Thomas}\ \bibnamefont {Dent}}, \bibinfo
  {author} {\bibfnamefont {Badri}\ \bibnamefont {Krishnan}}, \bibinfo {author}
  {\bibfnamefont {Grant~David}\ \bibnamefont {Meadors}}, \bibinfo {author}
  {\bibfnamefont {Alex~B.}\ \bibnamefont {Nielsen}}, \bibinfo {author}
  {\bibfnamefont {Alex}\ \bibnamefont {Nitz}}, \ and\ \bibinfo {author}
  {\bibfnamefont {Julian}\ \bibnamefont {Westerweck}},\ }\bibfield  {title}
  {\enquote {\bibinfo {title} {{Comments on: "Echoes from the abyss: Evidence
  for Planck-scale structure at black hole horizons"}},}\ }\href@noop {} {\
  (\bibinfo {year} {2016})},\ \Eprint {http://arxiv.org/abs/1612.05625}
  {arXiv:1612.05625 [gr-qc]} \BibitemShut {NoStop}%
\bibitem [{\citenamefont {Nakano}\ \emph {et~al.}(2017)\citenamefont {Nakano},
  \citenamefont {Sago}, \citenamefont {Tagoshi},\ and\ \citenamefont
  {Tanaka}}]{Nakano:2017fvh}%
  \BibitemOpen
  \bibfield  {author} {\bibinfo {author} {\bibfnamefont {Hiroyuki}\
  \bibnamefont {Nakano}}, \bibinfo {author} {\bibfnamefont {Norichika}\
  \bibnamefont {Sago}}, \bibinfo {author} {\bibfnamefont {Hideyuki}\
  \bibnamefont {Tagoshi}}, \ and\ \bibinfo {author} {\bibfnamefont {Takahiro}\
  \bibnamefont {Tanaka}},\ }\bibfield  {title} {\enquote {\bibinfo {title}
  {{Black hole ringdown echoes and howls}},}\ }\href@noop {} {\  (\bibinfo
  {year} {2017})},\ \Eprint {http://arxiv.org/abs/1704.07175} {arXiv:1704.07175
  [gr-qc]} \BibitemShut {NoStop}%
\bibitem [{\citenamefont {Casals}\ and\ \citenamefont
  {Nolan}(2015)}]{Casals:2015oaa}%
  \BibitemOpen
  \bibfield  {author} {\bibinfo {author} {\bibfnamefont {Marc}\ \bibnamefont
  {Casals}}\ and\ \bibinfo {author} {\bibfnamefont {Brien~C.}\ \bibnamefont
  {Nolan}},\ }\bibfield  {title} {\enquote {\bibinfo {title} {{Geometric
  properties of a 2D spacetime arising in 4D black hole physics}},}\ }\href
  {\doibase 10.1103/PhysRevD.92.104030} {\bibfield  {journal} {\bibinfo
  {journal} {Phys. Rev.}\ }\textbf {\bibinfo {volume} {D92}},\ \bibinfo {pages}
  {104030} (\bibinfo {year} {2015})},\ \Eprint
  {http://arxiv.org/abs/1509.03911} {arXiv:1509.03911 [gr-qc]} \BibitemShut
  {NoStop}%
\bibitem [{\citenamefont {Frolov}\ and\ \citenamefont
  {Novikov}(1998)}]{Frolov:1998wf}%
  \BibitemOpen
  \bibinfo {editor} {\bibfnamefont {V.~P.}\ \bibnamefont {Frolov}}\ and\
  \bibinfo {editor} {\bibfnamefont {I.~D.}\ \bibnamefont {Novikov}},\ eds.,\
  \href@noop {} {\emph {\bibinfo {title} {{Black hole physics: Basic concepts
  and new developments}}}}\ (\bibinfo {year} {1998})\BibitemShut {NoStop}%
\bibitem [{\citenamefont {Berti}\ \emph {et~al.}(2009)\citenamefont {Berti},
  \citenamefont {Cardoso},\ and\ \citenamefont {Starinets}}]{Berti:2009kk}%
  \BibitemOpen
  \bibfield  {author} {\bibinfo {author} {\bibfnamefont {Emanuele}\
  \bibnamefont {Berti}}, \bibinfo {author} {\bibfnamefont {Vitor}\ \bibnamefont
  {Cardoso}}, \ and\ \bibinfo {author} {\bibfnamefont {Andrei~O.}\ \bibnamefont
  {Starinets}},\ }\bibfield  {title} {\enquote {\bibinfo {title} {{Quasinormal
  modes of black holes and black branes}},}\ }\href {\doibase
  10.1088/0264-9381/26/16/163001} {\bibfield  {journal} {\bibinfo  {journal}
  {Class. Quant. Grav.}\ }\textbf {\bibinfo {volume} {26}},\ \bibinfo {pages}
  {163001} (\bibinfo {year} {2009})},\ \Eprint {http://arxiv.org/abs/0905.2975}
  {arXiv:0905.2975 [gr-qc]} \BibitemShut {NoStop}%
\bibitem [{\citenamefont {Capon}(1998)}]{Caponthesis}%
  \BibitemOpen
  \bibfield  {author} {\bibinfo {author} {\bibfnamefont {Rachel}\ \bibnamefont
  {Capon}},\ }\emph {\bibinfo {title} {Radiation Reaction Near Black Holes}},\
  \href@noop {} {Ph.D. thesis},\ \bibinfo  {school} {University of Wales}
  (\bibinfo {year} {1998})\BibitemShut {NoStop}%
\bibitem [{\citenamefont {Wheeler}\ and\ \citenamefont
  {Klauder}(1972)}]{wheeler1972magic}%
  \BibitemOpen
  \bibfield  {author} {\bibinfo {author} {\bibfnamefont {J.A.}\ \bibnamefont
  {Wheeler}}\ and\ \bibinfo {author} {\bibfnamefont {J.R.}\ \bibnamefont
  {Klauder}},\ }\href {https://books.google.com/books?id=TQduOwAACAAJ} {\emph
  {\bibinfo {title} {Magic Without Magic: John Archibald Wheeler: A Collection
  of Essays in Honor of His Sixtieth Birthday}}}\ (\bibinfo  {publisher} {W. H.
  Freeman},\ \bibinfo {year} {1972})\BibitemShut {NoStop}%
\bibitem [{\citenamefont {Hadar}\ and\ \citenamefont
  {Kol}(2011)}]{Hadar:2009ip}%
  \BibitemOpen
  \bibfield  {author} {\bibinfo {author} {\bibfnamefont {Shahar}\ \bibnamefont
  {Hadar}}\ and\ \bibinfo {author} {\bibfnamefont {Barak}\ \bibnamefont
  {Kol}},\ }\bibfield  {title} {\enquote {\bibinfo {title} {{Post-ISCO Ringdown
  Amplitudes in Extreme Mass Ratio Inspiral}},}\ }\href {\doibase
  10.1103/PhysRevD.84.044019} {\bibfield  {journal} {\bibinfo  {journal} {Phys.
  Rev.}\ }\textbf {\bibinfo {volume} {D84}},\ \bibinfo {pages} {044019}
  (\bibinfo {year} {2011})},\ \Eprint {http://arxiv.org/abs/0911.3899}
  {arXiv:0911.3899 [gr-qc]} \BibitemShut {NoStop}%
\bibitem [{\citenamefont {Zimmerman}\ and\ \citenamefont
  {Chen}(2011)}]{Zimmerman:2011dx}%
  \BibitemOpen
  \bibfield  {author} {\bibinfo {author} {\bibfnamefont {Aaron}\ \bibnamefont
  {Zimmerman}}\ and\ \bibinfo {author} {\bibfnamefont {Yanbei}\ \bibnamefont
  {Chen}},\ }\bibfield  {title} {\enquote {\bibinfo {title} {{New Generic
  Ringdown Frequencies at the Birth of a Kerr Black Hole}},}\ }\href {\doibase
  10.1103/PhysRevD.84.084012} {\bibfield  {journal} {\bibinfo  {journal} {Phys.
  Rev.}\ }\textbf {\bibinfo {volume} {D84}},\ \bibinfo {pages} {084012}
  (\bibinfo {year} {2011})},\ \Eprint {http://arxiv.org/abs/1106.0782}
  {arXiv:1106.0782 [gr-qc]} \BibitemShut {NoStop}%
\bibitem [{mat()}]{matex}%
  \BibitemOpen
  \href@noop {} {}\bibinfo {howpublished}
  {\url{https://github.com/szhorvat/MaTeX}}\BibitemShut {NoStop}%
\bibitem [{\citenamefont {Lousto}\ and\ \citenamefont
  {Price}(1997)}]{Lousto:1997wf}%
  \BibitemOpen
  \bibfield  {author} {\bibinfo {author} {\bibfnamefont {Carlos~O.}\
  \bibnamefont {Lousto}}\ and\ \bibinfo {author} {\bibfnamefont {Richard~H.}\
  \bibnamefont {Price}},\ }\bibfield  {title} {\enquote {\bibinfo {title}
  {{Understanding initial data for black hole collisions}},}\ }\href {\doibase
  10.1103/PhysRevD.56.6439} {\bibfield  {journal} {\bibinfo  {journal} {Phys.
  Rev.}\ }\textbf {\bibinfo {volume} {D56}},\ \bibinfo {pages} {6439--6457}
  (\bibinfo {year} {1997})},\ \Eprint {http://arxiv.org/abs/gr-qc/9705071}
  {arXiv:gr-qc/9705071 [gr-qc]} \BibitemShut {NoStop}%
\bibitem [{\citenamefont {McKechan}\ \emph {et~al.}(2010)\citenamefont
  {McKechan}, \citenamefont {Robinson},\ and\ \citenamefont
  {Sathyaprakash}}]{McKechan:2010kp}%
  \BibitemOpen
  \bibfield  {author} {\bibinfo {author} {\bibfnamefont {D.~J.~A.}\
  \bibnamefont {McKechan}}, \bibinfo {author} {\bibfnamefont {C.}~\bibnamefont
  {Robinson}}, \ and\ \bibinfo {author} {\bibfnamefont {B.~S.}\ \bibnamefont
  {Sathyaprakash}},\ }\bibfield  {title} {\enquote {\bibinfo {title} {{A
  tapering window for time-domain templates and simulated signals in the
  detection of gravitational waves from coalescing compact binaries}},}\
  }\bibfield  {booktitle} {\emph {\bibinfo {booktitle} {{Gravitational waves.
  Proceedings, 8th Edoardo Amaldi Conference, Amaldi 8, New York, USA, June
  22-26, 2009}}},\ }\href {\doibase 10.1088/0264-9381/27/8/084020} {\bibfield
  {journal} {\bibinfo  {journal} {Class. Quant. Grav.}\ }\textbf {\bibinfo
  {volume} {27}},\ \bibinfo {pages} {084020} (\bibinfo {year} {2010})},\
  \Eprint {http://arxiv.org/abs/1003.2939} {arXiv:1003.2939 [gr-qc]}
  \BibitemShut {NoStop}%
\end{thebibliography}%

\end{document}